\documentclass[12pt]{article} 
\usepackage[T1]{fontenc}
\usepackage[latin1]{inputenc}
\usepackage{epsf}
\usepackage{amsmath,amsfonts}
\usepackage{amssymb,amsfonts}
\usepackage{hyperref}
\usepackage{float,array,multirow} 
\newcommand{\irrep}[1]{\ensuremath{\boldsymbol{#1}}}
\newcommand{\ie}{i.e.\ }

\newcommand{\Tr}{{\rm Tr}\ }

\newcommand{\A}{{\cal A}}
\newcommand{\C}{{\cal C}}
\newcommand{\E}{{\cal E}}

\newcommand{\K}{{\cal K}}
\newcommand{\M}{{\cal M}}
\newcommand{\R}{{\cal R}}
\newcommand{\T}{{\cal T}}
\newcommand{\tg}{\tilde g}
\newcommand{\Zint}{\mathbb{Z}}
\newcommand{\Real}{\mathbb{R}}
\def\a{\alpha}
\def\m{\mu}
\def\n{\nu}
\def\r{\rho}
\def\k{\kappa}

\def\Ra{\mathcal{R}}
\def\ra{\rightarrow}
\def\co{: \;\;\;}

\def\limit#1#2{\smash { \mathop{#1} \limits_{#2} }  }
\newcommand{\be}{\begin{equation}}
\newcommand{\ee}{\end{equation}}
\newcommand{\ben}{\begin{enumerate}}
\newcommand{\een}{\end{enumerate}}
\renewcommand{\sp}{\; , \; \; }
\makeatletter
\renewcommand{\@makefnmark}{\mbox{$^{\ddagger\@thefnmark}$}}
\makeatother
\numberwithin{equation}{section}
\numberwithin{table}{section}

\newcommand{\publititle}[8]
{ 
  \vspace*{-3cm}
  \begin{flushright} #1 \\ {\tt #2} \end{flushright}
  \vfill
  \begin{center}{\Large
    \bfseries #3}\end{center}
  \vskip 8mm
  \begin{center}{\large #4}\end{center}
  \begin{center}{\normalsize\sl #5}\end{center}
  \vskip 8mm     
  \nopagebreak 
  \noindent #6
  \vfill 
  \begin{flushleft} #7
  \end{flushleft}
  \hrule width 6.7cm \vskip.1mm
  {\small #8}
  \thispagestyle{empty}
  \clearpage
}

\begin{document}

\publititle{CERN--TH/98-282\\CPHT--S639-0898
}
{hep-th/9809039} 
{U-duality and M-Theory}
{N.A. Obers$^{a\ast}$ and B. Pioline$^{\,b\ddagger}$ }
{$^a$Theory Division, CERN, CH-1211 Geneva 23,
\\ 
$^b$Centre de Physique Th{\'e}orique, 
Ecole Polytechnique$^\dagger$,
{}F-91128 Palaiseau
\vskip 4mm
\centerline{\it to appear in Physics Reports}
}
{
This work is intended as a pedagogical introduction to M-theory
and to its maximally supersymmetric toroidal compactifications,
in the frameworks of 11D supergravity, type II string theory 
and M(atrix) theory. U-duality is used as the main tool and
guideline in  uncovering the spectrum of BPS states.
We review the 11D supergravity algebra and elementary 1/2-BPS solutions,
discuss T-duality in the perturbative and
non-perturbative sectors from an algebraic point of view,
and apply the same tools to the analysis of U-duality
at the level of the effective action and of the BPS spectrum,
with a particular emphasis on Weyl and Borel generators.
We derive the U-duality multiplets of BPS particles and strings,
U-duality invariant mass formulae for 1/2- and 1/4-BPS states
for general toroidal compactifications on
skew tori with gauge backgrounds, and U-duality 
multiplets of constraints for states to preserve
a given fraction of supersymmetry. A number of mysterious
states are encountered in $D\le 3$, whose
existence is implied by T-duality and 11D Lorentz invariance.
We then move to the M(atrix) theory point of view, give an introduction
to Discrete Light-Cone Quantization (DLCQ) in general 
and DLCQ of M-theory in particular. We
discuss the realization of U-duality as electric--magnetic
dualities of the Matrix gauge theory, display the Matrix gauge 
theory BPS spectrum in detail, and discuss the conjectured
extended U-duality group in this scheme.
}
{CERN--TH/98-282,\ CPHT--S639-0798,\\September 1998}
{$^{\dagger}${\small Unit{\'e} mixte CNRS UMR 7644} \\
$^{\ast}${\small {\tt obers@nordita.dk};
Address after Sept. 15: Nordita, Blegdamsvej 17, DK-2100
Copenhagen.  }  \\ 
$^{\ddagger}${ {\tt pioline@cpht.polytechnique.fr} }
}
\clearpage
\enlargethispage{1cm}
\tableofcontents
\clearpage
\enlargethispage{1cm}
\listoftables
\clearpage

\section{Introduction}
\subsection{Setting the scene}
Since its invention in the late sixties, string theory has
grown up in a tumultuous history of unexpected paradigm shifts
and deceptive lulls. Not the least of these storms was the 
discovery that the five anomaly-free perturbative superstring theories
were as many glances on a single eleven-dimensional {\it theoria
incognita}, soon baptized 
M-theory, awaiting a better name
\cite{Townsend:1995kk,Witten:1995ex}. 
The genus expansion of each
string theory corresponds to a different perturbative series in a
particular limit $g_s \rightarrow 0$ in the M-theory parameter space,
much in the same way as the genus expansion arises in 't Hooft 
large-$N$, fixed-$g^2_{\rm YM}N$ regime of Yang--Mills theory
\cite{'tHooft:1974jz}. M-theory can be defined
by the superstring expansions on each patch, and the superstring (perturbative
or non-perturbative) dualities allow a translation from one patch
to another, in a way analogous to the definition of a differential
manifold by charts and transition functions. This analogy 
overlooks the fact that string theories are only defined as
asymptotic series in $g_s \rightarrow 0$, and some analyticity is
therefore required to move into the bulk of parameter space.

This definition has been effective in uncovering a number of
features of M-theory, or rather its BPS sector, which behaves
in a controlled way under analytic continuation at finite-$g_s$.
In particular, M-theory is required to contain 
Cremmer, Julia and Scherk's eleven-dimensional supergravity 
\cite{Cremmer:1978km}
in order to account for 
the Kaluza--Klein-like tower of type IIA D0-branes
as excitations carrying momentum along the 
eleventh dimension of radius $R_s\sim g_s^{2/3}$, 
as shown by Townsend and Witten
\cite{Townsend:1995kk,Witten:1995ex};
it should also contain membrane and fivebrane states, in order to reproduce 
the D2- and D4-brane, as well as the NS5-brane and the type IIA
``fundamental'' string. Which of these states is elementary is
not decided yet, although M2-branes and D0-branes are favourite 
candidates \cite{deWit:1988ig,Banks:1997vh}.
It may even turn out that none of them may be required, and that
11D SUGRA may emerge as the low-energy limit of a non-gravitational
theory \cite{Horava:1997dd}.

While the {\it dualities} between string
theories relate different languages for the same physics, the
{\it symmetries} of string theory provide a powerful guide 
into M-theory, which is believed to hold beyond the BPS sector.
The best established of them is certainly T-duality, which
identifies seemingly distinct string backgrounds with isometries
(see for instance Refs. \cite{Giveon:1994fu,Alvarez:1995dn}
and references therein).
Throughout this review, we shall restrict ourselves to
maximally supersymmetric type II or M theories, and accordingly
T-duality will reduce to the inversion of a radius on a 
$d$-dimensional torus. To be more precise, a {\it T-duality}
maps to each other type IIA and type IIB string theories compactified
on circles with inverse radii, while a {\it T-symmetry} consists
of an even number of such inversions (together with Kalb--Ramond spectral
flows to which we shall return), and therefore corresponds to
a symmetry of type II string theories and of their M-theory 
extension\footnote{Having emphasized this point, we shall 
henceforth omit the distinction between dualities and symmetries.}.
As we shall recall, such T-symmetries on a torus $T^d$ generate
a $SO(d,d,\Zint)$ discrete symmetry group, the continuous version
of which $SO(d,d,\Real)$ appears as a symmetry of the low-energy
effective action.

On the other hand, the action of 11D or type IIA supergravity compactified 
on a torus $T^d$ as well as of the equations of motion 
of uncompactified type IIB supergravity 
have for long been known to exhibit continuous non-compact global symmetries, 
namely the exceptional symmetry $E_{d(d)}(\Real)$ of Cremmer and Julia
and the $Sl(2,\Real)$ symmetry of Schwarz and West respectively 
\cite{Cremmer:1980gs,Julia:1980gr,Julia:1997cy,Schwarz:1983wa}. These
symmetries transform the scalar fields and in general
do not preserve the weak coupling regime, which puts
them out of reach of perturbation theory, in contrast
to the well established target-space T-duality.

In analogy
with the electric--magnetic $Sl(2,\Zint)$ Montonen-Olive-Sen
duality of four-dimensional $N=4$ super Yang--Mills theory
\cite{Montonen:1977sn,Sen:1994yi}, Hull and Townsend have proposed
\cite{Hull:1995mz}
that a discrete subgroup $E_{d(d)}(\Zint)$ (resp. $Sl(2,\Zint)_B$)
remains as an exact quantum symmetry of M-theory
compactified on a torus $T^d$ (resp. of ten-dimensional
type IIB string theory and compactifications thereof)\footnote{The
first example of string duality actually appeared in the
context of heterotic string theory \cite{Font:1990gx}.}.
The two statements are actually equivalent, since after
compactification on a circle the type IIB string theory becomes
equivalent under T-duality on the (say) tenth direction
to type IIA, and the symmetry $E_{d(d)}(\Zint)$ can
be obtained by intertwining the $Sl(2,\Zint)_B$ non-perturbative
symmetry with the T-duality $SO(d-1,d-1,\Zint)$. Conversely,
the $Sl(2,\Zint)_B$ symmetry of type IIB theory can be
obtained from the M-theory description as the modular group 
of the 2-torus in the tenth and eleventh
directions \cite{Schwarz:1995dk,Aspinwall:1995fw}, 
and is a particular subgroup of the modular group
$Sl(d,\Zint)$ of the $d$-torus. This, being a remnant of
eleven-dimensional diffeomorphism invariance after compactification
on the torus $T^d$, has to be an exact symmetry as soon as
M-theory contains the graviton. The T-duality symmetry
$SO(d-1,d-1,\Zint)$ is however not manifest in the 
M-theory picture.
All in all, the U-duality group reads
\begin{equation}  
\label{udgr}
E_{d(d)} (\Zint) = Sl(d,\Zint)\bowtie SO(d-1,d-1,\Zint) \ , 
\end{equation} 
where the symbol $\bowtie$ denotes the group generated by the
two non-commuting subgroups. 

The structure of the group (\ref{udgr}) will
be discussed at length in this review, and a set of Weyl
and Borel generators will be identified. The former
preserve the rectangularity of the torus and the vanishing
of the gauge background, while the latter allow a move
to arbitrary tori. States are classified into
representations of the U-duality group $E_{d(d)}(\Zint)$, whether BPS or not,
and we will derive U-duality invariant mass and tension
formulae for 1/2- and 1/4-BPS states, as well as conditions
for a state to preserve a given fraction of the supersymmetries.
Besides the entertaining encounter
with discrete exceptional groups, this will actually teach
us about the spectrum of M-theory, since the
more M-theory is compactified, the more degrees of freedom
come into play. In particular, we will show the need to include
states with masses that behave as $1/g_s^n,\ n\ge 3$, which are
unconventional in perturbative string theory. 
An important application of these results is the exact determination of 
certain physical amplitudes in M-theory, such as the four-graviton
$R^4$ coupling, which can be interpreted as traces over M-theory BPS states
\cite{Bachas:1996bp,Green:1997as,Pioline:1997ix}. The weak coupling analysis
of these exact couplings provides a very useful insight into the
rules of semi-classical calculus in string theory 
\cite{Becker:1995kb,Ooguri:1996me,Green:1997tv,
Green:1997di,Bachas:1997mc,Kiritsis:1997hf,Pioline:1997pu}.

A proposal has recently been put forward by Banks, Fischler, Shenker and
Susskind to define M-theory
{\it ab initio} on the (discrete) light front, as the large-$N$ 
limit of a supersymmetric matrix model given by the dimensional  
reduction of 10D $U(N)$ super Yang--Mills (SYM) theory to $0+1$ dimension
\cite{Banks:1997vh,Susskind:1997cw}. This model
also describes low-energy interactions of D0-branes induced
by open string fluctuations 
\cite{Danielsson:1996uw,Kabat:1996cu,Douglas:1997yp}, and, as shown by 
Seiberg, arises from
considering M-theory on the light front as a particular 
limit of M-theory compactified on a circle, \ie type IIA
theory \cite{Seiberg:1997ad}. 
D0-branes are therefore identified as the partons of
M-theory in this framework. This proposal has passed numerous
tests, and has been shown to incorporate membrane and
(transverse) fivebrane solutions, and to reproduce 11D SUGRA
computations. 
The invariance under eleven-dimensional Lorentz
invariance remains, however, to be demonstrated
(see \cite{Lowe:1998wu} for a step in that direction).

Upon compactification of
$d$ dimensions, the D0-branes interact by open strings wrapped 
many times around the compactification manifold, and the
infinite-dimensional quantum mechanics can be rephrased as a 
gauge theory in $d+1$ dimensions
\cite{Taylor:1997ik,Ganor:1997zk}. This dramatic increase of
degrees of freedom certainly removes part of the appeal of the
proposal, but becomes even more serious for $d \ge 4$, where the 
gauge theory loses its asymptotic freedom and becomes ill-defined
at small distances. We will briefly discuss the proposals 
for extending this definition to $d=4,5$.
We will also discuss the interpretation of M-theory BPS states
in the gauge theory, and show the occurrence of
unconventional states with energy $1/g_{\rm YM}^{2n},\ n\ge 2$. 
Despite these difficulties, 
the Matrix gauge theory gives a nice understanding of
U-duality as the electric--magnetic duality of the gauge theory,
together with the modular group of the torus on which it lives
\cite{Susskind:1996uh,Ganor:1997zk,Elitzur:1997zn}.
The interpretation of finite-$N$ matrix theory as the compactification
of M-theory on a light-like circle implies that the U-duality group
$E_{d(d)}(\Zint)$ be enlarged to $E_{d+1(d+1)}(\Zint)$~
\cite{Hull:1997jc,Blau:1997du,Obers:1997kk,Hull:1998kb}; we will show
that this extra symmetry mixes the rank $N$ of the gauge group
with charges in a way reminiscent of Nahm duality. All these features
are guidelines for a hypothetical fundamental definition
of Matrix gauge theory.

\subsection{Sources and omissions}
This review is intended as a pedagogical introduction to M-theory,
from the point of view of its 11D SUGRA low-energy limit, its
strongly coupled type II string description, and its purported M(atrix)
theory definition. It is restricted to maximally supersymmetric
toroidally compactified M-theory, and uses U-duality as the main tool 
to uncover the part of the spectrum that is annihilated by half 
or a quarter of the
32 supersymmetries. The exposition mainly follows 
\cite{Elitzur:1997zn,Pioline:1997pu,Obers:1997kk},
but relies heavily on \cite{Witten:1995zh,Witten:1995ex,Townsend:1997wg,
Seiberg:1997ad,Dijkgraaf:1997hk}. 
It is usefully supplemented by other presentations
on supergravity solutions \cite{Stelle:1996tz,Townsend:1996xj,
Townsend:1997wg},
M(atrix) theory \cite{Banks:1997mn,Bigatti:1997jy,Dijkgraaf:1997ku},
D-branes \cite{Polchinski:1996na,Russo:1997ez,Taylor:1997dy,Bachas:1998rg},
string dualities \cite{Douglas:1996vj,Schwarz:1996bh,Vafa:1997pm,
deWit:1997sz,Nicolai:1997hi,Sen:1998kr}
and perturbative string theory \cite{Ooguri:1996ik,Kiritsis:1997hj}
and general introductions \cite{Pioline:1998qq,Argurio:1998cp,Schwarz:1998mm}.
The following topics are beyond the scope of this work:
\begin{itemize}
\item{Black hole entropy}: the modelisation of extremal black-holes by
D-brane bound states has allowed a description of their microscopic
degrees of freedom and a derivation of their Bekenstein--Hawking
entropy \cite{Strominger:1996sh} (see
\cite{Maldacena:1996ky,Peet:1997es} for reviews). 
The latter can be related to a U-duality invariant 
of the black hole charges
\cite{Horowitz:1996ac,Cvetic:1996zq,Kallosh:1996uy,
Andrianopoli:1997hb,Andrianopoli:1997wi}, 
and U-duality can even allow the 
control of non-extremal states \cite{Sfetsos:1998xs}.

\item{Gauge dynamics}: the study of D-brane configurations has also
led to a qualitative understanding of gauge theories dynamics
as world-volume dynamics of these objects; see
\cite{Giveon:1998sr} for a thorough review. We will mainly consider
configurations of parallel branes, as describing the Matrix
gauge theory description of M-theory on the light cone.

\item{BPS-saturated amplitudes}: A special class of terms 
in the effective actions of M-theory and string theory 
receives contributions from BPS states only. We will briefly discuss
an application of the M-theory mass formulae that we derived
to the computation of exact $R^4$ couplings in Subsection \ref{rfour},
and refer to the existing literature for more details
on the exact non-perturbative computation of these couplings,
and their interpretation at weak coupling as a sum of
instanton effects. Relevant references include 
\cite{Becker:1995kb,Ooguri:1996me,Seiberg:1996ns} for two-derivative
terms in $N=2$ type II strings, \cite{Harvey:1996ir,Harvey:1996gc,
Gregori:1997hi,Antoniadis:1997zt} for 
four-derivative terms in type II/heterotic theories,
\cite{Bachas:1996bp,Bachas:1997mc,Kiritsis:1997hf}, for
$F^4$ (and related) terms in type I/heterotic theories,
and \cite{Green:1997tv,Green:1997di,Green:1997as,Kiritsis:1997em,
Antoniadis:1997zt,Berkovits:1997pj,Kehagias:1997cq,Pioline:1997pu,
Pioline:1997ix,Green:1998me,Green:1998by} for 
$R^4$ (and related) terms in type IIB/M-theory. Infinite series
of higher-derivative BPS-saturated $R^2 F^{2g-2}$ or $R^4 H^{4g-4}$, 
and $R^{n}$ terms
have also been computed or conjectured in Refs.
\cite{Antoniadis:1995zn,Marino:1998pg} and
\cite{Berkovits:1998ex,Russo:1997mk,Russo:1997fi,Kehagias:1997jg}.

\item{Scattering amplitudes}: in order to validate the M(atrix)
theory conjecture of BFSS, a number of scattering computations have been
carried out both in the Matrix model and in 11D supergravity; they
have shown agreement up to two loops, see for instance 
\cite{Becker:1997wh,
Becker:1997xw,Polchinski:1997pz,Chepelev:1997fk,Dine:1997sz,
Keski-Vakkuri:1997wr,Taylor:1998yt,Okawa:1998pz,McCarthy:1998uw,Fabbrichesi:1998qn}.
This agreement is better than naively expected, and indicates
the existence of non-renormalization theorems 
\cite{Paban:1998qy} for these interactions.

\item{D-instanton matrix model}: an alternative formulation of
M-theory as a statistical matrix model 
has been proposed by Ishibashi, Kawai, Kitazawa
and Tsuchiya \cite{Ishibashi:1996xs}. It has so far not
been developed to the same extent as the BFSS proposal,
and in particular the origin of U-duality has not been
explicited. See Refs. \cite{Chepelev:1997ug,Fayyazuddin:1997zx,
Ho:1998rd,Fayyazuddin:1997yf,Sathiapalan:1997ik,Kristjansen:1997mx,
Chekhov:1997cs,Chepelev:1997av,Fukuma:1997en,Kitsunezaki:1997iu,
Biswas:1998be,Mandal:1997kj,Suyama:1998ig,Aoki:1998vn,Ambjorn:1998ba,
Fukuma:1998jc} for further discussion.

\item{Twelve dimensions and beyond}: the structure of U-duality
symmetry has led to speculate on the existence of a 12D
\cite{Vafa:1996xn,Bars:1996dz,Bars:1997bz,Nishino:1996wp} or higher
\cite{Bars:1997ug,Bars:1997ab,Bars:1996hr,Nishino:1997hk,Sezgin:1997gr,
Rudychev:1997ui,Nishino:1998qn}
dimensional parent of M-theory,
with extra time directions. The $N=2$ heterotic strings suggest
an appealing construction of this theory (see \cite{Martinec:1997cw} 
for a review).
However, the full
higher-dimensional Lorentz symmetry is partially reduced
to its U-duality subgroup, and its usefulness remains
unclear at present. We shall, however, encounter in Subsection
\ref{wgwru} a tantalizing hint for an 
extra time-like direction with ``length'' $l_p^3$.

\item{String networks}: a construction of 1/4-BPS states based
on three-string junctions 
\cite{Schwarz:1996bh,Callan:1997kz,Dasgupta:1997pu}
has been suggested \cite{Sen:1997xi},
that reproduces the U-duality invariant mass formula in 8 dimensions 
\cite{Bhattacharyya:1998vr,Bhattacharyya:1998vr,Kumar:1998te}. 
These solutions have been constructed
from the M2-brane \cite{Krogh:1998dx,Matsuo:1998jw}
and their dynamics discussed in \cite{Rey:1997sp,Callan:1998sf}, 
but their supergravity 
description is still unclear.

\item{Non-commutative geometry}: it has been argued that
non-commutative geometry \cite{Connes:1994} is the appropriate framework to
discuss D-brane dynamics, and is even required in the presence
of Kalb--Ramond two-form background
\cite{Connes:1998cr,Douglas:1998fm,Cheung:1998nr}.
This description incorporates T-duality 
\cite{Reiffel:1998,Schwarz:1998qj} and even U-duality
in its Born--Infeld generalization \cite{Hofman:1998pd}. 
It should in particular (see Subsection \ref{nahm}) extend
Nahm's duality of ordinary two-dimensional Maxwell theory
to higher-dimensional cases \cite{Nekrasov:1998ss}.
Related discussions can be found in Refs.
\cite{Ho:1998xh,Li:1998ks,Kawano:1998re,Berkooz:1998st,Morariu:1998qm,
Leigh:1998jk,Connes:1998qv}.

\item{Gauged supergravity}: 11D SUGRA possesses 
maximally supersymmetric backgrounds other than tori, namely
compactifications on products of spheres and anti-deSitter
spaces \cite{Freund:1980xh}. These correspond to the near-horizon
geometry of M2- and M5-branes, and have been argued to provide
a dual description to the gauge theory on these extended objects
\cite{Maldacena:1997re}. They will be ignored in this review.

\item{String theories with non maximal supersymmetry}: 
the $E_8 \times E_8$ heterotic string and type I string
can be obtained from M-theory by
orbifold compactification \cite{Horava:1996ma,Horava:1996qa}, while
the $SO(32)$ heterotic string is related to $E_8 \times E_8$ by a T-duality,
or to type I string theory by a non-perturbative duality 
\cite{Polchinski:1996df}. M(atrix) theory descriptions have been
proposed both in the heterotic
\cite{Banks:1997it,Motl:1997tb,Lowe:1997sx,Kim:1997uv,Rey:1997hj,Horava:1997ns,
Govindarajan:1997iw,Kabat:1997za,Lowe:1997uf,
Krogh:1998vb,Krogh:1998rw,Diaconescu:1998cy}
and the type II 
\cite{Fischler:1997ts,Kim:1997gh,Berkooz:1997pe,
Govindarajan:1997ba,Kim:1997qc,
Diaconescu:1998rs,Diaconescu:1998cy}
cases, as well as on non-orientable surfaces
\cite{Kim:1997aj,Zwart:1998kr}, and will not be treated here.

\item{Non-BPS states}: The study of stable non-BPS states has
been iniated in \cite{Sen:1998rg} and further examined 
\cite{Sen:1998ii,Sen:1998sm,Sen:1998tt,Bergman:1998xv}.
It would be interesting to investigate the implications of U-duality
symmetry on the spectrum of non-BPS states.

\end{itemize}
\vskip -5mm
\subsection{Outline}
\enlargethispage{1cm}
Section \ref{mtheory} introduces the superalgebra and fundamental BPS states of M-theory
in the context of 11D SUGRA and type IIA/B superstring theories.
T-duality is recalled and revisited in Section \ref{tdua} from an algebraic
point of view, at the level of the effective action and of the
perturbative and non-perturbative BPS spectrum.
The same techniques are used in Section \ref{udua} to
introduce U-duality and its action on the spectrum of particles
and strings, restricting to Weyl generators.
Borel generators are incorporated in Section \ref{sskew}, where
U-duality invariant mass and tension formulae
for general toroidal compactification
with arbitrary gauge backgrounds are derived,  
as well as U-duality multiplets of BPS constraints.
Section \ref{smgt} introduces Matrix gauge theory as the Discrete
Light-Cone Quantization of M-theory following an argument
by Seiberg, and discusses the dictionary between M-theory and 
Matrix gauge theory. The U-duality symmetry is finally discussed 
in Section \ref{usymmat}
from the perspective of the Matrix gauge theory, as 
well as the extended U-duality symmetry arising from
the extra light-like direction.

\clearpage
\section{M-theory and BPS states \label{mtheory}}
\subsection{M-theory and type IIA string theory  }

M-theory was originally introduced as the strong coupling limit of 
type IIA superstring theory. The latter has been argued 
\cite{Witten:1995zh,Townsend:1995kk} to dynamically
generate an extra compact dimension at finite coupling of radius 
$R_s\sim g_s^{2/3}$
in units of an eleven-dimensional Planck length $l_p$:
\begin{equation}
\label{matching}
R_s/l_p = g_s^{2/3}\ ,\quad l_p^3 = g_s l_s^3
\end{equation}
where $1/l_s^2=\alpha'$ denotes the string tension and
$g_s$ its coupling constant. The strong coupling limit
$g_s\rightarrow\infty$ should  therefore 
exhibit eleven-dimensional $N=1$ super-Poincar{\'e} invariance. 

While a consistent eleven-dimensional theory of quantum gravity
is still missing, it has been known for a long time
that type IIA supergravity can
be obtained from the eleven-dimensional $N=1$ 
supergravity of Cremmer, Julia and Scherk, by dimensional reduction
on a circle. M-theory is therefore required to reduce at energies
much smaller than $1/l_p$ to 11D SUGRA, in the same way
as type IIA (or type IIB) superstring theory reduces to 
type IIA \cite{Campbell:1984zc,Huq:1985im,Giani:1984wc} 
(or type IIB \cite{Schwarz:1983wa,Howe:1984sr,Schwarz:1983qr}) supergravity
at energies much smaller than $1/l_s$ (which is also smaller than
both the ten-dimensional Planck mass $g_s^{-1/4}/l_s$ 
and the eleven-dimensional Planck mass $g_s^{-1/3}/l_s$ at weak
coupling). This is summarized in the following diagram:
\begin{center} 
\begin{tabular}{ccc}
\label{diag} 
M-theory  & $\limit{\longrightarrow}{S^1}$ & type IIA string theory
\\
$\downarrow$   &  & $\downarrow$
\\
11D supergravity  & $\limit{\longrightarrow}{S^1}$ & 10D 
type IIA supergravity
\\
\end{tabular}
\end{center} 
where compactification on a circle occurs from left to right
and the energy decreases from top to bottom.

The matching relations \eqref{matching} can be easily obtained by
studying the Kaluza--Klein reduction of 11D SUGRA, described by
the action 
\begin{equation} 
\label{11da} 
S_{11} = \frac{1}{l_p^9} \int {\rm d }^{11} x
 \sqrt{-g} \left(R - \frac{l_p^6}{48} (d\C)^2\right)
+ \frac{\sqrt{2}}{2^7 \cdot 3^2} \int  \C \wedge d\C \wedge d\C
\end{equation}
up to fermionic terms that we will ignore in the following.
In addition to the usual Einstein--Hilbert term involving the
scalar curvature  $R$ of the metric $g_{MN}$, the action 
contains a kinetic term for the 3--form gauge potential $\C_{MNR}$
(which we shall often denote by $\C_3$)
as well as a topological Wess--Zumino term required by supersymmetry. 
The action \eqref{11da} does not contain any dimensionless parameter,
and the normalization of the Wess--Zumino term with respect to the
Einstein term is fixed by supersymmetry. 
The dependence
on the Planck length $l_p$ has been reinstated by dimensional analysis,
with the following conventions:
\begin{equation}
[{d}x]=0,\ [g_{MN}]=2,\ [\sqrt{-g}]=11,\ [R]=-2,\ [\C_{MNR}]=0,\ [d]=0\ ,
\end{equation} 
relegating the dimension to the metric only.
In particular, the relation between the eleven-dimensional Planck
length and Newton's constant is given by $ \k^2_{11} =  l_p^9/(2 (2 \pi)^8)$. 
We will generally ignore all numerical factors.

Dimensional reduction is carried out by substituting an ansatz
\begin{equation}  
\label{u1f}
{d}s_{11}^2 = R_{s}^2  ( {d} x^s + \A_{\mu} {d} x^\mu )^2
+ {d} s_{10}^2 
\end{equation}  
for the metric, where $R_{s}$ stands for the fluctuating radius of
compactification (as measured in the eleven-dimensional metric)
and $\A$ describes the Kaluza--Klein $U(1)$ gauge field arising
from the isometry along $x^s$ \footnote{The subscript $s$ is used to indicate
that $s$tring theory is obtained in this way.},
and splitting the three-form $\C_{MNR}$ in a two-form
$B_{\mu\nu}=\C_{\mu\nu s}$ and a 3-form $\C_{\m\n\r}$.
Only the zero Fourier component (\ie the zero Kaluza--Klein
momentum part) of these fields along $x^s$ is kept. 
On dimensional grounds the scalar curvature becomes
\begin{equation}
R(g_{MN}) = R(g_{\mu\nu}) + \left(\frac{\partial R_s}{R_s}\right)^2
+ R_s^2 (d\A)^2\ ,
\end{equation}
so that the reduced action reads
\begin{multline}
\label{11dkk}
S_{10} = \frac{1}{l_p^9} \int {\rm d }^{10} x
 R_s \sqrt{-g} \left[R + \left(\frac{\partial R_s}{R_s}\right)^2
+ R_s^2 (d\A)^2 + l_p^6 (d\C)^2 + \frac{l_p^6}{R_s^2} (dB)^2 \right]
\\+ \int B \wedge d\C \wedge d\C\ .
\end{multline}
On the other hand, the low-energy limit of type IIA string theory is
given (in the string frame) by the action 
\begin{multline} 
\label{10da}
S_{{\rm IIA}} = \frac{1}{l_s^8}\int {d}^{10} x \sqrt{-g} \left[ 
e^{- 2 \phi} \left(
R + 4(\partial \phi )^2  - \frac{l_s^4 }{12} (d B)^2 \right) 
  - \frac{l_s^2}{4 } (d \A)^2 - \frac{l_s^6}{48} (d \C)^2
\right]
\\
+ \int B \wedge d\C \wedge d\C\ ,
\end{multline} 
which describes the dynamics of the (bosonic) massless sector
\begin{subequations}
\begin{eqnarray} 
\label{nsf}
\mbox{NS-NS}\ &  :\ &    g_{\m \n} \sp B_{\m \n} \sp \phi \\  
\mbox{R-R} \ &  :\ & \A_\mu \sp \C_{\m\n\r}
\end{eqnarray} 
\end{subequations}
denoting the metric, antisymmetric tensor and dilaton from the
Neveu--Schwarz square sector, and the one- and three-form gauge
potentials from the Ramond square sector 
(indices $\mu $ run over  $1 \ldots 10$).
Ramond $p$-form gauge fields  will be generically denoted by $\mathcal{R}_p$.
The dependence on the string length $l_s$ is again instated
on dimensional grounds, while the dependence on the coupling 
\begin{equation} 
\label{strc}
g_s^2 = e ^{2 \phi} 
\end{equation} 
stems from the fact that the two-derivative action originates
from string tree level (hence the $e ^{-2\phi}$ factor), with
each Ramond field coming with an additional power of $e ^{\phi}$,
ensuring the correct Maxwell and Bianchi identities 
(see \cite{Polyakov:1996bn} for a recent discussion).
In particular, the ten-dimensional Newton's constant is given by
$\kappa_{10}^2 =  g_s^2 l_s^8$. Identifying the dilaton $\phi$
with the scalar modulus $\ln R_s$ up to a numerical factor,
and matching the two actions
\eqref{11dkk} and \eqref{10da} leads to the relations
\begin{equation} 
\frac{ R_{s}}{l_p^9} = \frac{1}{g_s^2 l_s^8}
\sp \;\;\; 
\frac{ 1}{R_{s} l_p^3} = \frac{1}{g_s^2 l_s^4}
\sp \;\;\; 
\frac{ R_{s}^3}{l_p^9} = \frac{1}{l_s^6}
\sp \;\;\; 
\frac{ R_{s}}{l_p^3} = \frac{1}{l_s^2}\ ,
\end{equation} 
obtained by comparing the terms $R$, $dB$, $d\A$ and $d \C$
respectively in Eqs. \eqref{11dkk} and \eqref{10da}.
Two of these four relations turn out to be redundant 
as a consequence of supersymmetry,
and they can be reduced to the matching relations 
already stated in Eq. \eqref{matching},
or equivalently
\begin{equation}  
\label{msr}
R_s = l_s g_s 
\sp
\frac{R_s}{l_p^3} = \frac{1}{l_s^2} \ ,
\end{equation} 
which summarize the relation between the M-theory parameters $\{l_p,R_s\}$
and the string theory parameters $\{l_s,g_s\}$.

Using \eqref{msr} and \eqref{strc}
 in the metric \eqref{u1f} we find the alternative form
\begin{equation}
\label{u1fb}
\frac{{d}s_{11}^2}{l_p^2} 
 = e^{4 \phi/3} 
( {d} x^s + \A_{\mu} {\rm d} x^\mu )^2
+ e^{-2 \phi/3} \frac{{\rm d} s_{10}^2 }{l_s^2} \ ,  
\end{equation}  
which will be  used to relate low-energy solutions of M-theory
and type IIA string theory.

\subsection{M-theory superalgebra and BPS states\label{bpss}} 

While M-theory has to reduce to 11D SUGRA in the low-energy limit,
little is known about its microscopic degrees of freedom.
It is however postulated that the $N=1$ supersymmetry of 11D SUGRA
should be valid at any energy, and the spectrum is therefore
organized into representations of the super-Poincar{\'e} 
algebra \cite{Townsend:1995gp}:
\begin{subequations}
\label{susy}
\begin{eqnarray}
\label{susy1}
\left\{ Q_\alpha, Q_\beta \right\} 
&=& (C\Gamma^{M})_{\alpha\beta} Z_M
+ \frac{1}{2} (C\Gamma_{MN})_{\alpha\beta} Z^{MN} \\
&&+ \frac{1}{5!} (C\Gamma_{MNPQR})_{\alpha\beta} Z^{MNPQR} \nonumber\\
\label{susy2} 
\left[ Q_\alpha , Z^{M\dots} \right] &=& 0\ .
\end{eqnarray}
\end{subequations}
Here $Q_\alpha$ denotes the 32-component Majorana spinor
generating the supersymmetry (see \cite{Kugo:1983bn} for
a general account on spinorial representations), and 
$\Gamma_{MN\dots}$ the
antisymmetric product of $\Gamma$ matrices, 
\ie $\Gamma_M \Gamma_N \dots$ for distinct indices and zero otherwise. 
See Appendix \ref{gamma} for our gamma matrix conventions.

In addition to the usual translation operator $P_M$, which
we denoted by $Z_M$ for uniformity, the right-hand
side of Eq. \eqref{susy1} contains  ``central charges'' $Z^{MN}$
and $Z^{MNRST}$ in non-trivial representations of the Lorentz group.
These charges appear as irreducible representations (irreps) 
in the decomposition $528 = \irrep{11}+ \irrep{55} + \irrep{462}$ 
of the symmetric tensor product $\{Q_\alpha,Q_\beta\}$, and
the simplest assumption is that they should commute with
the SUSY charges $Q_\alpha$ \footnote{It is possible to introduce
non-Abelian relations while still preserving the Jacobi identity
\cite{Sezgin:1997cj},
but the status of this possibility is still unclear.}
(their commutation properties with
the Lorentz generators are encoded in their index structure).
They can be identified as the electric and magnetic charges
of extended objects \cite{Hughes:1986dn,deAzcarraga:1989gm}
with respect to the gauge potential $\C_{MNP}$ and the metric
$g_{MN}$ and their Kaluza--Klein descendants. 

The various components
of the central charges, their corresponding potentials, as well as the
nature of the solution, are 
summarized
in Table \ref{zcharge}. Here, $\E_{6}$ denotes the six-form
dual to $\C_{3}$ and $\K_{I;IMNPQRST}$ the 7-form\footnote{No
antisymmetry is assumed for indices separated by a semi-colon.
The peculiar index structure $\K_{1;8}=\K_{I;IMNPQRST}$ 
ensures that the seven-form indices $M,\dots,T$ are distinct from
the compact direction $I$, and the double occurrence of $I$ has the
same origin as the square radius $R^2$ in Eq. \eqref{11dt}.}
dual to the Kaluza--Klein gauge potential $g_{IM}$ after 
compactifying the direction $I$. This hints toward the existence
of extended states charged under these gauge fields, namely
2-branes, 5-branes, 6-branes and 9-branes. 
The 9-branes, which are not charged under a gauge potential, are not
dynamical and correspond to the ``end-of-the-world'' branes in
compactifications of M-theory with lower supersymmetry
\cite{Horava:1996ma}. They will not be further considered in this review,  
but we will shortly return to the M2, M5 and KK6-brane. 

\begin{table}[h] 
\begin{center}
\begin{tabular}{|l||l|l|l|l|l|}
\hline 
$Z_0$ & $Z_I$ & $Z^{IJ}$ & $Z^{IJKLM}$ & $Z^{0I}$ & $Z^{0IJKL}$ \\
\hline
$g_{00}$ & $g_{0I}$ & $\C_{0IJ}$ & $\E_{0IJKLM}$ & none &
$\K_{M;MNPQRST
0}$ \\
\hline
mass & momentum & M2-brane & M5-brane & 9-brane  & KK6-brane \\
\hline
\end{tabular}
\end{center} 
\caption{M-theory central charges, gauge fields and extended objects. 
\label{zcharge}}  
\end{table}

The generic representation of the superalgebra \eqref{susy} 
is generated by the action of 16
fermionic creation operators on a vacuum
$|0\rangle$ in a given representation of the Lorentz
group; it is therefore $2^{16}$-dimensional,
\ie contains 32768 bosonic states and 32768 fermionic states.
The positivity of the matrix $\langle 0 | \{ Q_\alpha,Q_\beta \} | 0
\rangle$ implies a bound on the rest mass $Z_0$ known as the 
Bogomolny bound. When this bound is saturated, part of the supersymmetries
annihilate the vacuum $|0\rangle$:
\begin{equation}
\label{bps}
\sum_Z Z^{MN\dots} (C\Gamma_{MN\dots})^{\a\beta} Q_\beta | 0 \rangle =
0\ ,
\end{equation}
resulting in a reduced degeneracy. 
Equation \eqref{bps} requires that the $32\times 32$ matrix
$Z^{MN\dots} (\Gamma_{MN\dots})^{\a\beta}$ has at least one
zero eigenvalue, and implies in particular 
the BPS condition
\begin{equation}
\det_{\alpha\beta} \left( \sum_Z (Z\cdot \Gamma)^{\a\beta} \right)=0\ ,
\end{equation}
which determines the rest mass $Z_0$ in terms of the other charges.

The dimension can be further reduced if the zero eigenvalue 
is degenerate, and this requires more relations between
the various charges. Since only $Z_0$ contributes to the trace on  
the right-hand side of Eq. \eqref{susy1}, the maximum number of
zero eigenvalues is 16, corresponding to a state annihilated
by half the supersymmetries, or in short a 1/2-BPS state.
Because of its reduced dimension,
a BPS state with smallest charge cannot decay, except if
it can pair up with another state of opposite charge to make a representation
twice as long \cite{Witten:1978mh}. These states can therefore be followed
at strong coupling (in the M-theory language, this means 
for arbitrary geometries of the compactification manifold)
and serve as the basis for many duality checks.

As an illustration, we wish to investigate the case where, besides
the mass $\M=Z_0$, only the two-form central charges $Z^{IJ}$ 
do not vanish. This will be later interpreted as an arbitrary superposition of
M2-branes.  We therefore have to solve the eigenvalue equation:
\begin{equation}
\label{eig1}
\Gamma \epsilon = \M \epsilon \sp \;\;\; 
\Gamma \equiv  Z^{IJ} \Gamma_{0IJ} 
\end{equation}
Squaring this equation yields
\begin{equation}
\label{eig2}
\Gamma ^2 =  Z^{IJ} Z^{IJ} + Z^{IJ} Z^{KL} \Gamma_{IJKL} \circeq \M^2
\ ,
\end{equation}
where the symbol $\circeq$ denotes the equality when acting on
$\epsilon$. 
The space of solutions now depends on the value of 
$k^{IJKL}\equiv Z^{[IJ} Z^{KL]}=Z \wedge Z$. If $k=0$, Eq. \eqref{eig2}
implies $(Z^{IJ})^2=\M^2$ and $\Gamma^2=\M^2$. Since $\Tr \Gamma=0$,
the $32\times 32$ matrix $\Gamma$ has 16 eigenvalues $\M$ and
16 eigenvalues $-\M$, and therefore Eq. \eqref{eig1} is satisfied for
a dimension-16 space of vectors $\epsilon$. The state with
charges $Z^{IJ}$ is therefore annihilated by half the supersymmetry
generators $Q_\alpha$, and has a mass
\begin{equation}
\label{bps2}
\M_0^2 = Z^{IJ} Z^{IJ}\ .
\end{equation}
The condition $Z\wedge Z=0$ means that the antisymmetric 
charge matrix $Z^{IJ}$ has rank 2, \ie that only parallel
M2-branes are superposed.

If on the other hand $Z\wedge Z\ne 0$, we may rewrite Eq. \eqref{eig2} as
\begin{equation}
\label{eig3}
\Gamma' \epsilon = \left(\M^2 - \M_0^2\right) \epsilon \sp
\Gamma' =  k^{IJKL} \Gamma_{IJKL} \ ,
\end{equation}
and we are lead back
to an equation similar to Eq. \eqref{eig1}. Squaring again yields
\begin{multline}
\label{eig4}
\Gamma^{'2} =  (k^{IJKL})^2 + (k\cdot k)^{IJKL} \Gamma_{IJKL} 
+ (k\wedge k)^{IJKLMNPQ} \Gamma_{IJKLMNPQ}  \\ \circeq  \left(
\M^2-\M_0^2\right)^2 \ , 
\end{multline}
where $(k\cdot k)^{IJKL}=k^{IJMN}k^{KLMN}$. As before,
if $k\cdot k=k \wedge k=0$, this equation implies
$(k^{IJKL})^2=(\M^2-\M_0^2)^2=\Gamma^{'2}$. Since $\Tr \Gamma'=0$,
Eq. \eqref{eig3} is satisfied by half the supersymmetries,
but Eq. \eqref{eig1} by a quarter only. We therefore get
a 1/4-BPS state with mass squared:
\begin{subequations}
\begin{eqnarray}
\label{bps4}
\M^2 = Z^{IJ} Z^{IJ} + \sqrt{ k^{IJKL} k^{IJKL} } \\
k^{IJKL} = Z^{[IJ} Z^{KL]}\ .
\label{bps2c}
\end{eqnarray}
\end{subequations}
This expression reduces to Eq. \eqref{bps2} for a 1/2-BPS
state, \ie when $k^{IJKL}=0$. On the other hand, if
$k\cdot k$ or $k\wedge k\ne 0$ do not vanish, the state is 
at most 1/8-BPS and we have to carry the same analysis one
step further. Note that the conditions
$k\cdot k\ne 0$ (resp. $k\wedge k\ne 0$) can only be satisfied
when $d\geq 6$ ( resp. $d\geq 8$), in agreement with the
absence of 1/8-BPS states in more than five space-time dimensions.

\subsection{BPS solutions of 11D SUGRA}
In want of a microscopic formulation of M-theory (or of
non-perturbative type IIA string theory), it is certainly
difficult to determine what representations of the
eleven-dimensional Poincar{\'e} superalgebra actually
occur in the spectrum. However, this is achievable for
BPS states, since supersymmetry protects these from quantum effects
and in particular determine their exact mass formula.
They can be studied at arbitrarily low energy, and in particular
in the 11D SUGRA limit of M-theory.
Instead of describing the equations implied by
the BPS condition on the supergravity configuration,
we refer the reader to existing reviews in the literature
\cite{Duff:1995an,Duff:1996zn,Townsend:1996xj,Stelle:1996tz,Townsend:1997wg},
and content ourselves with recalling the four 1/2-BPS 
standard solutions:
the $pp$-wave and three extended solutions, the 
membrane (or M2-brane), fivebrane (M5-brane) and the Kaluza--Klein monopole,
also known as the KK6-brane. 

The eleven-dimensional metric describing the extended solutions splits
into two parts: the world-volume, denoted by $E^{1,p}$, including
the time and $p$ world-volume directions, 
and the transverse Euclidean part $E^{10-p}$.  
These solutions are given in terms of a harmonic function $H$
on the transverse space, which we choose as a single pole
\begin{equation}
\label{harm}
H(r) = 1 + \frac{k}{r^{8-p}}\ ,
\end{equation}
although any superposition of such poles would do (this is 
stating the no-force condition between static BPS states;
the constant shift in Eq. \eqref{harm} ensures the asymptotic
flatness of space-time, required for a soliton interpretation).
The constant $k$ depends
on Newton's constant $\kappa_{11} $ and on the $p$-brane tension,
and is quantized by the requirement that the space-time be
smooth (we will henceforth choose the smaller quantum).

The $pp$-wave\footnote{The name {\it $pp$-wave} stands for
plane fronted wave  with parallel rays \cite{Brinkmann:1923}.
The solution \eqref{ppw} was generalized in 
\cite{Bergshoeff:1995as} to include excitations of
the three-form potential.} and KK6-brane solutions only involve the metric,
and read \cite{Hull:1984vh,Townsend:1995kk}
\begin{subequations}
\label{ppw} 
\begin{equation}
\mbox{pp-wave} \co {\rm d}s_{11}^2 = 
-{\rm d}t^2 + {\rm d}\rho^2 + (H-1) ({\rm d}t + {\rm d}\rho)^2 + 
 {\rm d}s^2 (E^9) 
\end{equation} 
\begin{equation} 
H= 1 + \frac{k}{r^7}
\end{equation}
\end{subequations}
\begin{subequations}
\label{kk6}
\begin{equation} 
\mbox{KK6-brane} \co {\rm d}s_{11}^2 = 
{\rm d}s ^2 (E^{1,6})  + {\rm d}s_{\rm TN}^2 (y) 
\end{equation} 
\begin{equation} 
{\rm d}s_{\rm TN}^2  = H {\rm d}  y^i {\rm d} y^i +  
 H^{-1} \left({\rm d} \psi_{\rm TN} + V_i(y) {\rm d} y^i \right)^2  
\sp i =1,2,3 
\end{equation} 
\begin{equation} 
\nabla \times V = \nabla \cdot H 
\sp
H = 1 + \frac{k}{|y|} \ .
\end{equation} 
\end{subequations}
The KK6-brane solution is analogous to the five-dimensional
Kaluza--Klein monopole \cite{Sorkin:1983ns}, and is
built out from the four-dimensional Taub--NUT
gravitational instanton (see Ref. \cite{Eguchi:1980jx} for a review
of this topic), which is asymptotically of the form $\Real^3 \times S^1$,
where $\psi_{\rm TN}$ is the compact coordinate of $S^1$ with period
$2\pi R$.
Consequently, this solution only arises when at least one direction is 
compact. It is localized in the four Taub--NUT directions, as should
be the case for a 6-brane, and magnetically charged under the
graviphoton $g_{\mu \rm TN}$.  It can be considered as
the electromagnetic dual of a $pp$-wave, electrically charged under 
the graviphoton arising after compactification on a circle of
radius $R$. $pp$-waves in compact directions will be 
called indifferently Kaluza--Klein excitations or momentum states.

The corresponding solutions for the M2- and M5-brane read
\cite{Duff:1991xz,Gueven:1992hh}: 
\begin{subequations}
\begin{equation} 
\mbox{M2-brane} \co {\rm d}s_{11}^2 = H^{-2/3} {\rm d} s^2 (E^{1,2})
+ H^{1/3}  {\rm d} s^2 (E^8)
\end{equation} 
\begin{equation} 
d\C_3 = {\rm Vol} (E^{1,2}) \wedge d   H^{-1}
\end{equation} 
\begin{equation} 
H  = 1 + \frac{k}{r^6}
\sp
k = \frac{\kappa_{11}^2 \T_2 }{3 \Omega_7}
\end{equation} 
\end{subequations}
\begin{subequations}
\begin{equation} 
 \mbox{M5-brane} \co 
{\rm d}s_{11}^2 = H^{-1/3} {\rm d} s^2 (E^{1,5})
+ H^{2/3} {\rm d} s^2 (E^5)
\end{equation} 
\begin{equation} 
d\C_3 = \star_5 d H
\end{equation} 
\begin{equation} 
H  = 1 + \frac{k}{r^3}
\sp
k = \frac{\kappa_{11}^2 \T_5 }{3 \Omega_4}\ ,
\end{equation} 
\end{subequations}
which also show that the M2-brane (resp. M5-) is 
electrically (resp. magnetically) charged under the 3-form gauge potential.
The symbol $\star_q$ denotes Hodge duality in $q$ dimensions, and 
$\Omega_n$ the volume of the sphere $S^n$ with unit radius:
\begin{equation}
\Omega_n =
\frac{2\pi^{\frac{n+1}{2}}}{\Gamma\left(\frac{n+1}{2}\right)}\ .
\end{equation}

The tensions (or mass per unit world-volume) of these four basic BPS 
configurations can be easily evaluated from ADM boundary integrals
and Dirac quantization,
or more easily yet by dimensional analysis:
\begin{equation} 
\label{11dt}  
\begin{split} 
\mbox{KK-state} \co &  \T_0 = \frac{1}{R} \sp 
 \mbox{KK6-brane} \co  \T_6 = \frac{R^2}{l_p^9} \ ,\\ 
\mbox{M2-brane} \co &  \T_2 = \frac{1}{l_p^3}
\sp 
\mbox{M5-brane} \co   \T_5 = \frac{1}{l_p^6} \ .  
\end{split}
\end{equation} 
The tension (\ie mass) of the $pp$-wave 
with momentum along a compact direction of
radius $R$ (occasionally denoted as $R_{\rm TN}$)
 is the one expected for a massless particle in eleven
dimensions; the tension of the KK6-brane is easily obtained from the latter
by electric--magnetic duality, after reading off from 
Eq. \eqref{11dkk} the Kaluza--Klein 
gauge coupling $1/g_{\rm KK}^2 = R^3/l_p^9$:
\begin{equation}
\T_6 = \frac{\T_0}{g_{\rm KK}^2} = \frac{R^2}{l_p^9}\ .
\end{equation}

All these BPS states have been inferred from a classical analysis of
11D supergravity. They should in principle arise from a microscopic
definition of M-theory, which would allow a full account of
their interactions.
Nevertheless, it is still possible to formulate their dynamics
in terms of their collective coordinates which result from
the breaking of global symmetries in the presence of the soliton
\cite{Gervais:1975yg}.
Supersymmetry gives an important guideline, since (the unbroken) half of the 32
supercharges has to be realized linearly on the world-volume, while the
other half is realized non-linearly. This fixes the dynamics of the
M5-brane to be described in terms of the chiral $(2,0)$ six-dimensional
tensor theory
\cite{Callan:1991ky}, while the membrane is described by the 2+1
supermembrane action  \cite{Bergshoeff:1987cm,deWit:1988ig}. 
Unfortunately, the quantization of these
two theories remains a challenge.
As for the KK6-brane, the description of its dynamics is still 
an unsettled problem \cite{Hanany:1997xc}.

\subsection{Reduction to type IIA BPS solutions\label{rediia}}
Upon compactification on a circle (with periodic boundary conditions
on the fermion fields), the supersymmetry algebra is unaffected
and the generators merely decompose under the reduced
Lorentz group. The 32-component Majorana spinor $Q_\alpha$
decomposes into two 16-component Majorana--Weyl spinors
of $SO(1,9)$ with opposite chiralities, and the $N=1$ 
supersymmetry in 11D gives rise to non-chiral $N=2$
supersymmetry in 10D. However, it is convenient not to
separate the two chiralities explicitly, and rewrite
the supersymmetry algebra as
\begin{eqnarray}
\label{susy1a}
\left\{ Q_\alpha, Q_\beta \right\} 
&=& (C\Gamma^{\mu})_{\alpha\beta} P_\mu
+ (C\Gamma_{s})_{\alpha\beta} Z \nonumber\\
&&+ \frac{1}{2} (C\Gamma_{\mu\nu})_{\alpha\beta} Z^{\mu\nu}
+ (C\Gamma_{\mu}\Gamma_{s})_{\alpha\beta} Z^{\mu} \\
&&
+ \frac{1}{5!} (C\Gamma_{\mu\nu\rho\sigma\tau})_{\alpha\beta} 
  Z^{\mu\nu\rho\sigma\tau}
+ \frac{1}{4!} (C\Gamma_{\mu\nu\rho\sigma}\Gamma_s)_{\alpha\beta} 
  Z^{\mu\nu\rho\sigma}\nonumber\ ,
\end{eqnarray}
where the eleventh Gamma matrix $\Gamma_s$
is identified with the 10D chirality operator
$\Gamma_0\Gamma_1\dots\Gamma_9$. The eleven-dimensional central
charges give rise to the charges $Z$, $Z^\mu$, $Z^{\mu\nu}$,
$Z^{\mu\nu\rho\sigma}$, $Z^{\mu\nu\rho\sigma\tau}$ whose
interpretation is summarized in Table \ref{zcharii},
where we omitted the momentum charge $P_\mu$. In this table,
$\K_{m;mnpqrst}$ denotes the 6-form dual to $g_{\mu m}$ after
compactification of the direction $m$. 

\begin{table}[h] 
\begin{center}
\begin{tabular}{|l|l|l|l|l|l|l|l|l|}
\hline 
$Z$ & $Z^{ij}$ & $Z^i$ & $Z^{ijklm}$ & $Z^{ijkl}$& $Z^{0i}$ 
& $Z^0$ & $Z^{0ijkl}$ & $Z^{0ijk}$ \\
\hline
$\A_0$ & $\C_{0ij}$ & $B_{0i}$ & $\E_{0ijklm}$ & $\mathcal{R}_{0ijkl}$& 
none&none & $\K_{m;mpqrs0}$ & $\mathcal{R}_{0lmnpqr}$ \\
\hline
D0     & D2         & F1       & NS5           & D4 & 
D8  & 9-brane & KK5 & D6\\
\hline
\end{tabular}
\end{center} 
\caption{Type IIA central charges, gauge fields and extended objects
\label{zcharii}}  
\end{table}
 
Under Kaluza--Klein reduction, the BPS solutions of 
11D SUGRA yield BPS solutions of type IIA supergravity.
This reduction can, however, be carried out only if the 
eleventh dimension is a Killing vector of the configuration.
This is automatically obeyed if the eleventh direction
is chosen along the world-volume $E^{1,p}$, and reduces the
eleven-dimensional $p$-brane to a ten-dimensional $(p-1)$-brane
with tension $\T_{p-1} = R \T_{p}$;
this procedure is called {\it diagonal} or {\it double} reduction
\cite{Duff:1987bx}, 
and we shall call the resulting solutions {\it wrapped}
or {\it longitudinal} branes. One may also
want to choose the eleventh direction transverse to the
brane, but this is not an isometry, since the 
dependence of the harmonic function $H$ on the transverse
coordinates is non-trivial. However, this can 
be easily evaded by using the superposition property of BPS
states, and constructing a continuous stack of parallel $p$-branes
along the eleventh direction. The harmonic function on $E^{10-p}$
turns into an harmonic function on $E^{9-p}$:
\begin{equation}
\int_{-\infty}^{\infty} \frac{dx^s}{\left[ (x^s)^2 + \rho^2
  \right]^{\frac{8-p}{2}}} \sim \frac{1}{\rho^{7-p}}\ .
\end{equation}
We therefore obtain an {\it unwrapped} or {\it transverse} 
$p$-brane in ten dimensions with the same tension $\T_{p}$
as the one we started with. This procedure is usually called
{\it vertical} or {\it direct} reduction. It has also been proposed
to reduce along the isometry that arises when the sphere
$S^{9-p}$ in the transverse space $E^{10-p}$ is odd-dimensional,
hence given as a $U(1)$ Hopf fibration
\cite{Duff:1998us}, but the status of 
the solutions obtained by this {\it angular} reduction is
still unclear.

Applying this procedure to the four M-theory BPS configurations, with tensions
given in Eq. \eqref{11dt}, we find, after using the relations \eqref{msr},
the set of BPS states of type IIA string theory listed in Table \ref{tma}:  

\begin{table}[h] 
\begin{center}
\begin{tabular}{|l|lcccr|l|}
\hline 
M-theory & \multicolumn{5}{c|}{mass/tension} &type IIA \\ \hline \hline   
 longitudinal M2-brane &  
$ \T_1  $&=&$  \frac{R_s}{l_p^3} $&=&$ \frac{1}{l_s^2} $ & 
 F-string  \\ \hline  
transverse M2-brane &  
$ \T_2 $&=&$  \frac{1}{l_p^3} $&=&$ \frac{1}{g_s l_s^3}$ &  
 D2-brane \\ \hline  
 longitudinal M5-brane & 
$\T_4 $&=&$ \frac{R_s}{l_p^6} $&=&$ \frac{1}{g_s l_s^5} $ & 
 D4-brane \\ \hline  
 transverse M5-brane & 
$\T_5 $&=&$ \frac{1}{l_p^6} $&=&$ \frac{1}{g_s^2 l_s^6}$ & 
 NS5-brane \\ \hline 
longitudinal KK mode &  
$\T_0 $&=&$ \frac{1}{R_s} $&=&$ \frac{1}{g_s l_s}$ &  
D0-brane \\ \hline  
transverse KK mode &  
$\T_0 $&=&$ \frac{1}{R_i} $&=&$ \frac{1}{R_i}$ &  
KK mode  \\ \hline  
longitudinal KK6-brane &  
$\T_5 $&=&$ \frac{R_s R_{\rm TN}^2 }{l_p^9} $&=&$ \frac{R_{\rm TN}^2}{g_s^2 l_s^8 }$ &  KK5-brane  \\ \hline  
 KK6-brane with $R_{\rm TN} = R_s$ &  
$\T_6$&=&$ \frac{R_s^2}{l_p^9} $&=&$ \frac{1}{g_s l_s^7 }$ & 
 D6-brane \\ \hline  
 transverse KK6-brane &  
$\T_6$&=&$ \frac{R_{\rm TN}^2}{l_p^9} $&=&$ \frac{R_{\rm TN}^2}{g_s^3 l_s^9 }$ & 
 6${}_3^1$-brane  \\ \hline  
\end{tabular}
\end{center} 
\caption{Relation between M-theory and type IIA BPS states.  
\label{tma}}  
\end{table}

As the table shows, we recover the set of all
1/2 BPS solutions of type IIA string theory,
which include the KK excitations, the fundamental string and the
set of solitonic states comprised by the NS5-brane, KK5-brane
and the D$p$-branes\footnote{The letter D stands for the Dirichlet 
boundary conditions in the $9-p$ directions orthogonal to the
world-volume of the D$p$-brane, which force the open strings to move
on this $(p+1)$-dimensional hyperplane.} 
with $p=0,2,4,6$
\footnote{There is also an 8-brane coupling to a nine-form, whose
expectation value is related to the cosmological constant
\cite{Polchinski:1995mt,Bergshoeff:1995vh,Bergshoeff:1996ui}.}.
The NS5-brane is a solitonic solution that is magnetically charged under the
Neveu--Schwarz $B$-field \cite{Callan:1991ky}. 
The D$p$-branes are solitonic solutions, 
electrically charged under
the RR gauge potentials $\Ra_{p+1}$ (or magnetically under 
$\Ra_{7-p}$)
\cite{Polchinski:1995mt}. 
The tension of these BPS states does not receive any quantum 
corrections perturbative or non-perturbative, which is why
these objects are useful when considering non-perturbative dualities. 
States electrically (resp. magnetically) 
charged under the Neveu--Schwarz gauge fields have tensions that scale
with the string coupling constant as $g_s^0$ 
(resp. $1/g_s^2$), whereas states charged under the
Ramond fields have tensions that scale as $1/g_s$.

The last line in Table \ref{tma} is an unconventional solution, which we
call a 6${}_3^1$-brane,   
obtained by 
vertical reduction of the KK6-brane in a
direction in the $\Real^3$ part of the Taub--NUT 
space \cite{Blau:1997du}. The integration
involved in building up the stack is, however, logarithmically
divergent, and, if regularized, yields a non-asymptotically flat space.
However, as we will see in more detail in Subsection \ref{exotic}, 
at the algebraic level
this solution is required by U-duality symmetry. At that point we will
also explain our nomenclature for this (and other) non-conventional solutions. 
It is also interesting to note that all the tensions obtained above
are not independent, since they follow from the basic relations \eqref{msr}. 
This already hints at the presence of 
a larger structure that relates all these states, a fact that
we will establish using the conjectured U-duality symmetry of
compactified M-theory.

The dimensional reduction can also be carried out at the level of the 
supergravity configuration itself. For example, using
the  relation \eqref{u1fb} between the 11D metric and 10D string metric, 
one finds that a solution with 11D metric of the form
\begin{equation}
{\rm d}s_{11}^2 = H^\kappa 
{\rm d} s^2 (E^{1,p}) 
+ H^{\lambda} {\rm d} s^2 (E^{10-p}) 
\end{equation}
yields two 10D solutions with metric and dilaton 
\begin{subequations}
\begin{equation}
{\rm d}s_{10}^2 = H^\alpha  
  {\rm d} s^2 (E^{1,p'}) 
+ H^{\beta} {\rm d} s^2 (E^{9-p'}) \sp e^{-2 \phi} = H^{\gamma}  
\end{equation}
where
\begin{equation}
\mbox{diagonal}: \;\;\; p' = p- 1 \sp 
\alpha = \frac{3 \kappa}{2} \sp
\beta = \lambda + \frac{\kappa}{2} \sp  
\gamma = - \frac{3 \kappa}{2}\ ,
\end{equation} 
\begin{equation}
\mbox{vertical} : \;\;\; p' = p \sp 
\alpha = \kappa + \frac{ \lambda }{2} \sp
\beta = \frac{3 \lambda}{2} \sp  
\gamma = - \frac{3 \lambda}{2}\ ,
\end{equation} 
\end{subequations} 
for diagonal and vertical reduction respectively. 
As explained in the beginning of this subsection, in the first case 
the harmonic function is the same as the original one, and in the second case 
it is an harmonic function on a transverse space with one dimension less.
The reduction of the gauge potentials can be worked out similarly. 

The resulting 10D type IIA configurations are then
described by the following solutions:
\begin{subequations}
\begin{equation}
\mbox{F-string} \co {\rm d}s_{10}^2 = H^{-1} {\rm d} s^2 (E^{1,1}) 
+ {\rm d} s^2 (E^8) 
\end{equation}
\begin{equation}
B_{01} = H^{-1} \sp e^{-2 \phi} = H \sp H = 1 + \frac{k}{r^6}  
\end{equation}
\end{subequations}
\begin{subequations}
\begin{equation}
\mbox{NS5-brane} \co {\rm d}s_{10}^2 =  {\rm d} s^2 (E^{1,5}) 
+ H {\rm d} s^2 (E^4) 
\end{equation}
\begin{equation}
d B = \star_4 d H  \sp  e^{-2 \phi} = H^{-1}  \sp H = 1 + \frac{k}{r^2}  
\end{equation}
\end{subequations}
\begin{subequations}
\begin{equation}
\mbox{D$p$-brane} \co 
{\rm d}s_{10}^2 =  H^{-1/2} {\rm d} s^2 (E^{1,p}) 
+ H^{1/2} {\rm d} s^2 (E^{9-p}) 
\end{equation}
\begin{equation}
 e^{-2 \phi} = H^{(p-3)/2}  \sp H = 1 + \frac{k}{r^{7-p}}  
\end{equation}
\begin{equation}
F_e^{(p+2)} = {\rm Vol} (E^{1,p}) \wedge d H^{-1} \sp p = 0,1,2 
\end{equation}
\begin{equation}
F_m^{(8-p)} = \star_{9-p} d H \sp p = 4,5,6 
\end{equation}
\begin{equation}
F^{(5)} = F_e^{(5)} + F_m^{(5)} \sp p =3 
 \end{equation}
\end{subequations}
where, for completeness, we have included the D$p$-brane
solutions for all $p=0 \ldots 6$,
although we note that only even $p$ occurs in type IIA. The subscripts
$e$ and $m$ indicate whether the $p$-branes are electrically or
magnetically charged under the indicated fields.  
One also finds the ten-dimensional gravitational solutions, consisting of
the $pp$-waves and KK5-brane, which have a metric analogous to the 
eleven-dimensional case (see \eqref{ppw} and \eqref{kk6}), 
with harmonic functions on a transverse space
with one dimension less. Of course, one may explicitly verify that
all of  these solutions
are indeed solutions of the  tree-level action \eqref{10da}.

In contrast to the M2-brane and M5-brane, the dynamics of D$p$-branes 
has a nice and tractable description as $(p+1)$-dimensional
hyperplanes on which open strings can end and exchange momentum with
\cite{Polchinski:1995mt}.
The integration of open string fluctuations around a single D-brane 
at tree level yields the Born--Infeld 
action \cite{Callan:1988wz,Leigh:1989jq,Bachas:1996kx},
\begin{equation} 
\label{borninfeld}
 S_{\rm BI} = \frac{1}{l_s^{p+1}}  \int {\rm d}^{p+1} \xi
e^{-\phi} \sqrt{ \hat{g} + \hat{B} + l_s^2 F } \ .
\end{equation} 
Here, the hatted fields $\hat{g}, \hat{B}$  stand for the pullbacks of the 
bulk metric and antisymmetric tensor to the world-volume of the brane,
and $F$ is the field strength of the $U(1)$ gauge field living on the brane.  
The coupling to the RR gauge potentials is
given by the topological term
\cite{Douglas:1995bn,Green:1996bh}\footnote{There is also a gravitational
term required for the cancellation of anomalies \cite{Green:1997dd},
but it does not contribute on flat backgrounds.}
\begin{equation} 
\label{rrcoup}
S_{\rm RR} = i \int e^{\hat B + l_s^2 F } \wedge  \Ra    \ ,
\end{equation} 
where $\Ra = \sum_p \Ra_p$ denotes the total RR potential. 

In the zero-slope limit, the Born--Infeld action becomes the action of 
a supersymmetric Maxwell theory with 16 supercharges. 
In the presence of $N$ coinciding D-branes the world-volume gauge symmetry
gets enhanced from $U(1)^N$ to $U(N)$, as
a consequence of zero mass strings stretching between different 
D-branes \cite{Witten:1996im}.
The non-Abelian analogue of the Born--Infeld action is not known,
although some partial Abelianization is available
\cite{Tseytlin:1997cs},  
but its zero-slope limit is still given by $U(N)$ super-Yang--Mills theory.

\subsection{T-duality and type IIA/B string theory \label{tdua2a2b}} 

So far, we have discussed M-theory and its relation to type IIA string
theory. In this subsection, we turn to type IIB string theory and its relation,
via T-duality, to type IIA \cite{Dine:1989vu,Dai:1989ua}. 
We first recall that the massless sector
of type IIB consists of the same Neveu--Schwarz fields \eqref{nsf} 
as the type IIA
string, but the Ramond gauge potentials of type IIB now include 
a 0-form (scalar), a 2-form and a 4-form with self-dual field strength,  
\begin{equation} 
\label{2br}
a \sp \mathcal{B}_{\mu\nu} \sp \mathcal{D}_{\mu\nu\rho\sigma}    \ ,
\end{equation} 
with $*\mathcal{D}_4=\mathcal{D}_4$.
The low-energy effective action has a form similar to that 
in \eqref{10da}, with
the appropriate field strengths of the even-form RR potentials
in \eqref{2br}, as long as the 4-form is not included\footnote{A 
local covariant action for the self-dual four-form 
can be written with the help of auxiliary fields
\cite{Dall'Agata:1997ju}, but
for most purposes the equations of motion are sufficient.}.
The standard 1/2-BPS solutions of type IIB are the fundamental string,
NS5-brane, D$p$-branes with odd $p$, $pp$-waves and KK5-brane. 

In order to describe the precise T-duality mapping, we again
write the ten-dimensional metric as a $U(1)$ fibration
\begin{equation}
ds_{10}^2 = R^2 (dx^9 + A_\mu dx^\mu)^2 + g_{\mu \nu} dx^\mu dx^\nu
\quad \sp \mu,\nu = 0 \ldots 8 \ .
\end{equation}
T-duality on the direction $9$ relates the fields in the
type IIA and type IIB theories in the Neveu--Schwarz sector as
\begin{equation} 
\label{tid}
T_9 :\qquad R \leftrightarrow \frac{l_s^2}{R} 
\ ,\quad
g_s \leftrightarrow   g_s  \frac{l_s}{R} 
\ ,\quad
A_\mu  \leftrightarrow B_{9 \mu }\ ,\quad
B_{\mu\nu} \leftrightarrow B_{\mu\nu}-A_\mu B_{9 \nu }
+A_\nu B_{9\mu}\ ,
\end{equation} 
leaving $g_{\mu \nu}$ and the string length $l_s$ invariant. 
The Ramond gauge potentials are furthermore identified on both
sides according to 
\begin{equation} 
\label{trr}
T_9 :\ \qquad 
\Ra \leftrightarrow {\rm d} x^9 \cdot  \Ra + {\rm d} x^9 \wedge  \Ra
 \sp    \Ra = \sum_p  \Ra_p \ ,
\end{equation}
where $\cdot$ and $\wedge$ denote 
the interior and exterior products respectively.
In other words, the $9$ index is added to the antisymmetric indices
of $\mathcal{R}$ when absent, or deleted if it was already present. These
identifications actually receive corrections when
 $B \neq 0$, and the precise mapping is 
\cite{Bergshoeff:1995as,Green:1996bh,Eyras:1998hn}
\begin{equation}
e^B \Ra \rightarrow  {\rm d} x^9 \cdot (e^B \Ra) + {\rm d} x^9 \wedge  
(e^B \Ra)
\end{equation}
in accord with the T-duality covariance of the RR coupling in
\eqref{rrcoup}.
Whereas one T-duality maps the type IIA string theory to IIB
and should be thought of as a change of variable,
an even number of dualities can be performed and correspond
to actual global symmetries of either type IIA or type IIB theories.
This symmetry will be discussed in Section \ref{tdua}, and its
non-perturbative extension in Section \ref{udua}.

The action on the BPS spectrum can again be easily worked out, at the
level of tension formulae or of the supergravity solutions themselves.
As implied by the exchange of the Kaluza--Klein and Kalb--Ramond
gauge fields $A_\mu $ and $B_{9\mu }$, 
states with momentum along the 9th direction
are interchanged with fundamental string winding around
the same direction. On the other hand, T-duality exchanges
Neumann and Dirichlet boundary conditions on the open string world-sheet along
the 9th direction, mapping D$p$-branes to D$(p+1)$- or D$(p-1)$-branes,
depending on the orientation of the world-volume with
respect to $x^9$
\cite{Dai:1989ua,Bergshoeff:1996cy}.
This of course agrees with the mapping of Ramond gauge potentials
in Eq. \eqref{trr}. Similarly, NS5-branes are invariant or exchanged
with KK5-branes, according to whether they are wrapped or unwrapped,
respectively \cite{Eyras:1998hn,Papadopoulos:1998je} 
\footnote{
Whereas the worldvolume dynamics of type IIB NS5- and D5-branes
is described by a non-chiral $(1,1)$ vector multiplet, the type IIB
KK5-brane is chiral and supports a $(2,0)$ tensor multiplet. Indeed, it is
T-dual to the chiral type IIA NS5-brane \cite{Antoniadis:1997eg}. On the other
hand, the type IIA KK5-brane, dual to the type IIB NS5-brane, is nonchiral.
}. 
This can also be easily seen by applying
the transformation \eqref{tid} to the tension formulae, as
summarized in Table \ref{ttd} for a T-duality $T_i$ on an arbitrary
compact dimension with radius $R_i$.

\begin{table}[h] 
\hspace*{-.9cm}
\begin{tabular}{|l|l||l|l|}
\hline 
type IIA (B)  & tension  & $T_i$-dual tension & type IIB (A) \\ \hline \hline   
KK mode &  
$\M = \frac{1}{R_i}$ &  
$\M = \frac{R_i}{l_s^2}$ &  
winding mode \\ \hline   
wrapped D$p$-brane &  
$\T_{p-1}  = \frac{R_i}{g_s l_s^{p+1}}$ &  
$\T_{p-1}  = \frac{1}{g_s l_s^{p}} $ &  
unwrapped D$(p-1)$-brane \\ \hline  
wrapped NS5-brane &  
$\T_4 = \frac{R_i}{g_s^2 l_s^6}$ &  
$\T_4 = \frac{R_i}{g_s^2 l_s^6}$ &  
wrapped NS5-brane \\ \hline  
unwrapped NS5-brane  &  
$\T_5 = \frac{1}{g_s^2 l_s^6}$ &  
$\T_5 = \frac{R_i^2}{g_s^2 l_s^6}$ &  
unwrapped KK5-brane\\ \hline  
\end{tabular}
\caption[T-duality of type II BPS states]
{T-duality of type II BPS states.\label{ttd}}
\end{table} 

T-duality can then be used to translate the relation between
strongly coupled type IIA theory and M-theory in type IIB terms.
In this way, it is found that the
type IIB string theory is obtained by compactifying 
M-theory on a two-torus $T^2$, with vanishing area, 
and a complex structure $\tau$ equated to the type IIB
complex coupling parameter \cite{Schwarz:1995dk}:
\begin{equation}
\label{taudef}
\tau = a + \frac{i}{g_s} \ .
\end{equation}
Here, $a$ is the expectation value of the Ramond scalar and 
$g_s$ the type IIB string coupling.

We focus for simplicity on the case where the torus is rectangular, so
that $\tau$ is purely imaginary and hence the RR scalar $a$ vanishes.
In this case, the relation between the M-theory parameters and
type IIB parameters reads
\begin{equation}
\label{iibmatching}
g_s = \frac{R_s}{R_9} \sp \;\;\;  
l_s^2 = \frac{l_p^3}{R_s} \sp \;\;\;  
R_B = \frac{l_p^3}{R_s R_9}\ ,
\end{equation}
where $R_s, R_9$ are the radii of the M-theory torus and
$R_B$ the radius of the type IIB 9th direction. The uncompactified
type IIB theory is obtained in the limit $(R_s,R_9)\ra\infty$,
keeping $R_s/R_9$ fixed. From Eq. \eqref{iibmatching}, 
we can then identify the type IIB BPS states to those of M-theory
compactified on $T^2$. The results are displayed in Table
\ref{tmb} for states still existing in uncompactified type IIB
theory, and in Table \ref{tmbc} for states existing only for
finite values of $R_B$.

\begin{table}[h] 
\begin{center}
\begin{tabular}{|m{53mm}|c|l|}
\hline 
M-theory   & mass/tension & type IIB \\ \hline \hline   
 M2-brane wrapped around $  x^s$ &  
$\frac{R_s}{l_p^3} = \frac{1}{l_s^2}$ & 
 fundamental string \\ \hline 
M2-brane wrapped around $  x^9$ &  
$\frac{R_9}{l_p^3} = \frac{1}{g_s l_s^2}$ &  
 D1-brane (D-string) \\ \hline 
 M5-brane wrapped on $  x^s,x^9$ &
$\frac{R_s R_9}{l_p^6} = \frac{1}{g_s l_s^4} $ &   
  D3-brane \\ \hline  
  {KK6-brane wrapped on $  x^9$, charged under  $  g_{\mu s}$} &  
$\frac{R_s^2 R_9 }{l_p^9} = \frac{1}{g_s l_s^6}  $ &  
    D5-brane  \\ \hline
KK6-brane wrapped on $  x^s$, charged under $  g_{\mu 9}$ &  
$\frac{R_9^2 R_s}{l_p^9} = \frac{1}{g_s^2 l_s^6}  $ &  
     NS5-brane \\ \hline  
\end{tabular}
\end{center} 
\caption{Relations between M-theory and type IIB BPS states.\label{tmb}}
\end{table}
%
%
%
\begin{table}[H] 
\begin{center}
\begin{tabular}{|m{50mm}|c|l|}
\hline 
M-theory   & mass/tension & type IIB \\ \hline \hline   
M2-brane wrapped on $x^s,x^9$ &  
$\frac{R_9 R_s}{l_p^3} = \frac{1}{R_B}$ &  
KK mode  \\ \hline  
 unwrapped M5-brane & 
$\frac{1}{l_p^6} = \frac{R_B^2}{g_s^2 l_s^8}  $ &  
 KK5-brane with $R_{\rm TN} = R_s $ \\ \hline
unwrapped M2-brane &  
$ \frac{1}{l_p^3} = \frac{R_B}{g_s l_s^4}  $ &  
wrapped D3-brane\\ \hline  
 M5-brane wrapped on $  x^s$ &  
$\frac{R_s}{l_p^6} = \frac{R_B}{g_s l_s^6}  $ &  
wrapped D5-brane \\ \hline  
 M5-brane wrapped on $  x^9$ &  
$\frac{R_9}{l_p^6} = \frac{R_B}{g_s^2 l_s^6}  $ &  
wrapped NS5-brane \\ \hline  
 unwrapped KK6-brane, charged under  $  g_{\mu s}$ &  
$\frac{R_s^2}{l_p^9} = \frac{R_B}{g_s l_s^8}  $ &  
wrapped  D7-brane   \\ \hline  
 unwrapped KK6-brane,  charged under  $  g_{\mu 9}$& 
$\frac{R_s^2}{l_p^9} = \frac{R_B}{g_s^3 l_s^8}  $ &  
wrapped 7${}_3$-brane \\ \hline  
\end{tabular}
\end{center} 
\caption{More relations between M-theory and type IIB BPS states.  
\label{tmbc}}
\end{table}

As in Table \ref{tma}, we see in the last entry of Table \ref{tmbc} 
a non-standard
BPS state  with tension scaling as $g_s^{-3}$, which we have
called a 7${}_3$-brane. As this brane will turn out to be related
to the D7-brane by S-duality (see Subsection \ref{iibs}) it may also
be referred to as a (1,0) 7-brane.
This and other non-standard solutions will be discussed in more detail in 
Subsection \ref{exotic}.  

\clearpage
\section{T-duality and toroidal compactification \label{tdua}}

Having discussed how {\it dualities} of string theory lead to the
idea of a more fundamental eleven-dimensional M-theory, we now
turn to the {\it symmetries} that this theory should exhibit,
with the hope of getting more insight into its underlying structure.
For this purpose, it is convenient to consider compactifications
on tori, which have the advantage of preserving a maximal amount
of the original super-Poincar{\'e} symmetries, while bringing in
degrees of freedom from extended states in eleven dimensions
in a still manageable way. 

The approach here is similar to the one
that was taken for the perturbative string itself, where the
study of T-duality in toroidal compactifications revealed the
existence of spontaneously broken ``stringy'' gauge symmetries
(see \cite{Giveon:1994fu} for a review).
Given the analogy between the two problems, we shall first review in this
section how T-duality in string theory
appears at the level of the low-energy effective action 
and of the spectrum, with a particular emphasis on the brane spectrum.
We shall then apply the same techniques in Sections \ref{udua} and 
\ref{sskew}  in order to discuss U-duality in M-theory.

\subsection{Continuous symmetry of the effective action \label{cseas} }
Compactification of string theory on a torus $T^{d}$ can be 
easily worked out at the level of the low-energy effective action,
by substituting an ansatz similar to \eqref{u1f}
\begin{subequations}
\begin{equation}
{\rm d}s^2_{10} = g_{ij} \left( dx^i + A_\mu^i dx^\mu \right)
\left( dx^j + A_\nu^j dx^\nu \right)
+ g_{\mu \nu } dx^\mu  dx^\nu  
\end{equation}
\begin{equation}
i,j = 1 \ldots d \quad \sp \mu,\nu = 0 \ldots (9-d) 
\end{equation}
\end{subequations}
in the ten-dimensional action
\begin{equation}
S_{10} = \frac{1}{l_s^8}\int {\rm d}^{10} x \sqrt{-g} e^{- 2 \phi} \left(
R +4 (\partial\phi )^2  - \frac{l_s^4 }{12} (d B)^2 \right) \ ,
\end{equation}
where we omitted Ramond and fermion terms. We have also split the 
ten-dimensional two-form
$B$ into $d(d-1)/2$ scalars $B_{ij}$, $d$ vectors $B_{i \mu }$
and a two-form $B_{\mu \nu }$. 

Concentrating on the scalar sector,
and redefining the dilaton as $V e^{-2\phi}=l_s^d e^{-2\phi_d}$
where $V=\sqrt{\det g}$ \footnote{$g$ denotes the internal metric
$g_{ij}$, except in the space-time volume element $\sqrt{-g}$
multiplying the action density.}
is the volume of the internal metric, we obtain
\begin{equation}
S_{\rm scal}=
\frac{1}{l_s^{8-d}}\int {\rm d}^{10-d} x \sqrt{-g} e^{- 2 \phi_d} \left(
4 (\partial \phi_d )^2  + \frac{1}{4} \Tr \partial g \partial g^{-1}
+  \frac{1}{4} \Tr g^{-1}\partial B~ g^{-1} \partial B \right) \ .
\end{equation}
This can be rewritten as 
\begin{equation}
\label{dmdm}
S_{\rm scal}=
\frac{1}{l_s^{8-d}}\int {\rm d}^{10-d} x \sqrt{-g} e^{- 2 \phi_d} \left(
4 (\partial \phi_d )^2 + \frac{1}{8} \Tr \partial M \partial M^{-1}
\right)\ ,
\end{equation}
where $M$ is the $2d\times 2d$ symmetric matrix 
\begin{equation}
\label{mmat}
M=\begin{pmatrix}
g^{-1} & g^{-1}B \\
-B g^{-1} & g-B g^{-1} B\end{pmatrix}
\ ,\quad
M^t \eta M = \eta\ ,\quad
\eta= \begin{pmatrix}
 & \mathbb{I}_d \\
\mathbb{I}_d & \end{pmatrix}\ ,
\end{equation}
orthogonal for the signature $(d,d)$ metric $\eta$. The scalars 
$g_{ij}$ and $B_{ij}$ therefore parametrize a symmetric manifold
\begin{equation}
\mathcal{H} = \frac{SO(d,d,\Real)}{SO(d)\times SO(d)} \ni M\ ,
\end{equation}
where $SO(d)\times SO(d)$ is the maximal compact subgroup 
of $SO(d,d,\Real)$. The matrix $M$ is more properly thought of
as the $SO(d)\times SO(d)$ invariant $M=\mathcal{V}^t \mathcal{V}$
built out from the vielbein in $SO(d,d,\Real)$
\begin{equation}
\label{iwast}
\mathcal{V} = 
\left( \begin{array}{ccc|ccc}
1/R_1 &        &       &       &    & \\
    & 1/R_2    &       &       &    & \\
    &        &\ddots &       &    & \\
\hline
    &        &       & R_1 &    & \\
    &        &       &       & R_2 & \\
    &        &       &       &    &\ddots  
\end{array} \right)
.
\left( \begin{array}{ccc|ccc}
1 & -A_2^1    & \dots &B_{11} & B_{12} &\dots\\
    & 1      & \ddots&B_{21} & B_{22} &\dots\\
    &        & \ddots&\vdots & \vdots &     \\
\hline
    &        &       & 1     & A_2^1 &\dots\\
    &        &       &       & 1      &\ddots \\
    &        &       &       &        &\ddots \\ 
\end{array} \right)
\end{equation}
corresponding to the Iwasawa decomposition of $SO(d,d,\Real)$,
as will be discussed in more detail in Section \ref{contsymu}.
The two-derivative action for the scalars
$g_{ij},B_{ij},\phi_d$
is therefore invariant \cite{Maharana:1993my} under the action $M\rightarrow
\Omega^t M \Omega$ of $\Omega\in O(d,d,\Real)$, 
and so is the entire two-derivative action in the Neveu--Schwarz 
sector, if the $2d$ gauge fields $A^i_\mu$ and $B_{i \mu }$ transform
altogether as a vector under $O(d,d,\Real)$, 
the dilaton $\phi_d$, metric $g_{\mu\nu}$
and two-form $B_{\mu\nu}$ being invariant. 

The action 
on the Ramond sector is more complicated, since the Ramond scalars and 
one-forms transform as a spinor (resp. conjugate spinor)
of $SO(d,d,\Real)$, with the chirality
depending on whether we consider type IIA or IIB.
Elements of $O(d,d,\Real)$ with $(-1)$ determinant 
flip the chirality of spinors; they therefore are {\it not}
symmetries of the action in the Ramond sector, but
{\it dualities}, exchanging type IIA and type IIB theories. Indeed
it is easy to see that the $R\rightarrow 1/R$ dualities
that we discussed in Subsection \ref{tdua2a2b} belong to this class
of transformations. The tree-level effective action is therefore
invariant under the continuous symmetry $SO(d,d,\Real)$,
which extends the symmetry $Sl(d,\Real)$ that would be present
in the dimensional reduction of any Lorentz-invariant field theory.

\subsection{Charge quantization and T-duality symmetry}
Owing to the occurrence 
of particles charged under the gauge fields
$A ^i_\mu$ and $B_{i \mu}$, 
the continuous symmetry $SO(d,d,\Real)$ can, however, not
exist at the quantum level. For instance, perturbative string
states have integer momenta $m_i$ and winding numbers $m^i$ under
these gauge fields, lying in an even self-dual Lorentzian lattice
$\Gamma_p$. The 1/2-BPS states are obtained when the world-sheet
oscillators $\alpha^\dagger_{\mu n}$ 
and $\bar\alpha^\dagger_{\mu n}$ are not excited, and satisfy
the mass formula and matching condition
\begin{subequations}
\label{vectormass}
\begin{align}
\mathcal{M}^2 &= m^t M m \nonumber\\
&= (m_i + B_{ij} m^j)
g^{ik} (m_k + B_{kl} m^l) + m^i g_{ij} m^j \\
\|m\|^2 &=0\ ,
\end{align}
\end{subequations}
where $m=(m_i,m^i)$ is the vector of charges, $\|m\|^2=2 m_i m^i$
its Lorentzian square-norm and $M$ is the moduli matrix 
given in \eqref{mmat}. 

On the other hand, 1/4-BPS states are obtained when the 
world-sheet oscillators are excited on the holomorphic
(or antiholomorphic) side only, and have mass
\begin{equation}
\label{vectormass2}
\mathcal{M}^2 = m^t M m + \left| \| m \|^2 \right|\ ,
\end{equation}
where the norm $\left| \| m \|^2 \right|$ 
is equated to the left or right oscillator number
by the matching conditions. Only the discrete subgroup
preserving $\Gamma_p$ can be a quantum symmetry, and this group
is $O(d,d,\Zint)$, the set of integer-valued $O(d,d,\Real)$
matrices. In particular, the subgroup $Sl(d,\Real)$ of
$SO(d,d,\Real)$ is reduced to the modular group of the torus
$Sl(d,\Zint)$, an obvious consequence of momentum quantization
in compact spaces. 

In addition to this perturbative spectrum, type II string
theory also admits a variety of D-branes, which are charged under 
the Ramond gauge potentials.
Their charges take value in another lattice, $\Gamma_{D}$,
and transform  as a spinor under $SO(d,d,\Real)$.
Again, the determinant $(-1)$ elements of $O(d,d,\Zint)$ flip
the chirality of spinors, and therefore do not preserve
$\Gamma_{D}$. As we shall see shortly, $SO(d,d,\Zint)$
however does preserve the lattice of D-brane charges. This is
in agreement with the fact that this group can be seen as
the Weyl group of the extended gauge symmetries that appear
at particular points in the torus moduli space, and are
spontaneously broken elsewhere \cite{Giveon:1990er}.  

\subsection{Weyl and Borel generators}
In order to better understand the structure of the T-duality 
symmetry, it is useful to isolate a set of generating elements
of $SO(d,d,\Zint)$. We define {\it Weyl} elements as the ones
that preserve the conditions
\begin{equation}
\label{weylcond}
g_{ij} = R_i^2  \delta_{ij}\ , \quad B_{ij}=0\ ,
\end{equation}
that is square tori with vanishing two-form background,
and {\it Borel} elements as the ones that do not.
Weyl generators include the exchanges of radii 
$S_{ij}:R_i \leftrightarrow R_j$,
which belong to the $Sl(d,\Zint)$ modular group, as well as
the simultaneous inversions of two radii $T_{ij}:(R_i,R_j) \rightarrow
(1/R_j,1/R_i)$. 

We choose the following minimal set of Weyl
generators:
\begin{subequations}
\label{weylmin}
\label{tduag}
\begin{eqnarray}
S_{i}: R_i \leftrightarrow R_{i+1}\ ,\quad i=1\dots d-1\ , \\
T: (g_s,R_1,R_2) \leftrightarrow \left(\frac{g_s}{R_1 R_2},\frac{1}{R_2},
\frac{1}{R_1}\right)\ .
\end{eqnarray}
\end{subequations}
For convenience, we followed the double T-duality on directions 1 and
2 by an exchange of the two radii, included the action on the 
coupling constant and set the string length $l_s$ to 1. 
Altogether, the Weyl group of $SO(d,d,\Zint)$
is the finite group
\begin{equation}
\label{soddw}
\mathcal{W}\left(SO(d,d)\right)= \Zint_2 \bowtie \mathcal{S}_d
\end{equation}
generated by the T-duality transformation $T$ and the permutation group
$\mathcal{S}_d$ of the $d$ directions of the torus\footnote{The Weyl
group of $SO(d,d)$ can actually be written as the semi-direct product
$\mathcal{S}_d\ltimes(\Zint_2)^{d-1}$, where the commuting $\Zint_2$'s
are the double inversions of $R_i$ and, say, $R_1$.}.

On the other hand, Borel generators
include the Borel elements of the modular subgroup, acting as
$\gamma_i \rightarrow \gamma_i + \gamma_j$ on the homology lattice
of the lattice, as well as the integer shifts of the expectation
value of the two-form in the internal directions
$B_{ij} \rightarrow B_{ij}+1$. Any element in $SO(d,d,\Zint)$ can
be reached by a sequence of these transformations.

Weyl and Borel generators can be given a more precise  
definition as operators on the weight space of the Lie
group or algebra under consideration (see for instance
Ref. \cite{Humphreys:1972} for an introduction to the relevant
group theory)
\footnote{From this point of view,
Weyl generators are not properly speaking elements of the
group, but can be lifted to generators thereof, at the cost
of introducing $\Zint_2$ phases in their action on the
step operators $E_{\alpha}$. See for instance Appendix B in
Ref. \cite{Lerche:1989np},
for a discussion of this issue in the physics literature.}. 
Weyl generators correspond to orthogonal reflections
with respect to planes normal to any root and generate
a finite discrete group, while Borel generators
act on the weight lattice by translation by a positive root.
Any finite-dimensional irreducible representation (of the complex
Lie algebra) can then be obtained by action of the Borel group on a, 
so called, highest-weight vector, and splits into orbits of
the Weyl group with definite lengths.

\subsection{Weyl generators and Weyl reflections \label{wgwrt}  }
Weyl generators encode the simplest and most interesting part of 
T-duality. It is very easy to study the structure of the finite
group they generate, by viewing them as orthogonal reflections
in a vector space (the weight space)
generated by the logarithms of the radii. 
More precisely, let us represent the scalar moduli
$(\ln g_s,\ln R_1,\dots, \ln R_d)$ as a form $\varphi$
on a vector space $V_{d+1}$ with basis $e_0, e_1,\dots, e_d$, 
and associate to any
weight vector $ \lambda=x^0 e_0 + x^1 e_1 + \dots + x^d e_d$, its 
tension\footnote{One could omit the $x^0$
coordinate since $g_s$ can be absorbed by a power of the invariant
Planck length $\prod R_i/g_s^2$, but we include it for later
convenience.}
\begin{equation}
\label{dualb}
\mathcal{T}=e ^{\langle\varphi,\lambda\rangle} =
g_s^{x^0} R_1^{x^1} R_2^{x^2}\dots R_d^{x^d}\ .
\end{equation}
The vector $\lambda$ should be seen as labelling a state
in the BPS spectrum, with tension $\T$. The generators \eqref{tduag}
are then implemented as linear operators on $V_{d+1}$ with
matrices
\begin{equation}
\label{stmat}
 S_i =
\begin{pmatrix} 
1 &    &    &   \\
  &    &  1 &   \\
  &  1 &    &   \\
  &    &    & \mathbb{I}_{d-3}  \\
\end{pmatrix}
\ ,
\quad 
T = \begin{pmatrix} 
1 &    &   &\\
-1 &   & -1 &\\
-1 & -1 &   &\\
   &    &   & \mathbb{I}_{d-3}
\end{pmatrix} 
\end{equation}

These operators $S_i$ and $T$ in \eqref{stmat} are easily seen to be orthogonal
with respect to the signature 
$(-+\dots+)$ metric 
\begin{equation}
\label{tm}
{\rm d}s^2=-({\rm d}x^0)^2+({\rm d}x^i)^2 + {\rm d}x^0(dx^1 + \dots + 
{\rm d}x^{d})\ ,
\end{equation} 
and correspond to Weyl reflections
\begin{equation}
  \lambda \rightarrow \rho_\alpha (\lambda) =
\lambda - 2 \frac{\alpha\cdot~\lambda}{\alpha\cdot \alpha} \alpha 
\end{equation}
with respect to planes normal to the vectors
\begin{subequations} 
\begin{equation}
 \alpha_i = e_{i+1} - e_{i} \ \sp i=1\dots d-1
\end{equation}
\begin{equation}
 \alpha_0 = e_1 + e_2\ .
\end{equation}
\end{subequations}
The group generated by $S_i$ and $T$ is therefore a Coxeter group,
familiar from the theory of Lie algebras (see \cite{Humphreys:1972} 
for an introduction, and \cite{Humphreys:1990,Fuchs:1997} for a full account).
Its structure 
can be characterized by the matrix of scalar products of these roots:
\begin{subequations}
\begin{equation}
 (\alpha_i)^2 = (\alpha_0)^2 = 2 
\end{equation}
\begin{equation}
 \alpha_i\cdot \alpha_{i+1} = \alpha_2 \cdot \alpha_0 = -1\ .
\end{equation}
\end{subequations}
This precisely reproduces the Cartan matrix $D_d$
of the T-duality group $SO(d,d,\Real)$, summarized
in the Dynkin diagram:
\begin{equation}
\label{dyT}
\begin{array}{ccccccc}
\bigcirc_0&&&&&&\\
&\diagdown&&&&&\\
&         & \oplus_2&-&
\oplus_3&-\dots-&
\oplus_{d-1}\\
&\diagup  &&&&&\\
+_1&&&&&&
\end{array}
\end{equation}

The only delicate point is that the signature of the metric \eqref{tm}
on $V_{d+1}$ is not positive-definite. This can be easily evaded
by noting that the invariance of Newton's constant $\prod R_i /g_s^2$
implies that all roots are orthogonal to the vector 
\begin{equation}
\label{tdelta}
\delta = e_1+\dots +e_{d} - 2 e_0
\end{equation}
with negative proper length $\delta^2=-(d+4)$, 
so that the reflections actually restrict to the
hyperplane $V_{d}$ normal to $\delta$:
\begin{equation}
\delta\cdot x = x^0 = 0\ .
\end{equation}
The Lorentz metric on $V_{d+1}$ then restricts to a positive-definite metric
$g_{ij} = \delta_{ij}$ on $V_d$. The dualities $S_{i}$ 
and $T$ therefore generate the Coxeter group $D_d$,
which is the same as the Weyl group of the Lie algebra of
$SO(d,d,\Real)$. In order to distinguish
the various real and discrete forms of $D_d$, one needs to take into account
the Borel generators, which we defer to Subsection \ref{spf}.

The Dynkin diagram \eqref{dyT} allows a number of simple observations.
We may recognize the Dynkin diagram $A_{d-1}$ of the Lorentz group
$Sl(d,\Real)$ (denoted with $+$), extended with the root
$\bigcirc$ into the Dynkin diagram of the T-duality
symmetry $SO(d,d,\Real)$.
T-duality between type IIA and type IIB corresponds to the outer
automorphism acting as a reflection along the horizontal
axis of the Dynkin diagram.
The chain denoted with $\bigcirc$'s represents a dual
$Sl(d,\Real)$ subgroup, which is nothing but the Lorentz
group on the type IIB T-dual torus. The full T-duality group
is generated by these two non-commuting Lorentz groups
of the torus and the dual torus. 

Decompactification of
the torus $T^d$ into $T^{d-1}$ is achieved by dropping the rightmost
root, which reduces $D_d$ to $D_{d-1}$. When the root $\alpha_2$ 
is reached, the diagram disconnects into two pieces, corresponding
to the identity $SO(2,2,\Real)$ $=Sl(2,\Real)\times Sl(2,\Real)$,
or to the decomposition of the torus moduli space into the
$T$ and $U$ upper half-planes\footnote{The 
extra $\Zint_2$ exchanging the two $Sl(2,\Real)$ 
factors belongs to $O(2,2,\Real)$ but not to $SO(2,2,\Real)$.}.
Finally, for $d=1$ the T-duality group $SO(1,1,\Zint)$
becomes trivial, while the generator of $O(1,1,\Zint)$
corresponds to the inversion $R\leftrightarrow 1/R$, {\it not}
a symmetry of either type IIA or type IIB theories.

\subsection{BPS spectrum and highest weights}
Having proved that the transformations $S_i$ and $T$ indeed
generate the Weyl group of $SO(d,d,\Zint)$, we can use the
same formalism to investigate the orbit of the various BPS
states of string theory. According to \eqref{dualb}
the mass or tension can  be
represented  as a {\it weight vector} in $V_{d+1}$, and one should let
Weyl and Borel generators act on it to obtain the full orbit.
Each orbit admits a {\it highest weight}
{}from which all other elements can be reached by a sequence of Weyl and Borel
generators (Weyl generators alone are not sufficient, because they
preserve the length of the weight). 

All highest weights can be written
as linear combinations with positive integer coefficients of
the fundamental weights 
\begin{subequations} 
\begin{eqnarray}
\lambda^{(1)}  = e_1 -e_0 
& \rightarrow 
  & \mathcal{M}_{\rm wD}=\frac{R_1}{g_s} 
\\
\lambda^{(2)}  = e_1 + e_2 -2e_0
&\rightarrow 
  & \mathcal{M}_{\rm NS}=\frac{R_1 R_2}{g_s^2} 
\\
\lambda^{(d-2)}  = e_1 + \dots + e_{d-2} -2e_0
&\rightarrow
  & \mathcal{M}_{\rm w\dots wNS}=\frac{R_1 \dots R_{d-2}}{g_s^2} \;\;\;\;\;\;\;\; 
\\
\lambda^{(d-1)}  = e_1+\dots +e_{d-1}-2e_0\doteq -e_d 
&\rightarrow 
  & \mathcal{M}_{\rm wF}=\frac{1}{R_d} 
\\
\lambda^{(0)}  = -e_0  
&\rightarrow 
  & \mathcal{M}_{\rm D}=\frac{1}{g_s} 
\end{eqnarray}
\end{subequations}
dual to the simple roots, that is 
$ \lambda^{(i)}\cdot \alpha_{j} = -\delta_{ij}$ \footnote{The minus
sign shows that we are really considering {\it lowest}-weight vectors,
but we shall keep this abuse of language.}.We used 
the symbol $\doteq$ for equality modulo the
invariant vector $\delta$ in Eq. \eqref{tdelta},
and the notation $F$,$D$ and $NS$ for fundamental, Dirichlet
and Neveu--Schwarz states, respectively, depending on the power
of the coupling constant involved, and $w$ for each wrapped 
direction (the notation $wF$ is justified by the fact that the
Kaluza--Klein states are in the same multiplet as the string
winding states). This is summarized in the Dynkin diagram
\begin{equation}
\label{dynkt}
\begin{array}{ccccccc}
\frac{1}{g_s}&&&&&&\\
&\diagdown&&&&&\\
&         & \frac{R_1 R_2}{g_s^2}&-&
\frac{R_1 R_2 R_3}{g_s^2}&-\dots-&
\frac{1}{R_d}\\
&\diagup  &&&&&\\
\frac{R_1}{g_s}&&&&&&
\end{array}
\end{equation}
which shows the highest weights associated to each node of the Dynkin diagram.

In particular, we see from \eqref{dynkt} that the type IIA D-particle mass
($\mathcal{M}=1/g_s l_s$) lies in 
the spinor representation dual to $\alpha_1$, 
just as do the type IIB D-string tension ($\T_1=1/g_s l_s^2$)
and D-instanton action ($\T_{-1}=1/g_s$), whereas
the type IIB D-particle mass ($\mathcal{M}=R_i/g_s l_s^2$) and type IIA
D-string tension ($\T_1=R_i/g_s l_s^3$) and D-instanton action
($\T=R_i/g_s l_s$) transform in the spinor 
representation dual to $\alpha_0$, of opposite chirality.
On the other hand, the Kaluza--Klein states lie in a vector
representation. 
All highest-weight representations can be obtained from the tensor product
of these ``extreme'' (from the point of view of the Dynkin
diagram) representations. T-duality on a single radius exchanges the two
spinor representations, as it should.

\subsection{Weyl-invariant effective action}
In the previous subsections, we have discussed how the Weyl group of
$SO(d,d)$ arises as the finite group generated by the permutations and
double T-duality \eqref{weylmin}, whereas the low-energy action itself
is invariant under the continuous group $SO(d,d,\Real)$. This
has been checked in the scalar sector in Eq. \eqref{dmdm}, by direct
reduction of the 10D effective action on $T^d$. It is however
possible to rewrite the full action in a manifestly Weyl-invariant
way, by a step-by-step reduction from 10D, as was originally
developed in Ref. \cite{Lu:1996yn} in the context of 11D supergravity.
This procedure leads to a clear identification of ``dilatonic''
scalars, which appear through exponential factors in the action
and include the dilaton $g_s$ and the radii $R_i$ of the torus, versus 
``Peccei--Quinn'' scalars which have constant shift symmetries
and are better thought of as 0-forms with a 1-form field strength. 

Each field strength $F^{(p)}$
gives rise to field strengths of lower degree $F^{(q)}_{i_1\dots
i_{q}}$, with internal indices $i_1\dots i_{q}$ (given
by the exterior derivative of a $(q-1)$-form up to Chern--Simons
corrections), while the metric gives rise to Kaluza--Klein
two-form field strengths $\mathcal{F}^{(2)i}$ and one-form field strengths 
$\mathcal{F}^{i(1)}_{j}$, $i<j$,  of the vielbein components
in the upper triangular gauge
\begin{subequations}
\begin{equation}
g_{MN}=E_{M}^{P} E_{N}^{Q}\ \eta_{PQ}\sp
\end{equation}
\begin{equation}
\label{vbut}
E_{M}^N=
\left(\begin{array}{ccccc|c}
R_1 &        &        &       &        &           \\
    & R_2    &        &       &        &           \\
    &        & \ddots &       &        &           \\
    &        &        & R_{d-1}&       &           \\
    &        &        &       & R_d    &           \\
\hline
    &        &        &       &        & E_\mu ^\nu
\end{array}\right)
.
\left(\begin{array}{ccccc|c}
1   & \A_2^1 & \A_3^1 & \dots & \A_d^1 & \A_\mu^1  \\
    & 1      & \A_3^2 & \dots & \A_d^2 & \A_\mu^2  \\
    &        & \ddots &       &        & \vdots    \\
    &        &        & 1     & \A_d^{d-1} & \A_\mu ^{d-1} \\
    &        &        &       & 1      & \A_\mu ^d \\
\hline
    &        &        &       &        & \mathbb{I}_{11-d}
\end{array}\right)\ ,
\end{equation}
\end{subequations}
where $E_\mu ^\nu$ denotes the vielbein in the uncompactified
directions. The action \eqref{10da} in the Neveu--Schwarz
sector then takes the simple form:
\begin{multline}
\label{sgrt}
S_{{\rm NS},10-d}=\int {\rm d}^{10-d}x \sqrt{-g}~\frac{V}{g_s^2 l_s^8}
\left[ 
R + (\partial\phi)^2 + \sum_i \left(\frac{\partial R_i}{R_i}\right)^2 
+ \sum_{i<j} \left(\frac{R_i}{R_j}\mathcal{F}^{i(1)}_j\right)^2 
\right. \\
\left.
+ \sum_i \left(R_i \mathcal{F}^{(2)i} \right)^2  
+ \left( l_s^2 F^{(3)} \right)^2 
+ \sum_i \left(\frac{l_s^2}{R_i}F^{(2)}_i\right)^2 
+ \sum_{i<j} \left(\frac{l_s^2}{R_i R_j}F^{(1)}_{ij}\right)^2 \right]\
,
\end{multline}
where the first five terms  come from the reduction of the Einstein--Hilbert
term and the last three terms from the kinetic term of the two-form.

Putting together the forms of the same degree, we see that their
coefficients form the Weyl orbit $\Phi_{s}$, of the string tension 
($\mathcal{F}_\lambda^{(3)}$), the Weyl orbit $\Phi_{\rm KK}$ of the
Kaluza--Klein and winding states ($\mathcal{F}_\lambda^{(2)}$),
and the set of positive roots $\Phi_+=\{e_i\pm e_j, i<j\}$ 
($\mathcal{F}_\alpha^{(1)}$).
We can therefore rewrite the action in the Weyl-invariant form:
\begin{eqnarray}
\label{wit}
S_{{\rm NS},10-d}&=\int {\rm d}^{10-d}x \sqrt{-g}~\frac{V}{g_s^2 l_s^8}
\left[ 
R + \partial\varphi \cdot \partial\varphi 
+ \sum_{\alpha\in\Phi_+}   e^{-2 \langle\varphi,\alpha\rangle} 
            \left(\mathcal{F}^{(1)}_{\alpha}\right)^2 
\right.\nonumber\\
&\left. 
+ \sum_{\lambda\in\Phi_{\rm KK}} e^{-2 \langle\varphi,\lambda\rangle} 
            \left(\mathcal{F}^{(2)}_{\lambda}\right)^2  
+ \sum_{\lambda\in\Phi_s} e^{-2 \langle\varphi,\lambda\rangle}  
            \left( \mathcal{F}^{(3)}_{\lambda}\right)^2 
\right] \ , 
\end{eqnarray}
where $\varphi=(\ln g_s,\ln R_1,\dots,\ln R_d)$ is the vector of 
dilatonic scalars, $\langle\varphi,\lambda\rangle$ the 
duality bracket in Eq. \eqref{dualb} and $\partial\varphi\cdot\partial\varphi$
the Weyl-invariant kinetic term obtained from the non-diagonal
metric \eqref{tm}. A diagonal metric on the dilatonic scalars
is recovered upon going to the Einstein frame.

The Weyl group acts by permuting the various weights appearing
in Eq. \eqref{wit}, and the invariance in the gauge sector is 
therefore manifest. As for the scalars, the set of positive
roots $\Phi_+$ is not invariant under Weyl reflections,
but the Peccei--Quinn scalars undergo non-linear
transformations $\mathcal{A}^{(0)}\rightarrow e^{-2 \langle\varphi,
\alpha\rangle} 
\mathcal{A}^{(0)}$ that compensate the sign change \cite{Lu:1996ge}.
The Peccei--Quinn scalars therefore appear as displacements along
the positive (non-compact) roots. Together with the dilatonic
(non-compact) scalars $\varphi$, they generate the solvable Lie
subalgebra that forms the tangent space of the moduli
space $\mathcal{H}$ \cite{Andrianopoli:1996bq,Andrianopoli:1996ve,
Andrianopoli:1997zg,Trigiante:1998vu}.

We have so far concentrated on the Neveu--Schwarz sector, 
but the same reasoning can be applied to the full type II
action. The T-duality Weyl symmetry can, however, be
exhibited only by dualizing the $p$-form gauge fields 
$\mathcal{G}^{(p)}=d\mathcal{R}^{(p-1)}$ into lower
rank $(10-d-p)$-form gauge fields when possible, and keeping them
together when their dual when the self-duality condition $10-d-p=p$
is satisfied. We then  obtain, for the action of the Ramond fields 
\begin{eqnarray}
\label{witrr}
S_{\rm RR}&=\int {\rm d}^{10-d}x \sqrt{-g}~\frac{V}{g_s^2 l_s^8}
\left[ \sum_{\lambda\in\Phi_{\rm DI}} e^{-2 \langle\varphi,\lambda\rangle} 
            \left(\mathcal{G}^{(1)}_{\lambda}\right)^2  
+ \sum_{\lambda\in\Phi_{\rm D0}} e^{-2 \langle\varphi,\lambda\rangle}  
            \left( \mathcal{G}^{(2)}_{\lambda}\right)^2 \right.\nonumber\\
&\left.+ \sum_{\lambda\in\Phi_{\rm D1}} e^{-2 \langle\varphi,\lambda\rangle} 
            \left(\mathcal{G}^{(3)}_{\lambda}\right)^2  
+ \sum_{\lambda\in\Phi_{\rm D2}} e^{-2 \langle\varphi,\lambda\rangle}  
            \left( \mathcal{G}^{(4)}_{\lambda}\right)^2 
\right] \ , 
\end{eqnarray}
where $\Phi_{\rm DI}$,$\Phi_{\rm D0}$,$\Phi_{\rm D1}$,$\Phi_{\rm D2}$
denote the Weyl orbits with highest weight $1/g_s R_i$,
$1/g_s l_s$, $R_i/g_s l_s^2$, $1/g_s l_s^3$ respectively,
corresponding in turn to the two spinor representations.  

\subsection{Spectral flow and Borel generators\label{spf}}
Having discussed the structure of the Weyl group 
we now want to investigate the
full $SO(d,d,\Zint)$ symmetry. For this purpose, it is
instructive to go back to the perturbative multiplet
of Kaluza--Klein and winding states. The action of the
Weyl group on the highest weight $1/R_d$ of the vector representation
generates an orbit
of $2d$ elements, $1/R_i$ and $R_i$. However, a particle can
have any number of momentum excitation along each axis,
and wind along any cycle of the torus $T^d$. It is therefore
described by integer momenta $m_i$ and winding numbers $m^i$,
so that its mass on an arbitrary torus reads
\begin{equation}
\mathcal{M}^2= m_i g^{ij} m_j + m^i g_{ij} m^j \sp i,j = 1 \ldots d \ ,
\end{equation}
when $B_{ij}=0$.
This mass formula is then invariant under modular transformations
$\gamma^i \rightarrow \gamma^i + \Delta A_j^i \gamma^j$ of the torus, 
\ie integer shifts
$A_j^i \rightarrow A_j^i + \Delta A_j^i$ of the off-diagonal term of 
the metric (no sum on $i$) 
\begin{equation}
{\rm d}s^2_d = R_i ^2 ( dx ^i + A_j^i dx^j )^2 + g_{jk} dx^j dx^k \ ,
\end{equation}
upon transforming the momenta and winding as 
\begin{equation}
\label{sfla}
m_k \rightarrow m_k - \Delta A_{k}^i m_i \sp
m^k \rightarrow m^k + \delta_i^{k} \Delta A_{j}^i m^j \ .  
\end{equation}
This transformation generates a {\it spectral flow} on the lattice of charges
$m_i$ and $m^i$.

In addition, being charged under the gauge potential $B_{\mu i}$, the
momentum of the particle shifts according to $m_i \rightarrow \tilde m_i
= m_i + B_{ij} m^j$, yielding the mass \eqref{vectormass}. From this,
we see that 
the Borel generator $B_{ij}\rightarrow B_{ij}+\Delta
B_{ij}$ induces a spectral flow
\begin{equation}
\label{sflb}
m_k \rightarrow m_k + \Delta B_{jk}m^j \sp m^k \rightarrow m^k\ .
\end{equation}

The two spectral flows \eqref{sfla} and \eqref{sflb}
 can be understood in a unified way as
translations on the weight lattice by positive roots. Indeed,
the set of all positive roots of $SO(d,d)$ includes the $Sl(d)$
roots $e_j - e_i, i<j$, images of the simple roots
$\alpha_i=e_{i+1}-e_i, 1 \leq i \leq d-1$
under the Weyl group $\mathcal{S}_d$ of $Sl(d)$, as well as the
roots $e_i + e_j$, which are images of the T-duality simple root $\alpha_0=
e_1+e_2$. The translation by a root $e_j-e_i$ generates infinitesimal
rotations in the $(i,j)$ plane\footnote{The Borel generators 
$E_\alpha$ actually
either translate the weight vectors $\lambda$ or annihilate them
.}:
\begin{equation}
\Delta |-e_k\rangle = -\Delta A_k^i |-e_i\rangle \sp
\Delta |e_k\rangle = \delta^k_i  \Delta A_j^i |e_j \rangle 
\end{equation}
equivalent to the spectral flow in Eq. \eqref{sfla}, whereas
translations by a root $e_i+e_j$ generate an infinitesimal
$B_{ij}$ shift:
\begin{equation}
\Delta |-e_k\rangle = \Delta B_{jk} |e_j\rangle \sp
\Delta |e_k\rangle = 0
\end{equation}
as in Eq. \eqref{sflb}. 
The moduli $A_j^i$ and $B_{ij}$ can therefore be identified as
displacements on the moduli space $\mathcal{H}$ along the positive
roots $e_i -e_j$ and $e_i + e_j$. We note that the two displacements
do not necessarily commute and that only {\it
integer} shifts are symmetries of the charge lattice.

\subsection{D-branes and T-duality invariant mass\label{dbtd}}
In order to study the analogous properties of the D-brane states,
we may try to write down the moduli matrix $M_S \in SO(d,d,\Real)/
SO(d)\times SO(d)$ in the spinorial representation and look
for the transformations of charges that leave the mass
$m^t M_S  m $ invariant, when now $m$ is a spinor of D-brane charges.
It is in fact much easier to study the D-brane configuration
itself and compute its Born--Infeld mass
\cite{Pioline:1997pu,Ishikawa:1996aw}. 

BPS D-brane states
are obtained by wrapping D$p$-branes on a supersymmetric $p$-cycle
of the compactification manifold. In the case of a torus $T^d$,
this is simply a straight cycle, and in the static gauge
the embedding is 
specified by a set of integer (winding) numbers $N^i_\alpha$:
\begin{equation}
X^i = N^i_\alpha \sigma^{\alpha}\sp i=1\dots d\sp \alpha=1\dots p\ ,
\end{equation}
where $\sigma^\alpha$ and $X^i$ are the space-like world-volume and embedding
coordinates respectively. The numbers $N^i_\alpha$ can, however,
be changed by a world-volume diffeomorphism, and one should
instead look at the invariant
\begin{equation}
\label{n4def}
m^{ijkl} = \epsilon^{\alpha\beta\gamma\delta} N^i_\alpha
N^j_\beta N^k_\gamma N^l_\delta\ ,
\end{equation}
where we restricted to $p=4$ for illustrative purposes.
$m^{ijkl}$ is a four-form integer charge that specifies
the four-cycle in $T^d$.
In addition, the D-brane supports a $U(1)$ gauge field
that can be characterized by the invariants
\begin{equation}
\label{n2n0def}
m^{ij} = \frac{1}{2}\epsilon^{\alpha\beta\gamma\delta} N^i_\alpha
N^j_\beta F_{\gamma\delta}\ \sp 
m = \frac{1}{8}\epsilon^{\alpha\beta\gamma\delta} F_{\alpha\beta} 
F_{\gamma\delta} \ , 
\end{equation}
which are again integer-valued, because of the flux and instanton-number
integrality. The charges $\mathcal{N}=\{m,m^{ij},m^{ijkl},\dots\}$ 
constitute  precisely
the right number to make a spinor representation of $SO(d,d,\Zint)$
when $p=d$ or $p=d+1$ (depending on the type of theory and
dimensionality of the torus); indeed, the spinor representation of
$SO(d,d)$ decomposes under $Sl(d)$ as a sum of even or odd forms,
depending on the chirality of the spinor. The Chern--Simons coupling
\eqref{rrcoup} can be rewritten in terms of these charges (up
to corrections when $B\ne 0$) as
\begin{equation}
\int e ^{\hat B+\alpha' F} \R = m \R_0 + \frac{1}{2} m^{ij}
\R_{0ij} + \frac{1}{4!} m^{ijkl}
\R_{0ijkl} + \dots
\end{equation}
so that (for $p=4$) the instanton number $m$ can be identified as the
D0-brane charge, the flux $m^{ij}$ as the D2-brane charge and
$m^{ijkl}$ as the D4-brane charge. Configurations
with $m\ne 0$ exist in SYM theory on a torus, even for a $U(1)$ gauge group,
and correspond to torons \cite{'tHooft:1979uj,Guralnik:1997sy,
Guralnik:1997th}. 

The mass of the wrapped D-brane can be evaluated by using
the Born--Infeld action \eqref{borninfeld}, and depends only on the
parametrization-independent integer charges $m,m^{ij},m^{ijkl},\dots$
Explicitly, we obtain, for $p=d$, the T-duality invariant mass 
formula:\footnote{This expression was originally derived in Ref.
\cite{Pioline:1997pu} by a sequence of T-dualities and covariantizations.}
\begin{subequations}
\label{dparticle}
\begin{eqnarray}
\M^2 &=& \frac{1}{g_s^2 l_s^2}\tilde m^2 + 
\frac{1}{2~g_s^2 l_s^6} (\tilde m^{ij})^2 +
\frac{1}{4!~g_s^2 l_s^{10}}(\tilde m^{ijkl})^2 +\dots \\
\tilde m &=& m +  \frac{1}{2} m^{ij} B_{ij} + \frac{1}{8} m^{ijkl}
B_{ij} B_{kl}+\dots \\
\tilde m^{ij} &=& m^{ij} +  \frac{1}{2} m^{klij} B_{kl} + \dots\\
\tilde m^{ijkl} &=& m^{ijkl} + \dots
\end{eqnarray}
\end{subequations}
where the dots stand for the obvious extra terms when $d\ge 4$.
A similar expression holds for $p=d+1$ and yields the tension
of D-strings:
\begin{subequations}
\label{dstring}
\begin{eqnarray}
\T_1^2 &=& \frac{1}{g_s^2 l_s^6}(\tilde m^i)^2 + 
\frac{1}{3!~g_s^2 l_s^{10}} (\tilde m^{ijk})^2 +
\frac{1}{5!~g_s^2 l_s^{14}}(\tilde m^{ijkll})^2 +\dots \\
\tilde m^i &=& m^i +  \frac{1}{2} m^{jki} B_{jk} + \frac{1}{8} m^{jklmi}
B_{jk} B_{lm}+\dots \\
\tilde m^{ijk} &=& m^{ijk} +  \frac{1}{2} m^{lmijk} B_{lm} + \dots\\
\tilde m^{ijklm} &=& m^{ijklm} + \dots
\end{eqnarray}
\end{subequations}
where the integer charges read, e.g. for $p=5$,
\begin{subequations}
\begin{eqnarray}
m^{ijklm} &=& \epsilon^{\alpha\beta\gamma\delta\epsilon} N^i_\alpha
N^j_\beta N^k_\gamma N^l_\delta N^m_\epsilon
\\
m^{ijk} &=& \frac{1}{2}\epsilon^{\alpha\beta\gamma\delta\epsilon} N^i_\alpha
N^j_\beta N^k_\gamma F_{\delta\epsilon} \\
m^i &=& \frac{1}{8}\epsilon_{\alpha\beta\gamma\delta\epsilon} 
N^i_\alpha F_{\beta\gamma} F_{\delta\epsilon} \ . 
\end{eqnarray}
\end{subequations}

The mass formulae \eqref{dparticle} and \eqref{dstring} 
 hold for 1/2-BPS states only; they are the 
analogues of Eq. \eqref{vectormass} for the two spinor representations
of $SO(d,d)$. They can
be derived by analysing the BPS eigenvalue equation in a similar way
as in Subsection \ref{bpss}. This analysis is carried out 
in Appendix \ref{d0d2d4}, and yields, in addition, 
the conditions for the state to be 1/2-BPS, as well as the
extra contribution to the mass in the 1/4-BPS case. In the $d\le 6$ case,
we find a set of conditions:
\begin{subequations}
\label{4cond}
\begin{eqnarray}
k^{ijkl} & \equiv & m^{[ij} m^{kl]} + m~m^{ijkl} = 0 \\
k^{i;jklmn} & \equiv & m^{i[j} m^{klmn]} + m~m^{ijklmn} =  0\\
k^{ij;klmnpq} & \equiv &n^{ij} n^{klmnpq} + n^{ij[kl}n^{mnpq]}=  0
\end{eqnarray}
\end{subequations}
analogous to the level-matching condition $\|m\|^2=0$ 
on the perturbative states. In contrast to the latter,
they have a very clear geometric origin, since they
can be derived by expressing the charges $m$ in terms
of the integer numbers $N^i_\alpha$  (Eq. \eqref{dparticle}).
For $d=6$, they transform 
in a $\irrep{15}+\irrep{36}+\irrep{15}=\irrep{66}$ irrep of the T-duality group
$SO(6,6,\Zint)$. The last line in \eqref{4cond} drops when $d=5$, giving
a $\irrep{5} + \irrep{5}=\irrep{10}$ irrep of $SO(5,5,\Zint)$.
When $d=4$, only the 
$k^{1234} = m^2 \wedge m^2 + m~m^4 \equiv 0$ component remains,
which is a singlet under $SO(4,4,\Zint)$.

When the conditions $n=0$ in \eqref{4cond} are not met, 
the state is at most 1/4-BPS, and its mass receives an
extra contribution, e.g. for $d=5$:
\begin{equation}
\hspace*{-5mm}
\M^2 = \frac{1}{g_s^2 l_s^2} \left[
\tilde m^2 + \frac{1}{2l_s^4} \left(\tilde m^{ij}\right)^2 +
\frac{1}{4!l_s^8}\left(\tilde m^{ijkl}\right)^2 +
\sqrt{ \frac{1}{4! l_s^{12}}\left(\tilde k^{ijkl}\right)^2 + 
\frac{1}{5! l_s^{16}}\left(\tilde k^{i;jklmn}\right)^2 } \right] \  , 
\end{equation}
where the shifted charges are given by
\begin{equation}
\tilde k^{ijkl}= k^{ijkl} + B_{mn} k^{m;nijkl}\sp \;\;\; 
\tilde k^{i;jklmnp} = k^{i;jklmnp}\ .
\end{equation}
For $d=6$, there are still conditions to be imposed in order for
the state to be 1/4-BPS instead of simply 1/8-BPS, which are
now cubic in the charges $m$ and transform as a \irrep{32}
of $SO(6,6,\Zint)$ (see Appendix \ref{bpsmass} and Subsection 
\ref{bpscond}).

\clearpage
\section{U-duality in toroidal compactifications of M-theory
\label{udua} } 

T-duality is only a small part of the symmetries of toroidally
compactified string theory, namely the part visible in perturbation
theory. We shall now extend the  techniques of Section \ref{tdua} in order to
study the algebraic structure of the non-perturbative symmetries,
which go under the name of U-duality.
In this section, we focus on 
the subgroup of the U-duality symmetry that preserves compactifications on
rectangular tori with vanishing expectation values of the gauge potentials. 
The most general case of non-rectangular tori with gauge potentials,
for which the full U-duality symmetry can be exhibited, 
is discussed in the next section.

\subsection{Continuous R-symmetries of the superalgebra}
As in our presentation of uncompactified M-theory in Section
\ref{mtheory}, the superalgebra offers a convenient starting
point to discuss the symmetries of M-theory compactified
on a torus $T^d$. The $N=1$, 11D supersymmetry algebra
is preserved under toroidal compactification:
the generators $Q_\alpha$ merely decompose
as bispinor representations of the unbroken 
group $SO(1,10-d)\times SO(d)$, and form an
$N$-extended super-Poincar{\'e} algebra in dimensions
$D=11-d$. The first factor $SO(1,10-d)$
corresponds to the Lorentz group in the uncompactified
dimensions and is actually  part of the superalgebra,
while the second only acts as an automorphism
thereof, and is also known as an R-symmetry\footnote{The
R-symmetry is actually part of the {\it local} supersymmetry,
but we are only interested in its global flat limit.}.
There can be automorphisms beyond the obvious $SO(d)$
symmetry, however, and these are expected to be symmetries of the 
field theory.

This symmetry enhancement can be observed at the level of the
Clifford algebra itself \cite{Julia:1979fw,Lu:1996ge}. 
The Gamma matrices $\Gamma_M, M=0,d+1 \dots
10$ of eleven-dimensional supersymmetry can be kept to
form a (reducible) Clifford algebra  of $SO(1,10-d)$,
while the matrices $\Gamma_{I}$, $I=1\dots d$ form an
internal Clifford algebra. Note that we have chosen here, in contrast
to the notation of the rest of the review, the internal indices
running from 1 to $d$. 
 The generators $\Gamma_{IJ}$ generate 
the $SO(d)$ R-symmetry, but they can be supplemented by 
generators $\Gamma_I$ to form the Lie algebra of
a larger R-symmetry group $SO(d+1)$ \footnote{
This is the basis for the twelve-dimensional S-theory
 proposal \cite{Bars:1996dz}.
It is important that these generators commute with the
momentum charge $C\Gamma_\mu$.}. It was the attempt to exhibit
the $SO(8)$ symmetry of 11D SUGRA compactified on $T^7$
that led to the discovery of hidden symmetries \cite{Cremmer:1979up}.

The R-symmetry group is actually larger still.
Consider the algebra generated by
$\Gamma_{(2)},\Gamma_{(3)},\Gamma_{(6)},\Gamma_{(7)}$,
where the subscripts denote the number of antisymmetric internal
indices, and the corresponding generators are dropped when
the number of internal directions is insufficient:
\begin{itemize}
\item For $d=2$, the only generator $\Gamma_{IJ}=\Gamma_{12}$ generates a
$U(1)$ R-symmetry.
\item For $d=3$, $\Gamma_{(2)}$ and $\Gamma_{(3)}$ commute,
and generate an $SO(3)\times U(1)$ symmetry.
\item For $d=4$, $\Gamma_{(3)} = \Gamma_+ \Gamma_{(1)}$, where 
$\Gamma_+$ is the space-time or internal chirality (see Eq. \eqref{gamma})
and, together with $\Gamma_{(2)}$, generates an $SO(5)$ symmetry.
\item For $d=5$, $\Gamma_{(2)} \pm \Gamma_+ \Gamma_{(3)}$
generate two commuting $SO(5)$ subgroups.
\item For $d=6$, $\Gamma_{(6)}$ appears in the commutator 
$[\Gamma_{(3)},\Gamma_{(3)}]$ and a $USp(8)$ 
is generated. 
\item For $d=7$ (resp. $d=8$) the generator
$\Gamma_{(7)}$ comes into play and one obtains an $SU(8)\times U(1)$
(resp. $SO(16)$) R-symmetry group.
\end{itemize}
The various R-symmetry groups are summarized in the right column
of Table \ref{bars}, which furthermore gives the decomposition
of the 528 central charges on the right-hand side of Eq.
\eqref{susy1} under the Lorentz group $SO(1,10-d)$ in the
uncompact directions
and the R-symmetry group. The various columns correspond
to distinct $SO(1,10-d)$ representations, after dualizing (moving) central
charges into charges with less indices when possible.
In all these cases, the superalgebra can be recast in a
form manifestly invariant under the R-symmetry. Here we
collect the cases $D=4,5,6$, including the central charges,
which transform linearly under the R-symmetry:
\begin{itemize}
\item 
For $D=4$ ($d=7$), the 32 supercharges split into 8 complex Weyl
spinors transforming as an \irrep{8} $\oplus$ $\irrep{\bar{8}}$ of $SU(8)$:
\begin{subequations}
\begin{eqnarray}
\left\{ Q_{\alpha A},Q_{\dot{\beta}\bar{B}}\right\} &=&\sigma _{\alpha \dot{
\beta}}^\mu \,P_\mu \,\,\delta _{A\bar{B}}   \\
\left\{ Q_{\alpha A},Q_{\beta B}\right\} &=&\epsilon_{\alpha\beta }\,Z_{AB}\\
\left\{ Q_{\dot{\alpha}\bar{A}},Q_{\dot{\beta}\bar{B}}\right\} &=&
\epsilon_{\dot{\alpha}\dot{\beta}}\,Z_{\bar{A}
\bar{B}}^{*},  
\label{4dz}
\end{eqnarray}
\end{subequations}
where $\mu =0,1,2,3$ are $SO(3,1)$ vector indices, $\alpha ,\dot{\alpha}=1,2$
are Weyl spinor indices, and $A,\bar{A}=1,\cdots ,8$ are \irrep{8},
\irrep{{\bar 8}} indices of $SU(8)$. The central charges are
incorporated into a complex antisymmetric matrix $Z_{AB}$.

\item For $D=5$ ($d=6$), the 32 supercharges split into $8$ Dirac
spinors of $SO(4,1)$, transforming in the fundamental representation
of $USp(8)$. The $N=8$ superalgebra in a $USp(8)$ basis is  
\begin{equation}
\left\{ Q_{\alpha A},Q_{\beta B}\right\} = P_\mu \left( C\gamma ^\mu
\right)_{\alpha \beta }\Omega _{AB}+ C_{\alpha \beta }\,\,Z_{AB}  \label{5dz}
\end{equation}
where $\mu =0,1,2,3,4$ are $SO(4,1)$ vector indices, $\alpha =1,2,3,4$ are
Dirac spinor indices, $A=1,\cdots ,8$ are indices in the \irrep{8}
of $USp(8)$, and $\Omega_{AB}$ is the invariant symplectic form
and $Z_{AB}$ is the central charge matrix. 

\item For $D=6$ ($d=5$), the 32 supercharges form 4 
complex spinors 
transforming in the $(\irrep{4},\irrep{1})+(\irrep{1},\irrep{4})$ 
of $SO(5)\times SO(5)$ and the superalgbra
takes the form 
\newcommand{\pslash}{{p \hspace{-5pt} \slash}}
\begin{eqnarray}
\left\lbrace Q^a_\alpha,Q^b_\beta\right\rbrace& \!\! = \!\! &\omega^{ab}
\gamma^\mu_{\alpha\beta} p_\mu  ,\\[2mm] \left\lbrace
Q^a_\alpha,\bar Q^b_{\overline{\beta}}\right\rbrace & \!\! = \!\! &
\delta_{\alpha\overline\beta} Z^{ab} ,
\end{eqnarray}
where $a,b = 1,\ldots,4$ are $SO(5)$ spinor indices and $\omega^{ab}$
is an invariant antisymmetric matrix, from the local isomorphism
$SO(5)=USp(4)$. The 16 central charges are incorporated in a
matrix $Z^{ab}$ transforming as a
bispinor under the  R-symmetry $SO(5)\times SO(5)$ and 
satisfying the reality condition $Z^* =\omega Z\omega^t$.
\end{itemize}
The R-symmetries that we have discussed here will be of use
in the next section to determine the scalar manifold of
the compactified 11D SUGRA and hence the global symmetries.

\begin{table}[H] 
$$\hspace*{-1.5cm}
\setlength{\extrarowheight}{0pt}
\begin{array}{|c|c|c|c|c|c|c|c|c|}
\hline
d & Q_\alpha ^a &
\begin{matrix} p=0\\ Z^I,Z^{IJ}  \\ Z^{IJKLM} \end{matrix} & 
\begin{matrix} p=1\\ Z^\mu ,Z^{\mu I}  \\ Z^{\mu IJKL}\end{matrix} & 
\begin{matrix} p=2\\ Z^{\mu I}  \\ Z^{\mu \nu IJK} \end{matrix}& 
\begin{matrix} p=3\\ Z^{\mu \nu\rho IJ} \end{matrix} & 
\begin{matrix} p=4\\  Z^{\mu \nu\rho\sigma I}  \end{matrix} & 
\begin{matrix} p=5\\  Z^{\mu \nu\rho\sigma\tau}  \end{matrix} & 
H\\ \hline
1 & ({\bf \pm ,16}) & \begin{matrix} 1+0  \\ +0  \end{matrix} & 
\begin{matrix} 1+1  \\ +0  \end{matrix} & \begin{matrix}1  \\
+0\end{matrix}&
0 & 1 & \begin{matrix} 1^{+}  \\ +1^{-}  \end{matrix} & 1 \\ \hline
2 & \left( {\bf 2,16}\right) & 
\begin{matrix}2+1  \\ +0 \\ = {\bf 2+1} \end{matrix} &
\begin{matrix}1+2  \\ +0  \\ ={\bf 2+1}  \end{matrix}& 
\begin{matrix}1+0  \\ ={\bf 1} \end{matrix}& 1 &
\begin{matrix}\left[ 1\right]  \\ +2  \\ = {\bf 2}+{\bf 1} \end{matrix}&
\begin{matrix}\left( 1\right)  \\ _{move}  \end{matrix} &
SO(2) \\ \hline
3 & \begin{matrix} ({\bf 2},{\bf 8}^+) \\ + ({\bf 2},{\bf 8}^-)
\end{matrix}&
\begin{matrix}3+3  \\ +0 \\ = {\bf 3^+}+{\bf 3^-} \end{matrix} &
\begin{matrix}1+3  \\ +0 \\ = {\bf 3}+{\bf 1} \end{matrix} &
\begin{matrix}1+1  \\ = {\bf 1}+{\bf 1} \end{matrix} &
\begin{matrix}3+[1]  \\ = {\bf 3}+{\bf 1} \end{matrix} &
\begin{matrix}3^+  \\ +3^- \\ = {\bf 3^+}+{\bf 3^-} \end{matrix} &
\begin{matrix}\left( 1\right)  \\ _{move}  \end{matrix} &
\begin{matrix} SO(2)\\ \times U(1) \end{matrix} \\ \hline
4 & ({\bf 4},{\bf 8}) &
\begin{matrix}4+6  \\ +0 \\ = {\bf 10} \end{matrix} &
\begin{matrix}1+4  \\ +1 \\ = {\bf 5}+{\bf 1} \end{matrix} &
\begin{matrix}1+4\\+[1]  \\ = {\bf 5}+{\bf 1} \end{matrix} &
\begin{matrix}6+[4]  \\ = {\bf 10}\end{matrix} &
\begin{matrix}\left( 4\right)  \\ _{move}  \end{matrix} &
\begin{matrix}\left( 1\right)  \\ _{move}  \end{matrix} &
SO(5) \\ \hline
5 & 
\begin{matrix} ({\bf 4},{\bf \bar 4}) \\ + ({\bf \bar 4},{\bf 4})
\end{matrix}&
\begin{matrix}5+10  \\ +1 \\ = ({\bf 4},{\bf 4}) \end{matrix} &
\begin{matrix}1+5  \\ +5+[1] \\ = ({\bf 5},{\bf 1})\\+({\bf 1},{\bf
    5})\\+2({\bf 1},{\bf 1}) \end{matrix} &
\begin{matrix}1+10\\+[5]  \\ = ({\bf 4},{\bf 4}) \end{matrix} &
\begin{matrix}10^+ + 10^- \\ = ({\bf 10},{\bf 1})\\+({\bf 1},{\bf 10})
  \end{matrix} &
\begin{matrix}\left( 5\right)  \\ _{move}  \end{matrix} &
\begin{matrix}\left( 1\right)  \\ _{move}  \end{matrix} &
\begin{matrix}SO(5)\\ \times SO(5) \end{matrix}\\ \hline
6 & ({\bf 8},{\bf 4}) &
\begin{matrix}6+15  \\ +6 \\+[1]\\ = {\bf 27}+{\bf 1} \end{matrix} &
\begin{matrix}1+6  \\ +15+[6] \\ = {\bf 27}+{\bf 1}\end{matrix} &
\begin{matrix}1+20\\+[15]  \\ = {\bf 36} \end{matrix} &
\begin{matrix}\left( 15\right)  \\ _{move}  \end{matrix} &
\begin{matrix}\left( 6\right)  \\ _{move}  \end{matrix} &
\begin{matrix}\left( 1\right)  \\ _{move}  \end{matrix} &
USp(8) \\ \hline
7 & 
\begin{matrix} ({\bf 8^+},{\bf 2}) \\ + ({\bf 8^-},{\bf \bar 2}) 
    \end{matrix}&
\begin{matrix}7+21  \\ +21 \\+[7]\\ = {\bf 28}_c\end{matrix} &
\begin{matrix}1+7  \\ +35+[21] \\ = {\bf 63}+{\bf 1}\end{matrix} &
\begin{matrix}1^\pm+35^\pm\\ = {\bf 36}_c \end{matrix} &
\begin{matrix}\left( 21\right)  \\ _{move}  \end{matrix} &
\begin{matrix}\left( 7\right)  \\ _{move}  \end{matrix} & 0 &
SU(8) \\ \hline
8 & ({\bf 16},{\bf 2}) & 
\begin{matrix}8+28  \\ +56 \\+[28]\\ = {\bf 120}\end{matrix} &
\begin{matrix}1+8+70  \\ +[1+56] \\ = {\bf 135}+{\bf 1}\end{matrix} &
\begin{matrix}\left( 1+56\right)  \\ _{move}  \end{matrix} &
\begin{matrix}\left( 28\right)  \\ _{move}  \end{matrix} & 0 & 0 &
SO(16) \\ \hline
\end{array}
\setlength{\extrarowheight}{5pt}
$$
\caption[R-symmetry classification of super- and central
charges.]{Classification of the supercharges and central charges w.r.t
the Lorentz/R-symmetry group $SO(1,10-d)\times H$. 
Irreps of $H$ are in bold face. Charges
in parenthesis are Poincar{\'e}-dualized ({\it moved}) into charges in square
brackets. Adapted from Ref. \cite{Bars:1996dz}.
\label{bars}}
\end{table}


\subsection{Continuous symmetries of the effective action\label{contsymu}}
In our discussion
of the continuous symmetry of the effective
action of the toroidally compactified type IIA theory 
in Subsection \ref{cseas}, 
we have intentionnally focused our attention on the Neveu--Schwarz
sector, and have briefly described how the Ramond fields
would transform under the symmetries of the Neveu--Schwarz
scalar manifold. The distinction between Neveu--Schwarz
and Ramond sectors is however an artefact of perturbation theory
and, as we discussed in Section \ref{mtheory}, the two sets of fields
are unified in the 11D SUGRA description. They
mix under the eleven-dimensional Lorentz symmetries
unbroken by the compactification 
on $T^d$ \footnote{
Note that $d$ has been upgraded by one unit with
respect to the previous section.}, namely $Sl(d,\Real)$.
The low-energy effective action therefore admits a
continuous symmetry group $G_d$ containing 
\begin{equation}
SO(d-1,d-1,\Real)\bowtie Sl(d,\Real)\ ,
\end{equation}
where the symbol $\bowtie$ denotes the group generated by the
two non-commuting subgroups. As found by Cremmer and Julia 
\cite{Cremmer:1980gs,Julia:1980gr},
the groups $G_d$ turn out to correspond to the $E_{d(d)}$ series, 
listed in Table \ref{tud}. 

\begin{table}[h] 
\begin{center}
\begin{tabular}{|r|r||l|l|}
\hline
$D$ & $d$ & $G_d=E_{d(d)}$ & $H_d$\\ 
\hline
10& 1 &$\Real^+$                             & 1\\
9 & 2 &$Sl(2,\Real)\times \Real^+$          &$U(1)$\\
8 & 3 &$Sl(3,\Real)\times Sl(2,\Real)$&$SO(3)\times U(1)$\\
7 & 4 &$Sl(5,\Real)$                  &$SO(5)$\\
6 & 5 &$SO(5,5,\Real)$                &$SO(5)\times SO(5)$\\
5 & 6 &$E_{6(6)}$                     &$USp(8)$\\
4 & 7 &$E_{7(7)}$                     &$SU(8)$\\
3 & 8 &$E_{8(8)}$                     &$SO(16)$\\
\hline
\end{tabular}
\end{center}
\caption[Cremmer--Julia symmetry groups]{Cremmer--Julia symmetry groups and their maximal compact subgroups.
\label{tud}}
\end{table}

The notation $E_{d(d)}$ denotes a particular non-compact
form of the exceptional group $E_d$, 
namely its normal real form\footnote{
The normal real form has all its Cartan generators and positive roots
non-compact, and is the maximal non-compact real form of the complex
algebra $E_d(\mathbb{C})$ \cite{Helgason:1962,Gilmore:1974}.}, and 
from now on this distinction will be omitted. 
As evident from their Dynkin diagrams shown in Table
\ref{dynfigure}, the groups $E_d$ form an increasing family, 
whose members are related by a process of group disintegration
reflecting the decompactification of one compact direction in $T^d$.
This is displayed in Table \ref{dynfigure}, and will be 
discussed more fully in the next subsection.

\begin{table}
\begin{center}
\begin{picture}(200,490)(-30,-20)

\put(0,0){
\begin{picture}(190,50)
\thicklines
\put(-30,25){$E_9=\hat E_8$}
\multiput(0,0)(30,0){7}{\circle{8}}
\put(210,0){\circle*{8}}
\put(0,-10){\makebox(0,0){2}}
\put(30,-10){\makebox(0,0){4}}
\put(60,-10){\makebox(0,0){6}}
\put(90,-10){\makebox(0,0){5}}
\put(120,-10){\makebox(0,0){4}}
\put(150,-10){\makebox(0,0){3}}
\put(180,-10){\makebox(0,0){2}}
\put(210,-10){\makebox(0,0){1}}
\multiput(4,0)(30,0){7}{\line(1,0){22}}
\put(60,4){\line(0,1){22}}
\put(60,30){\circle{8}}
\put(52,30){\makebox(0,0){3}}
\end{picture}}

\put(0,70){
\begin{picture}(190,50)
\thicklines
\put(-30,25){$E_8$}
\multiput(0,0)(30,0){7}{\circle{8}}
\put(0,-10){\makebox(0,0){2}}
\put(30,-10){\makebox(0,0){4}}
\put(60,-10){\makebox(0,0){6}}
\put(90,-10){\makebox(0,0){5}}
\put(120,-10){\makebox(0,0){4}}
\put(150,-10){\makebox(0,0){3}}
\put(180,-10){\makebox(0,0){2}}
\multiput(4,0)(30,0){6}{\line(1,0){22}}
\put(60,4){\line(0,1){22}}
\put(60,30){\circle{8}}
\put(52,30){\makebox(0,0){3}}
\end{picture}}

\put(0,140){
\begin{picture}(190,50)
\thicklines
\put(-30,25){$E_7$}
\multiput(0,0)(30,0){6}{\circle{8}}
\put(0,-10){\makebox(0,0){2}}
\put(30,-10){\makebox(0,0){3}}
\put(60,-10){\makebox(0,0){4}}
\put(90,-10){\makebox(0,0){3}}
\put(120,-10){\makebox(0,0){2}}
\put(150,-10){\makebox(0,0){1}}
\multiput(4,0)(30,0){5}{\line(1,0){22}}
\put(60,4){\line(0,1){22}}
\put(60,30){\circle{8}}
\put(52,30){\makebox(0,0){2}}
\end{picture}}

\put(0,210){
\begin{picture}(190,50)
\thicklines
\put(-30,25){$E_6$}
\multiput(0,0)(30,0){5}{\circle{8}}
\put(0,-10){\makebox(0,0){1}}
\put(30,-10){\makebox(0,0){2}}
\put(60,-10){\makebox(0,0){3}}
\put(90,-10){\makebox(0,0){2}}
\put(120,-10){\makebox(0,0){1}}
\multiput(4,0)(30,0){4}{\line(1,0){22}}
\put(60,4){\line(0,1){22}}
\put(60,30){\circle{8}}
\put(52,30){\makebox(0,0){2}}
\end{picture}}

\put(0,280){
\begin{picture}(190,50)
\thicklines
\put(-30,25){$E_5=D_5$}
\multiput(0,0)(30,0){4}{\circle{8}}
\put(0,-10){\makebox(0,0){1}}
\put(30,-10){\makebox(0,0){2}}
\put(60,-10){\makebox(0,0){2}}
\put(90,-10){\makebox(0,0){1}}
\multiput(4,0)(30,0){3}{\line(1,0){22}}
\put(60,4){\line(0,1){22}}
\put(60,30){\circle{8}}
\put(52,30){\makebox(0,0){1}}
\end{picture}}

\put(0,350){
\begin{picture}(190,50)
\thicklines
\put(-30,25){$E_4=A_4$}
\multiput(0,0)(30,0){3}{\circle{8}}
\put(0,-10){\makebox(0,0){1}}
\put(30,-10){\makebox(0,0){1}}
\put(60,-10){\makebox(0,0){1}}
\multiput(4,0)(30,0){2}{\line(1,0){22}}
\put(60,4){\line(0,1){22}}
\put(60,30){\circle{8}}
\put(52,30){\makebox(0,0){1}}
\end{picture}}

\put(0,420){
\begin{picture}(190,50)
\thicklines
\put(-30,25){$E_3=A_2 \oplus A_1$}
\multiput(0,0)(30,0){2}{\circle{8}}
\put(0,-10){\makebox(0,0){1}}
\put(30,-10){\makebox(0,0){1}}
\multiput(4,0)(30,0){1}{\line(1,0){22}}
\put(60,30){\circle{8}}
\end{picture}}

\put(0,490){
\begin{picture}(190,50)
\thicklines
\put(-30,25){$E_2=A_1$}
\multiput(0,0)(30,0){1}{\circle{8}}
\end{picture}}

\end{picture}
\caption[Dynkin diagrams of $E_d$ and group-disintegrations]
{Dynkin diagrams of the $E_d$ series. The group
disintegration proceeds by omitting the rightmost node. The
integers shown are the Coxeter labels, that is the coordinates
of the highest root on all simple roots.\label{dynfigure}}
\end{center}
\end{table}

The occurrence of these groups 
can be understood by fitting the number of scalar fields (including
the duals of forms of higher degree) to
the dimension of a coset space $G_d/H_d$, where $H_d$ 
is the R-symmetry of the superalgebra
described in the previous section. In order to have
a positive metric for the scalars, it is necessary that
$H_d$ be the maximal compact subgroup of $G_d$.
Together with the dimension of the scalar manifold,
this suffices to determine $G_d$. 

Scalar fields
arise from the internal components of the metric $g_{IJ}$ of the torus
$T^d$, and from the expectation value of the three-form gauge field $\C_{IJK}$
on $T^d$; they also arise from the expectation value
$\mathcal{E}_{IJKLMN}$ on $T^d$ of the 
six-form dual to $\C_{MNP}$ in eleven dimensions, or equivalently
the expectation value of the scalar dual to the three-form $\C_{\mu \nu \rho}$
in $D=5$,  
the axion scalar dual to the
two-form $\mathcal{C}_{\mu \nu I}$ in $D=4$, or to the one-form
$\mathcal{C}_{\mu IJ}$ in $D=3$; similarly, the Kaluza--Klein
gauge potentials $g_{\mu I}$ can be dualized in $D=3$ into 
scalars $\mathcal{K}_I$, which can be interpreted as the expectation
value $\mathcal{K}_{I;JKLMNPQR}$ on $T^d$ of the magnetic
gauge potential dual to $g_{MN}$ in eleven dimensions.
The counting is summarized in Table \ref{tsc}. The factor $\Real^+$
appearing in $D=10$ and $D=9$ corresponds to the type IIA dilaton,
and generates a scaling symmetry of the effective action, called
{\it trombonne symmetry} in Ref. \cite{Cremmer:1997xj}.
Note that a quite different U-duality group would be inferred
if one did not dualize the Ramond fields into fields with
less indices \cite{Lu:1997df,Cremmer:1997ct},
or if one would considerer Euclidean supergravities
\cite{Hull:1998br,Cremmer:1998em}.

An analogous counting has been performed in Tables
\ref{tvec} and \ref{ttf} for one-form 
and two-form potentials, inducing particle and string
electric charges, respectively. 
The latter can be put in one-to-one correspondence to the
central charges of the supersymmetry algebra
discussed in the previous section, {\it with two exceptions}.
Firstly, the Lorentz-invariant central charge $Z^{01234}$ in five dimensions,
where $0\dots 4$ denote the five space-time dimensions, does not
correspond to any one-form potential 
\cite{Bars:1996bv,Bars:1996dz}\footnote{Equivalently,
the central charges $Z^{01234}$,$Z^{02345}\dots$ transform as
a vector in six space-time dimensions. These charges could
be attributed to a KK6-brane, if only the KK6-brane did not need
six compact directions to yield a string, and seven to yield a particle
state.}. This truncation of the superalgebra is consistent
with U-duality and is of no concern, except for the twelve-dimensional
origin of M-theory. 
Secondly, there are only 120 Lorentz singlet 
central charges in $D=3$ for 128 gauge potentials
(equivalently, there are only 64 Lorentz vector charges in $D=4$
for 70 two-form gauge fields). As we shall see shortly,
U-duality implies that there should in fact be $248$
electric charges in $D=3$ ($133$ string charges in $D=4$),
yielding a linear representation of the duality group
$E_8$ (resp. $E_7$). Of course, the notion of electric
charge is ill-defined in $D=3$, where a one-form (or a two-form in
$D=4$) is Poincar{\'e}-dual to a zero-form and a particle (or a
string) to an instanton. Another manifestation of the pathology
of the $D=3$ case is the non-asymptotic flatness of the
point-like solitons (or string-like in $D=4$), and the logarithmic
divergence of the kernel of the Laplacian in the transverse
directions. In spite of
these difficulties, we shall pursue the algebraic analysis of
these cases in the hope that they can be resolved.

If the charges $m$ 
under the gauge fields can be put in one-to-one correspondence
with the central charges $Z$, they are nevertheless not equal: the gauge
charges are integer-quantized, as we will discuss in the next subsection,
whereas the central charges are moduli-dependent linear combinations of the
latter:
\begin{equation}
\label{zvq}
Z=\mathcal{V}\cdot m \ ,
\end{equation}
where $\mathcal{V}$ is an element in the group $G_d$ containing
the moduli dependence; it is defined up to the left action
of the compact subgroup $K=H_d$, inducing an R-symmetry transformation
on $Z$. 

The local $H_d$ gauge invariance can be conveniently gauge-fixed  
thanks to the Iwasawa decomposition
(see for instance \cite{Knapp:1986,Moore:1993zc})
\begin{equation}
\label{iwa}
\mathcal{V}= k\cdot a\cdot n \in  K\cdot A \cdot N
\end{equation}
of $G_d$ into the maximal compact $K$, Abelian $A$ 
and nilpotent $N$.
A natural gauge is obtained by taking $K=1$, in which case 
the ``vielbein'' $\mathcal{V}$ becomes a
(generalized) upper triangular matrix $\mathcal{V}=a\cdot n$.
The Abelian factor $A$ is parametrized by the ``dilatonic scalars'',
namely the radii of the internal torus, whereas the nilpotent factor $N$
incorporates the ``gauge scalars'', namely the expectation values
of the gauge fields (including the off-diagonal metric, three-form
and their duals) on the torus.
$G_d$ acts on 
the charges $m$ from the left and on $\mathcal{V}$
from the right. The transformed $\mathcal{V}$ can then
be brought back into an upper triangular form by
a moduli-dependent R-symmetry compensating transformation on the
left. This implies that the central charges $Z$
transform non-linearly under the continuous U-duality
group $G_d$. For the case of T-duality in  type II string theory
this decomposition is given in Eq. \eqref{iwast}. 
 In Section \ref{sskew}, we shall
obtain an explicit parametrization of $\mathcal{V}$
in terms of the shape of the torus and the various
gauge backgrounds.

\begin{table}[h] 
\begin{center}
\begin{tabular}{|r|r||r|r|r|r||r|r|}
\hline
$D$& $d$  & $g$ & $\C_3$ & $\E_6$ &  $\K_{1;8}$ &
total & scalar manifold   \\ \hline      
10& 1 & 1  &     &    &   & 1 &$\Real^+$  \\    
9 & 2 & 3  &     &    &   & 3 &$Sl(2,\Real)/U(1) \times \Real^{+}$ \\     
8 & 3 & 6  & 1   &    &   & 7 &$Sl(3,\Real)/SO(3) \times Sl(2,\Real)/U(1)$ \\ 
7 & 4 & 10 & 4   &    &   &14 &$Sl(5,\Real)/SO(5)$ \\     
6 & 5 & 15 & 10  &    &   &25 &$SO(5,5,\Real)/SO(5) \times SO(5) $  \\     
5 & 6 & 21 & 20  & 1  &   &42 &$E_{6(6)}/USp(8)$  \\     
4 & 7 & 28 & 35  & 7  &   & 70 &$E_{7(7)}/SU(8)$\\     
3 & 8 & 36 & 56  & 28 & 8 & 128&$E_{8(8)}/SO(16)$\\     
\hline
\end{tabular}
\end{center}
\caption[Scalar counting and scalar manifolds]{Scalar counting and scalar manifolds in compactified M-theory.\label{tsc}}
\end{table}

\begin{table}[h] 
\begin{center}
\begin{tabular}{|r|r||r|r|r|r||r|r|}
\hline
$D$& $d$  & $g$ & $\C_3$ & $\E_6$ &  $\K_{1;8}$ &
total & charge representation   \\ \hline      
10& 1 & 1  &     &    &   & 1 &\irrep{1}  \\    
9 & 2 & 2  & 1   &    &   & 3 &\irrep{3} of $Sl(2)$ \\     
8 & 3 & 3  & 3   &    &   & 6 &\irrep{(3,2)} of $Sl(3)\times Sl(2)$ \\ 
7 & 4 & 4  & 6   &    &   &10 &\irrep{10} of $Sl(5)$ \\     
6 & 5 & 5  & 10  & 1  &   &16 &\irrep{16} of $SO(5,5)$  \\     
5 & 6 & 6  & 15  & 6  &   &27 &\irrep{27} of $E_{6(6)}$  \\     
4 & 7 & 7  & 21  & 21 & 7 &56  &\irrep{56} of $E_{7(7)}$\\     
3 & 8 & 8  & 28  & 56 & 36& 128&\irrep{248} of $E_{8(8)}$\\     
\hline
\end{tabular}
\end{center}
\caption[Vectors and particle charge representations]
{Vectors and particle charge representations in 
compactified M-theory.
\label{tvec}}
\end{table}

\begin{table}[h] 
\begin{center}
\begin{tabular}{|r|r||r|r|r|r||r|r|}
\hline
$D$& $d$  & $g$ & $\C_3$ & $\E_6$ &  $\K_{1;8}$ &
total & charge representation   \\ \hline      
10& 1 &    & 1   &    &   & 1 &\irrep{1}  \\    
9 & 2 &    & 2   &    &   & 2 &\irrep{2} of $Sl(2)$ \\     
8 & 3 &    & 3   &    &   & 3 &\irrep{(3,1)} of $Sl(3)\times Sl(2)$ \\ 
7 & 4 &    & 4   & 1  &   & 5 &\irrep{5} of $Sl(5)$ \\     
6 & 5 &    & 5   & 5  &   &10 &\irrep{10} of $SO(5,5)$  \\     
5 & 6 &    & 6   & 15 &6  &27 &\irrep{\bar{27}} of $E_{6(6)}$  \\     
4 & 7 &    & 7   & 35 &28 &70 &\irrep{133} of $E_{7(7)}$\\     
\hline
\end{tabular}
\end{center}
\caption[Two-forms  and string charge representations]
{Two-forms  and string charge representations in 
compactified M-theory.
\label{ttf}}
\end{table}

\subsection{Charge quantization and U-duality}
As in the case of T-duality, the continuous symmetry $E_{d(d)}(\Real)$
of the two-derivative effective action cannot be a symmetry
of the quantum theory: the gauge potentials transform 
non-trivially under $E_{d}$, and the continuous symmetry is
therefore broken by the existence of states charged
under these potentials. At best there can remain a
discrete subgroup $E_{d(d)}(\Zint)$, 
which leaves the lattice of charges invariant.
For one thing, a subset of the charges corresponds to the Kaluza--Klein
momentum along the internal torus, and are therefore constrained to
lie in the reciprocal lattice of the torus. Another subset of charges
corresponds to the wrapping numbers of extended objects around
cycles of $T^d$, and are then constrained to lie in the homology
lattice of $T^d$.

A way to determine the remaining discrete subgroup is to consider M-theory
compactified to $D=4$ dimensions, in which case Poincar{\'e} duality
exchanges gauge one-forms with their magnetic duals
\cite{Hull:1995mz}. In this
dimension, Dirac--Zwanziger charge quantization takes the usual form
\begin{equation}
\label{dzq}
m^i n'_{i} - m^{'i} n_i \in \Zint
\end{equation}
for two particles of electric and magnetic charges $m^i$ and $n_i$ 
respectively, and $i$ runs from 1 to 28, as read off from
Table \ref{tvec}. This condition is invariant
under the electric--magnetic duality $Sp(56,\Zint)$, under which
$(m^i,n_i)$ transforms as a vector. The exact symmetry group
is therefore at most
\begin{equation}
\label{e7z}
E_{7(7)}(\Zint) \subset E_{7(7)}(\Real) \cap Sp(56,\Zint)\ ,
\end{equation}
This translates into a condition on $E_{d(d)}(\Zint)$ for $d\le 7$
by the embedding $E_{d(d)}(\Zint)\subset E_{7(7)}(\Zint)$. A similar
condition can be obtained in $D=3$, where all one-forms
are dual to scalars.

The condition \eqref{e7z} requires a precise knowledge of the
embedding of $E_{7(7)}(\Real)$ in $Sp(46,\Real)$. 
Instead, we shall take another approach, and {\it postulate}
that the U-duality group of M-theory compactified
on a torus $T^d$ is generated by the T-duality 
$SO(d-1,d-1,\Zint)$ of type IIA string theory compactified on $T^{d-1}$,
{\it and} by the modular group $Sl(d,\Zint)$ of the torus $T^d$:
\begin{equation}
\label{uduagr}
E_{d(d)}(\Zint) = SO(d-1,d-1,\Zint) \bowtie Sl(d,\Zint)\ .
\end{equation}
The former was argued to be a non-perturbative symmetry of type 
IIA string theory, as discussed in the previous section, while 
the latter is the remnant of eleven-dimensional general 
reparametrization invariance, after compactification on a torus
$T^d$: it is therefore guaranteed to hold, as long as M-theory,
whatever its formulation may be, contains the graviton in its spectrum.
The above construct is therefore the minimal U-duality group, and 
since it preserves the symplectic condition \eqref{dzq}\footnote{A
verification of this statement requires a precise knowledge of the
branching functions of $Sp(56)$ into $E_7$.} also the maximal one.

In the $d=2$ case, the U-duality group \eqref{uduagr} 
is the modular group $Sl(2,\Zint)$
of the M-theory torus, which in particular contains the exchange
of $R_s$ and $R_9$; translated in type IIB variables,
this is simply the $Sl(2,\Zint)$ S-duality of type IIB
theory (in 9 or 10 dimensions), which contains the strong-weak
coupling duality $g_s \ra 1/g_s$, as can be seen from Eq. 
\eqref{iibmatching}. Note that we do not expect any
quantum symmetry from the trombonne symmetry factor $\Real^+$.
For $d=3$, the T-duality group
splits into two factors $Sl(2,\Zint)\times Sl(2,\Zint)$,
one of which is a subgroup of the modular group $Sl(3,\Zint)$
of the M-theory torus $T^3$. The definition \eqref{uduagr} therefore yields
$E_{3(3)}(\Zint) = Sl(3,\Zint) \times Sl(2,\Zint)$ and is the
natural discrete group of $E_3$. For $d=4$, $SO(3,3,\Zint)$
is isomorphic to a $Sl(4,\Zint)$ (in the same
way as $SO(6)\sim SU(4)$) which does not commute with the
modular group $Sl(4,\Zint)$ of M-theory on a torus $T^4$. Altogether, 
they make the $Sl(5,\Zint)$ subgroup of $E_{4(4)}(\Real)=Sl(5,\Real)$. For
$d=5$, we obtain the $SO(5,5,\Zint)$ subgroup of $E_{5(5)}(\Real)=
SO(5,5,\Real)$.
For $d\ge 6$, this provides a {\it definition} of 
the discrete subgroups of the
exceptional groups $E_{d(d)}(\Real)$\footnote{This is particularly
interesting in the $d\ge 9$ case, where we obtain discrete versions
of affine and hyperbolic groups, see Section \ref{wgwru}.}. 
These groups are summarized
in the rather tautological Table \ref{tds}. We note that it is crucial that
the groups $E_{d(d)}(\Real)$ be non-compact in order for an infinite
discrete group to exist. The maximal non-compact form is also required
in order that all representations be real (\ie that the mass of
a particle and its anti-particle be equal, see Section \ref{spfm}).

\begin{table}[h]
\begin{center}
\begin{tabular}{|r|l||c|c|} \hline 
$D$& $d$ & $E_{d(d)}(\Real)$ & $E_{d(d)}(\Zint)$ \\ \hline  
10 & 1 & 1 & 1 \\
9  & 2 & $Sl(2,\Real)$ & $Sl(2,\Zint)$\\
8  & 3 & $Sl(3,\Real)\times Sl(2,\Real)$ & $Sl(3,\Zint)\times Sl(2,\Zint)$\\
7  & 4 & $Sl(5,\Real)$ & $Sl(5,\Zint)$\\
6  & 5 & $SO(5,5,\Real)$ & $SO(5,5,\Zint)$\\
5  & 6 & $E_{6(6)}(\Real)$ & $E_{6(6)}(\Zint)$ \\
4  & 7 & $E_{7(7)}(\Real)$ & $E_{7(7)}(\Zint)$ \\
3  & 8 & $E_{8(8)}(\Real)$ & $E_{8(8)}(\Zint)$ \\ \hline
\end{tabular}
\end{center}
\caption{Discrete subgroups of $E_d$.
\label{tds}}
\end{table}

\subsection{Weyl and Borel generators \label{ssweyl} }
A set of generators of the U-duality group can easily be obtained by 
conjugating
the T-duality generators under $Sl(d,\Zint)$. The Weyl
generators now include the exchange of the eleven-dimensional
radius $R_s$ with any radius of the string-theory torus $T^{d-1}$,
in addition to the exchange of the string-theory torus
directions among themselves and T-duality on two directions
thereof. It is interesting to rephrase the latter in
M-theory variables, using relations \eqref{matching} and \eqref{tduag}:
\begin{equation}
T_{ij}:
R_i \ra \frac{ l_p^3}{R_j R_s}\ ,\quad
R_j \ra \frac{ l_p^3}{R_s R_i}\ ,\quad
R_s \ra \frac{ l_p^3}{R_i R_j}\ ,\quad
l_p^3 \ra \frac{ l_p^6 }{R_i R_j R_s}
\end{equation}
These relations are symmetric under permutation of $i,j,s$
indices, and using an $R_k \leftrightarrow R_s$ transformation, we
are free to choose $i,j,s$ along any direction of
the M-theory torus $T^d$. The M-theory T-duality therefore
reads
\begin{equation}
\label{mthz2}
T_{IJK}:
R_I \ra \frac{ l_p^3}{R_J R_K}\ ,\quad
R_J \ra \frac{ l_p^3}{R_K R_I}\ ,\quad
R_K \ra \frac{ l_p^3}{R_I R_J}\ ,\quad
l_p^3 \ra \frac{ l_p^6 }{R_I R_J R_K}
\end{equation}
and in particular involves {\it three} directions, contrary
to the naive expectation. We emphasize that the
above equation summarizes the non-trivial part of U-duality, 
and arises as a mixture of T-duality and S-duality transformations.
It can in particular be used 
to derive [Antoniadis:1999rm] the duality between 
the heterotic string compactified on $T^4$ and type IIA compactified 
on $K_3$ in the Horava-Witten picture [Horava:1996ma], 
and thus unify all vacua with 16 supersymmetries. We 
however restrict ourselves to the maximally supersymmetric case
in this review.

The Weyl group can be written
in a way, similar to Eq. \eqref{soddw}\footnote{This equation holds
for $d\geq 3$ only; when $d < 3$ the $\Zint_2$ symmetry \eqref{mthz2}
collapses
and only the permutation group $\mathcal{S}_d$ remains.}: 
\begin{equation}
\mathcal{W}\left(E_d\right)= \Zint_2 \bowtie \mathcal{S}_d
\end{equation}
but it should be borne in mind
that the algebraic relations between the $\Zint_2$ symmetry
$T_{123}$ and the permutations $S_{IJ}$ are different from those
of the T-duality generators $T_{12}$ and $S_{ij}$; in addition
$d$ differs by one unit from the one we used there.
We also note that the transformations $T_{IJK}$ and $S_{IJ}$
preserve the Newton's constant
\begin{equation}
\label{uduainv}
\frac{1}{\kappa_d^2} = \frac{\prod{R_I}}{l_p^9}
 = \frac{V_R }{l_p^9}
\ ,
\end{equation}
where we have defined  $V_R$ to be the volume of the M-theory compactification
torus. 

On the other hand, the Borel generators now include a 
generator $\gamma_i \rightarrow \gamma_i+\gamma_s$ that mixes 
the eleven-dimensional direction with the other ones,
as well as the T-duality spectral flow $B_{ij}\rightarrow
B_{ij}+1$, from which, by an $R_s\leftrightarrow R_i$ conjugation, we
can reach the more general {\it M-theory spectral flow}\footnote{As
discussed in Subsection \ref{uduaspfl}, the $\C$ shift actually
has to be accompanied by $\E$ and $\K$ shifts to be
a symmetry of the equations of motion.}
\begin{equation}
C_{IJK}: \mathcal{C}_{IJK} \rightarrow  \mathcal{C}_{IJK}+1\ .
\end{equation}
We should also include a set of generators shifting 
the other scalars from the dual gauge potentials, as explained in Section
\ref{contsymu}:  
\begin{subequations}
\label{eksh}
\begin{eqnarray}
E_{IJKLMN}&:& \mathcal{E}_{IJKLMN} \rightarrow
\mathcal{E}_{IJKLMN}+1\\
K_{I;JKLMNPQR}&:& \mathcal{K}_{I;JKLMNPQR} \rightarrow
\mathcal{K}_{I;JKLMNPQR}+1\ .
\end{eqnarray}
\end{subequations}
These scalars and corresponding shifts are needed for $d \geq 6$ and $d \geq 8$ respectively. 
For $ d \geq 9$, as will become clear in Section \ref{wgwru}, the enlargement
of the symmetry group to an affine or Kac-Moody symmetry requires an
infinite number of such Borel generators. 
As we shall see in Subsection \ref{uduaspfl}, the Borel generators
\eqref{eksh} can be obtained from commutators of $\C_{IJK}$ transformations. 

\subsection{Type IIB BPS states and S-duality\label{iibs}}
Before studying the structure of the U-duality group, we shall
pause and briefly discuss the action of the extra Weyl generator
$R_{s} \leftrightarrow R_{9}$ on the type IIB side. Using the
identification \eqref{iibmatching} to convert to type IIB
variables, this action inverts the coupling constant and
rescales the string length as
\begin{equation}
\label{bsdua}
g_s \leftrightarrow \frac{1}{g_s}   
\sp 
l_s^2 \leftrightarrow l_s^2 g_s\ , 
\end{equation} 
in such a way that Newton's constant $1/(g_s^2 l_s^8)$ is invariant.
Its action on the BPS spectrum can be straightforwardly obtained
by working out the action on the masses or tensions, and is 
summarized in Table \ref{tsd}.

\begin{table}[h] 
\begin{center}
\begin{tabular}{|l|c||c|l|}
\hline
state & tension & S-dual & dual state \\ \hline
D1-brane & $\frac{1}{g_s l_s^2}$ & $\frac{1}{l_s^2}$ & F-string \\  
D3-brane & $\frac{1}{g_s l_s^4}$ & $\frac{1}{g_s l_s^4}$ & D3-brane \\
D5-brane & $\frac{1}{g_s l_s^6}$ & $\frac{1}{g_s^2 l_s^6}$ & NS5-brane \\ 
KK5-brane & $\frac{R^2}{g_s^2 l_s^{8}}$ & $\frac{R^2}{g_s^2 l_s^{8}}$
& KK5-brane \\ 
D7-brane & $\frac{1}{g_s l_s^8}$ & $\frac{1}{g_s^3 l_s^8}$ & 7${}_3$-brane \\
D9-brane & $\frac{1}{g_s l_s^{10}}$ & $\frac{1}{g_s^4 l_s^{10}}$ & 9${}_4$-brane
 \\ 
\hline
\end{tabular}
\end{center}
\caption{S-dual type IIB BPS states. 
\label{tsd}}
\end{table}
In this table, we have displayed the action of the $\Zint_2$ Weyl
element only. Under more general duality transformations, the 
fundamental string and the NS5-brane generate orbits of 
so called $(p,q)$ strings and $(p,q)$ five-branes. The former
can be seen as a bound state of $p$ fundamental strings
and $q$ D1-branes, or (in the Euclidean case) 
as a coherent superposition of $q$ D1-branes
with $p$ instantons \cite{Kiritsis:1997em}. The $(p,q)$ five-branes
similarly correspond to bound states of $p$ NS5-branes and  $q$ D5-branes.

On the other hand, the action of S-duality on the D7 and D9-brane yields
states with tension $1/g_s^3$ and $1/g_s^4$ respectively. 
These exotic states will be discussed in Subsection
\ref{exotic}, where our nomenclature will be explained as well. 
Again, such states
have less than three transverse dimensions, and do not preserve
the asymptotic flatness of space-time and the asymptotic constant
value of the scalar fields. In particular, the D7-brane generates 
a  monodromy $\tau\rightarrow \tau+1$ in 
the complex scalar $\tau$ at infinity. Its images under S-duality
then generate a more general $Sl(2,\Zint)_B$ monodromy 
\begin{equation}
M=\begin{pmatrix} 1 - pq & p^2 \\ -q^2 & 1+pq \end{pmatrix}
\end{equation}
ascribable to a $(p,q)$ 7-brane\footnote{It has also been proposed
that the IIB 7-branes transform as a triplet of $Sl(2,\Zint)$
\cite{Meessen:1998qm}.}.
We finally remark  that the relations in Table  \ref{tsd}
 can also be verified directly using the $R_s \leftrightarrow R_9$ flip
and the 
 M-theory/IIB identifications as (un)wrapped M-theory branes,
given in Tables \ref{tmb} and \ref{tmbc}.

\subsection{Weyl generators and Weyl reflections \label{wgwru}}
In order to understand the occurrence of the $E_{d(d)}$ U-duality group, 
we shall now apply the same technique as in the T-duality
case and investigate the group generated by the Weyl
generators. We choose as a minimal set of
Weyl generators the exchange of the M-theory torus
directions
$S_{I}: R_I \leftrightarrow R_{I+1}$, where $I=1\dots d-1$,
as well as the T-duality $T=T_{123}$ on directions 1,2,3
of the M-theory torus. 
Adapting the construction of Ref. \cite{Elitzur:1997zn}\footnote{In 
Ref. \cite{Elitzur:1997zn}, the discussion was carried out
from the gauge theory side, and  the U-duality invariant \eqref{uduainv}
was used to eliminate the vector $e_0$, except when $d=9$. This vector
can, however, be kept for any $d$, and, as we shall momentarily see,
appears as an extra time-like direction.}and Subsection
\ref{wgwrt}, we represent the monomials
$\varphi=(\ln l_p^3, \ln R_1, \ln R_2, \dots, \ln R_d)$ 
as a form on a vector space $V_{d+1}$ with basis
$e_0, e_1, e_2, \dots, e_d$, and associate to any weight vector
$ \lambda=x^0 e_0 + x^1 e_1 + \dots + x^d e_d$ its 
``tension''\footnote{$\T$ actually has the dimension of a $p$-brane
tension $\T_{p}$, with $p=-(3 x^0 + x^1+\dots+x^d +1)$.}
\begin{equation}
\label{dualbu}
\mathcal{T}=e ^{\langle\varphi,\lambda\rangle} =
l_p^{3x^0} R_1^{x^1} R_2^{x^2}\dots R_d^{x^d}\ .
\end{equation}
The generators $S_I$ and
$T$ can then be implemented as linear operators on
$V_{d+1}$, with matrix
\begin{equation}
\label{sit}
S_I =
\begin{pmatrix} 
1 &    &    &   \\
  &    &  1 &   \\
  &  1 &    &   \\
  &    &    & \mathbb{I}_{d-3}  \\
\end{pmatrix}
\ ,\quad
T = \begin{pmatrix} 
2 & 1  & 1 & 1 &\\
-1 &   & -1 & -1 &\\
-1 & -1 &   & -1 &\\
-1  &- 1  & -1 &  &\\
   &    &   &    & \mathbb{I}_{d-3}
\end{pmatrix} \ .  
\end{equation}

The operators $S_I$ and $T$ in \eqref{sit}
are easily seen to be orthogonal
with respect to the Lorentz metric 
\begin{equation}
\label{um}
{\rm d}s^2 = -({\rm d}x^0)^2+({\rm d}x^I)^2\ ,
\end{equation} 
and correspond to Weyl reflections
\begin{equation} 
\lambda \rightarrow \rho_{\alpha} (\lambda) =
\lambda - 2 \frac{\alpha\cdot \lambda}{\alpha\cdot \alpha} \alpha 
\end{equation}
along planes orthogonal to the vectors
\begin{equation} 
\alpha_I = e_{I+1} - e_{I} \ ,I=1\dots d-1\ ,\quad
\alpha_0 = e_1 + e_2 + e_3 - e_0 \ .
\end{equation}
It is very striking that $l_p^3$ appears on the same footing as the
other radii $R_I$, but with a minus sign in the metric: it can be 
interpreted as the radius of an extra time-like direction, much in
the spirit of certain proposals about F-theory
\cite{Vafa:1996xn,Bars:1996dz}.
The only non-vanishing (Lorentzian) scalar products of these roots
turn out to be
\begin{equation} 
(\alpha_I)^2 = (\alpha_0)^2 = 2 \ ,\quad
\alpha_I\cdot \alpha_{I+1} = \alpha_3 \cdot \alpha_0 = -1 
\end{equation}
summarized in the Dynkin diagram:
\begin{equation}
\label{dynkinu}
\begin{array}{ccccccccc}
&&&&\bigcirc_{0}&&&&\\
&&&&\mid& &&&\\
+_1 &-&
\oplus_2 &-&
\oplus_3 &-&
\oplus_4&-\dots-&
\oplus_{d-1}
\end{array}
\end{equation}
This is precisely the Dynkin diagram of $E_{d}$ as
shown in Table \ref{dynfigure}, in agreement
with the analysis based on moduli counting. 

In Eq. \eqref{dynkinu} it is easy to recognize
the diagrams of the $SO(d-1,d-1,\Zint)$ (denoted by $\bigcirc$'s) 
and $Sl(d,\Zint)$ (denoted by $+$'s) subgroups. 
The branching of the $Sl(d)$ diagram on the third
root reflects the action of T-duality on three directions.
The full diagram can be built from the M-theory Lorentz group
$Sl(d,\Zint)$ denoted by $+$'s, and from the type IIB Lorentz
group $Sl(d-1,\Zint)$ generated by the roots $\alpha_0,\alpha_3,\dots,
\alpha_{d-1}$ \footnote{From
this point of view, the U-duality is a consequence of general
coordinate invariance in M and type IIB theories \cite{Lavrinenko:1998hf}.}.
Under decompactification, the rightmost root has to be
dropped, so that $E_d$ disintegrates into $E_{d-1}$ \footnote{There
is a notable exception for $d=8$, where $E_8$ disintegrates 
into $E_7\times Sl(2)$. This is because
the extended Dynkin diagram of $E_8$ has an extra root
connected to $\alpha_8$. Only $Sl(2)$ singlets remain in
the spectrum, however. The same happens in 
$d=4$, where $E_3=Sl(3)\times Sl(2)$ in $E_4=Sl(5)$ is not a maximal 
embedding.}. When the
root at the intersection is reached, the diagram falls
into two pieces, corresponding to the two $Sl(2)$ and
$Sl(3)$ subgroups in $D=8$. The root $\alpha_0$ itself 
disappears for $d=2$, leaving only the root $\alpha_1$
of $Sl(2,\Real)$. 

Again, the action of the Weyl group on $V_{d+1}$ is reducible,
at least for $d\le 8$. Indeed, the invariance of Newton's 
constant $\prod R_I / l_p^9$
implies that the roots are all orthogonal to the vector 
\begin{equation}
\label{delta}
\delta = e_1+\dots +e_{d} - 3 e_0\ ,
\end{equation} 
with proper length $\delta^2=d-9$, 
so that the reflections 
actually restrict to the
hyperplane $V_{d}$ normal to $\delta$:
\begin{equation} x^1 + \dots + x^d + 3 x^0 = 0 \ .
\end{equation}
The Lorentz metric on $V_{d+1}$ restricts to a metric
$g_{IJ} = \delta_{IJ} - 1/9$ on $V_d$, which is 
positive-definite for $d\le 8$, so that $S_I$ and $T$ indeed
generate the Weyl group of the Lie algebra $E_d(\Real)$.
The order and number of roots of these groups are
recalled in Table \ref{twg} \cite{Humphreys:1990}.

When $d=9$, however, the invariant vector $\delta$ becomes
null, so that $V_{d+1}$ no longer splits into $\delta$ and
its orthogonal space; the generators act on
the entire Lorentzian vector space $V_{d+1}$, and the
generators $S_I$ and $T$  no longer span a finite group.
Instead, they correspond to the Weyl group of the 
{\it affine} Lie algebra $E_9=\hat E_8$. This is in agreement
with the occurrence of infinitely many conserved currents
in $D=2$ space-time dimensions. This case requires  
a specific treatment and will be discussed in Subsection
\ref{e9sec}. For $d>9$, that is
compactification to a line or a point, the situation
is even more dramatic, with the occurrence of the hyperbolic
Kac--Moody algebras $E_{10}$ and $E_{11}$, about which
very little is known. The reader should go to
\cite{Julia:1983,Julia:1981wc,Nicolai:1998gi,Gebert:1994zq} 
for further discussion and references.

\begin{table}[h]
\begin{equation*}
\begin{array}{|l|cccccccc|}
\hline
d & 2 & 3 & 4 & 5 & 6 & 7 & 8 & 9\\
\hline
E_d  & A_1 & A_2 \times A_1 & A_4 & D_5 & E_6 &
E_7 & E_8 & \hat E_8 \\ \hline
\mbox{order} & 2 & 6 \times 2& 5! & 2^4 5! & 2^7 3^4 5 & 
2^{10}3^4 5\ 7 & 2^{14} 3^5 5^2 7 & \infty \\ \hline
\mbox{roots} & 2 & 6 + 2 & 20 & 40 & 72 & 126 & 240 & \infty \\
\hline
\end{array} 
\end{equation*}
\caption{Order and number of roots of $E_d$ Weyl groups.
\label{twg}}
\end{table}

\subsection{BPS spectrum and highest weights}
Pursuing the parallel with our presentation on T-duality,
we now discuss the representations of the U-duality Weyl group.
The fundamental weights 
dual to the roots $\alpha_1,\dots, \alpha_{d-1},\alpha_0$
are easily computed:
\begin{subequations}
\label{hwu}
\begin{eqnarray}
\lambda^{(1)} = e_1 -e_0 
\ra& \mathcal{T}_1=&\frac{R_1}{l_p^3}\\
\lambda^{(2)} = e_1+ e_2 -2 e_0 
\ra& \T_3=&\frac{R_1 R_2}{l_p^6}\\
\lambda^{(3)} = e_1+ e_2 + e_3 -3 e_0
\ra& \T'_5=&\frac{R_1 R_2 R_3}{l_p^9}\\
\lambda^{(4)} = e_1+ \dots + e_4 -3 e_0
\ra& \T'_4=&\frac{R_1 R_2 R_3 R_4}{l_p^9}\\
\dots & \nonumber\\
\lambda^{(d-2)} = e_1+ \dots + e_{d-2} -3 e_0
\ra& \T'_{10-d}=&\frac{R_1 \dots R_{d-2}}{l_p^9}\\
\lambda^{(d-1)} = e_1+ \dots + e_{d-1} -3 e_0 \doteq -e_d 
\ra& \mathcal{M}=&\frac{1}{R_d} \label{doteq}\\ 
\lambda^{(0)} = - e_0
\ra& \T_2=&\frac{1}{l_p^3}
\end{eqnarray}
\end{subequations} 
where the symbol $\doteq$ in Eq. \eqref{doteq} denotes equality
modulo $\delta$, that is up to a power of the invariant
Planck length. In the above equations, we have translated
the weight vectors into monomials, and interpreted it
as the tension $\T_{p+1}$ of a $p$-brane:
\begin{itemize}
\item The weight $\lambda^{(d-1)}$ corresponds to the Kaluza--Klein
states, with mass $1/R_I$, as well as its U-duality
descendants. We shall name its orbit the {\it particle multiplet},
or {\it flux multiplet}, for reasons that will become apparent in
Subsection \ref{symmass}.
\item 
 The weight $\lambda^{(1)}$ on the other hand
has dimension $1/L^2$, and corresponds to the tension of
a membrane wrapped on the direction $1$: it will go
under the name of {\it string multiplet}, or {\it momentum
multiplet}. The latter name will also become clear in Subsection \ref{symmass}.
\item
The weight $\lambda^{(0)}$ is the highest weight of the {\it membrane
multiplet} containing the fundamental membrane with tension
$1/l_p^3$, together with its descendants. 
\item
The weights $\lambda^{(2)}$ and $\lambda^{(5)}$ both correspond
to threebrane tensions $\T_3$ and $\T_3'$. Even though they
are inequivalent under the Weyl group, it turns out that 
$\lambda^{(5)}$ is a descendant of $\lambda^{(3)}$ under
the full U-duality group. The U-duality orbit of the state
with tension $\T'_3$ is therefore a subset of the orbit of the state
with tension $\T_3$, and $\lambda^{(3)}$ is the true highest-weight
vector of the {\it threebrane multiplet}. 
\item
The same holds for $\lambda^{(6)}$ associated to a membrane tension
$\T_2'$ and descendant of the highest weight $\lambda^{(0)}$ of the
membrane multiplet under U-duality, as well as for $\lambda^{(7)}$
and $\lambda^{(1)}$. 
\item 
The weight $\lambda^{(3)}$ 
corresponds to a fivebrane tension $\T_5'$,
but is again not the highest weight of the {\it fivebrane 
multiplet}, which is instead a non-fundamental weight:
\begin{equation}
\T_5 = \frac{1}{l_p^6} \rightarrow \lambda = - 2 e_6 = 
2 \lambda^{(0)} \ .  
\end{equation}
Similarly, the weight $\lambda^{(4)}$ corresponds to a fourbrane tension
$\T_4'$,
and is not the highest weight of the {\it fourbrane 
multiplet}, which is instead a non-fundamental weight:
\begin{equation}
\T_4 = \frac{R_1}{l_p^6} \rightarrow \lambda = e_1 - 2 e_6 = 
\lambda^{(1)} + \lambda^{(0)} \ .  
\end{equation}
\item 
Finally, the instanton multiplet does not appear in Eq. \eqref{hwu}.
An instanton configuration can be obtained by wrapping a 
membrane on a three-cycle\footnote{We should, however, warn the reader that
it is not the representation arising in non-perturbative
couplings, as we shall discuss in Subsection \ref{rfour}.}, 
and corresponds to a weight vector
\begin{equation}
\T_{-1} = \frac{R_1 R_2 R_3}{l_p^3} \rightarrow \lambda = \alpha_0 \ .  
\end{equation}
Since this vector is a simple root, it corresponds to a multiplet
in the adjoint representation. It is, however, {\it not} the 
highest weight of the U-duality multiplet, which is instead 
the highest root $\psi$ whose expansion coefficients on the
base of the simple roots are given by the Coxeter labels
in Table \ref{dynfigure}. An explicit computation gives
\begin{subequations}
\label{psi}
\begin{eqnarray}
d=4: \  \psi &=& \delta - \lambda^{(1)} - \lambda^{(0)} \\ 
d=5: \  \psi &=& \delta - \lambda^{(2)} \\  
d=6: \  \psi &=& \delta - \lambda^{(0)} \\  
d=7: \  \psi &=& \delta - \lambda^{(1)} \\  
d=8: \  \psi &=& \delta - \lambda^{(7)} \  . 
\end{eqnarray}
\end{subequations}
Since the fundamental weights $\lambda^{(i)}$ are dual to the
simple roots $\alpha_I$, it is clear that 
$\psi\cdot \alpha_I = \delta_{i,I}$, where $I$ is the index
appearing on $\lambda$ in Eq. \eqref{psi} at a given $d$, and
moreover it can be easily checked that $\psi ^2=2$.
The highest root can therefore be added as an extra root
in the Dynkin diagrams in Table \ref{dynfigure},
and turns them into extended Dynkin diagrams.
\end{itemize}

The previous considerations are summarized in the diagram
\begin{equation}
\begin{array}{ccccccccc}
&&&&\frac{1}{l_p^3}&&&&\\
&&&&\mid& &&&\\
\frac{R_1}{l_p^3}&-&
\frac{R_1 R_2}{l_p^6}&-&
\frac{R_1 R_2 R_3}{l_p^9}&-&
\frac{R_1 R_2 R_3 R_4}{l_p^9}&-\dots-&
\frac{1}{R_d}
\end{array}
\end{equation}
where we have indicated the highest weight associated to each node
of the Dynkin diagram.
For simplicity, we shall henceforth focus
our attention on the {\it particle} and {\it string}
multiplets, corresponding to the rightmost node with weight $\lambda^{(d-1)}$ 
and leftmost node with weight $\lambda^{(1)}$ respectively. 

\subsection{The particle alias flux multiplet \label{spfm} }
The full particle multiplet can be obtained by acting
with Weyl and Borel transformations on the Kaluza--Klein state with
mass $1/R_I$. Instead of working out the precise transformation
of the supergravity configurations\footnote{See Ref. \cite{Lu:1998sx}
for the construction of U-duality multiplets of $p$-brane solutions,
and Ref. \cite{Ferrara:1997ci} for a discussion of the continuous U-duality
orbits of $p$-brane solutions.}, 
we can restrict ourselves to
considering the masses of the various states in the multiplet.
We note that the action of $S_{IJ}$ and 
$T_{IJK}$ on the dilatonic scalars $R_I$ is independent of the dimension
$d$ of the torus, so that we can work out the maximally
compactified case $D=3$, and obtain the 
higher-dimensional cases by simply deleting states that
require too many different directions on $T^d$ to exist.

The results are displayed in Table \ref{tfm}, where distinct
letters stand for distinct indices. The states are organized
in representations of the $Sl(8,\Zint)$ modular group of the torus
$T^8$. These representations arrange themselves 
in shells with increasing power 
of $l_p^3$; since $l_p$ is invariant under $Sl(8,\Zint)$, 
this corresponds to the grading with respect to the simple 
root $\alpha_0$. Generalized T-duality $T_{IJK}$ 
may move from one shell to the next or
previous one, whereas $S_{IJ}$ acts within each shell. 
Eight states with 
mass $V_R/l_p^9$ have been added in the middle line,
corresponding to zero-length weights that 
cannot be reached from the length-2 highest state. These 
states are, however, necessary in order
to get a complete representation of the modular group
$Sl(8,\Zint)$, and can be reached by a Borel
transformation in $Sl(8,\Zint)$. They can be thought
of as the eight ways to resolve the radius that appears square
in the mass of
the other states on the same line,
into a product of two distinct radii. 
This is not
required for the other lines, since all squares
can be absorbed with a power of Newton's constant.

In the last column of Table \ref{tfm}, we have indicated the representation
of $Sl(8,\Zint)$ that yields the same dimension. The superscripts
denote the number of antisymmetric indices, and no
symmetry property is assumed across a semicolon. In other
words, $m^{1;7}$ correspond to the $V \otimes \wedge^7 V$
where $V$ is the defining representation of $Sl(d)$. These
representations are precisely dual to those under which the
various gauge vectors transform (see Table \ref{tvec}); they
actually correspond to the charges of the BPS state under
these $U(1)$ gauge symmetries (see also Subsection \ref{contsymu}). 
They generalize
the D-brane charges we discussed in Section \ref{tdua}. 
Altogether, these states sum up to $\irrep{248}$,
the adjoint representation of $E_8$, which indeed decomposes
in the indicated way under the branching $Sl(8)\subset E_8$.
The occurrence of the adjoint representation simply follows from
the last equality of Eq. \eqref{psi} identifying the fundamental 
weight $\lambda^{(7)}$ with the highest root of $E_8$.

\begin{table}[h]
\begin{center}
\begin{tabular}{|c|l|l|}  
\hline
mass $\M$ & $Sl(8)$ irrep & charge\\ \hline  
$\frac{1}{R_I}$ & \irrep{8}& $m_1$ \\  
$\frac{R_I R_J}{l_p^3}$ & \irrep{28} &  $m^2 $\\ 
$    \frac{R_I R_J R_K R_L R_M}{l_p^6}$ & \irrep{56} & $m^5$ \\  
$\frac{R_I^2 R_J R_K R_L R_M R_N R_P}{l_p^9}$, 8 $\frac{V_R}{l_p^9}$
& \irrep{1}+\irrep{63} & $m^{1;7}$ \\
$ \frac{R_I^2 R_J^2 R_K^2 R_L R_M R_N R_P R_Q}{l_p^{12}}$
&  \irrep{56} & $ m^{3;8} $\\   
$ \frac{R_I^2 R_J^2 R_K^2 R_L^2 R_M^2 R_N^2 R_P R_Q}{l_p^{15}}$ & \irrep{28} &
$m^{6;8}$ \\ 
$  \frac{R_I^3 R_J^2 R_K^2 R_L^2 R_M^2 R_N^2 R_P^2 R_Q^2}{l_p^{18}} $ & 
\irrep{8} &$m^{1;8;8}$ \\  
\hline
\end{tabular} 
\end{center}
\caption{Particle/flux multiplet \irrep{248} of $E_8$.
\label{tfm}}
\end{table}

The first three lines in Table \ref{tfm} have an obvious interpretation.
The state with mass $\frac{1}{R_I}$ is simply the Kaluza--Klein
excitation on the dimension $I$, and $m_I$ denotes the vector
of integer momentum charges. The state with mass $R_I R_J/l_p^3$
is the membrane wrapped on a two-cycle $T^2$ of the compactification
torus $T^d$, and the two-form $m^{IJ}$ labels the precise two-cycle,
just as in the D-brane case of the previous section. The third
line corresponds to the fivebrane wrapped on the five-cycle labelled
by $m^{IJKLM}$. The states on the fourth line are more interesting.
The first of them involves one square radius, and therefore does
not exist in uncompactified eleven dimensions. It is simply
the KK6-brane with Taub--NUT direction along $R_I$ and wrapped
along the directions $J$ to $P$. The second state with mass
$V_R/l_p^9$, however, does exist in eleven uncompactified directions,
and has the tension of a would-be 8-brane. Its asymptotic space-time
is however not flat, but logarithmically divergent. The status of
this solution is unclear at present, together with that of the following
lines of the table. These states only appear as
particles in $D=3$, with the peculiarities that we have already
mentioned.

Upon decompactification, the last two lines in Table \ref{tfm} 
disappear since they require eight distinct radii, and the
particle multiplet reduces to a representation of the corresponding
U-duality group, as indicated in Table \ref{tfs}. When $d\ge 4$,
the representation remains the one dual to the rightmost root.
For $d=3$, the U-duality group disconnects into $Sl(3)$
and $Sl(2)$, and $-e_d$ becomes equal to $\lambda^{(2)}+\lambda
^{(3)}$ instead of being equal to $\lambda^{(3)}$, as in other
cases. Consequently, the particle multiplet transforms
as a $\irrep{(3,2)}$ representation of U-duality.

The full particle multiplet on $T^d$ can be easily decomposed
in representations of the U-duality group $E_{d-1}(\Zint)$ in one dimension
higher by separating the states in Table \ref{tfm} according
to their dependence on the decompactified radius $R_{d}$
(which gives a gradation with respect to the simple
root $\alpha_{d-1}$). We obtain the general decomposition\footnote{For
$d=8$, this is 
$\irrep{248} =\irrep{1} \oplus \irrep{56} \oplus (\irrep{133}
\oplus \irrep{1}) \oplus \irrep{56}  \oplus
\irrep{1}$.}
\begin{equation}
\label{partdec}
\M^{(d)} = 1 \vert_{-1}
\oplus \M\vert_0  \oplus (\T_1 \oplus \T'_1)\vert_1 \oplus
\T'_2  \vert_2  \oplus (\T_1' )^{2} \vert_3\ ,
\end{equation}
where we have denoted the multiplets as in Eq. \eqref{hwu}
and specified the power of $R_{d}$ in subscript. The notation
$(\T_1')^{2}$ means twice the fundamental weight associated to 
$\T_1'$.
The multiplets  on the right-hand side of \eqref{partdec} become empty 
as $d$ decreases. 
In particular, we note that the particle multiplet on $T^d$
decomposes into a singlet, corresponding to the Kaluza--Klein
excitation around the decompactified direction $x^d$, as well 
as a particle and a string multiplet on $T^{d-1}$,
depending on whether the state was wrapped around $x^d$.
There are also a number of additional states that appear for $d\geq 6$,
to which we shall come back in Subsection \ref{uduaext}.

\begin{table}[h]
\begin{center}
\begin{tabular}{|c|c||c|c|l|}
\hline
$D$&$d$ & U-duality group & irrep & $Sl(d)$ content \\ \hline  
10&1 & 1 & \irrep{1} & \irrep{1} \\
9 &2 & $Sl(2,\Zint)$ & \irrep{3}  & \irrep{2} + \irrep{1}  \\
8 &3 & $Sl(3,\Zint)\times Sl(2,\Zint)$ & \irrep{(3,2)} & \irrep{3} + 
   \irrep{3} \\
7 &4 & $Sl(5,\Zint)$ & \irrep{10} & \irrep{4} + \irrep{6} \\
6 &5 & $SO(5,5,\Zint)$ & \irrep{16} & \irrep{5} + \irrep{10} + \irrep{1} \\
5 &6 & $E_{6(6)}(\Zint)$ & \irrep{27} & \irrep{6} + \irrep{15} + \irrep{6} \\
4 &7 & $E_{7(7)}(\Zint)$ & \irrep{56} & \irrep{7} + \irrep{21} +
\irrep{21} +\irrep{7}  \\
3 &8 & $E_{8(8)}(\Zint)$ & \irrep{248} & 2(\irrep{8} + \irrep{28} + 
\irrep{56}) + \irrep{63}  + \irrep{1}\\
\hline
\end{tabular}
\end{center}
\caption{Particle/flux multiplets of $E_d$.
\label{tfs}}
\end{table}

As a side remark, we note that Table \ref{tfm} is symmetric
under reflection with respect to the middle line: 
for each state with mass $\M$ there is a state with mass $\M'$
satisfying
\begin{equation}
\label{mir8}
\M \M' = \left( \frac{V_R}{l_p^9} \right)^2 \ ,  
\end{equation}  
where $V_R$ is the volume of the eight-torus. In particular,
the lowest weight is equal to minus the highest weight,
modulo the invariant vector $\delta$. This is a general 
property of real representations of compact group, and indeed
\irrep{248} is the adjoint representation of $E_8$, therefore real.
The same also holds for the \irrep{56} representation of $E_7$ in the 
$d=7$ case. However, whether real or not with respect
to the {\it compact} real form of the group $E_d(\mathbb{C})$, 
all the representations
appearing in Table \ref{tfs} are real {\it as representations of
the non-compact group $E_{d(d)}$}, as is required by the existence
of an anti-particle for each particle. This is obvious for 
$d\le 4$; for $d=5$, it is equivalent to the statement that
the spinor of $SO(8)$ is real, since the reality properties
of spinors of $SO(p,q)$ depend only on $p-q \mod 8$. This property
is a characteristic feature of the representations of the maximally
non compact real form.
 
\subsection{T-duality decomposition and exotic states \label{exotic} }
In order to make contact with string theory solutions,
it is useful to decompose the particle multiplet into
irreducible representations of the T-duality group $SO(d-1,d-1,\Zint)$.
This can be simply carried
out by distinguishing whether the indices lie along the eleventh
dimension or not, and substituting the matching relations
\eqref{matching}. Since T-duality commutes with the grading in powers
of the string coupling $g_s$, the various irreps are then sorted out according
to the dependence of the mass of the states on $g_s$.
Table \ref{partbranch8} summarizes the decomposition of the 
particle/flux multiplet for M-theory on $T^8$ into irreducible representations
of the $SO(7,7)$ T-duality symmetry group of type IIA string theory on $T^7$,
as well as the $Sl(8)$ (resp. $Sl(7)$) modular group of the M-theory 
(resp. string theory) torus.
 
The masses of the states in the $a$-th column depend on the string coupling
constant as $1/g_s^{a-1}$, and are given by
\begin{subequations}
\label{2a7}
\begin{eqnarray}
V: & \frac{1}{R_i},\frac{R_i}{l_s^2} \\ 
S_A: & \frac{1}{g_s} \left( \frac{1}{l_s}, \frac{R_iR_j}{l_s^3}, 
  \frac{R_iR_j R_k R_l}{l_s^5}, 
  \frac{R_iR_j R_k R_l R_m R_n}{l_s^7} \right) \\  
S+ AS : & \frac{1}{g_s^2} \left( \frac{R_iR_j R_k R_l R_m }{l_s^6}, 
  \frac{V'_R }{l_s^8}, 
  \frac{R_i^2 R_j R_k R_l R_m R_n}{l_s^8},   
  \frac{V'_R R_i R_j }{l_s^{10} } \right) \\ 
S_B : & \frac{V'_R}{g_s^3 l_s^8}  \left( \frac{R_i}{l_s}, 
  \frac{R_i R_j R_k }{l_s^3}, 
  \frac{R_i R_j R_k R_l R_m }{l_s^5},  
  \frac{V'_R }{l_s^{7} } \right) \\ 
V' : & \left( \frac{V'_R}{g_s^2 l_s^8} \right)^2  \left( \frac{l_s^2}{R_i},
R_i \right) 
\end{eqnarray}
\end{subequations}
where $V'_R$ denotes the volume of the string-theory seven-torus.
At level $1/g_s^0$ we observe the usual KK and winding states of the string
and the level $1/g_s$ reproduces the D0-,D2-,D4- and D6-branes.
At level $1/g_s^2$, the NS5 and KK5-brane appear together with two new
types of state, a $7_{2}$-brane and a $5_{2}^2$-brane. 
Our nomenclature displays on-line the number of spatial world-volume
directions, \ie the number of radii appearing linearly in the mass; 
the superscript specifies the number of directions (if non-zero) that appear 
quadratic, cubic, etc., listed from the right to the left.  
the subscript denotes the inverse power of the string
coupling appearing in the mass formula;
for example, in this convention the KK5-brane is a $5_2^1$-brane.
According to this notation, we find at level $1/g_s^3$ a 
$6_3^1$-, $4_3^3$-, $2_3^5$- and $0_3^7$-brane. 
Their masses are related to those of
the even D$p$-branes, by the type IIA (on $T^7$) mirror symmetry 
 \begin{equation}
\label{mirs7a}
\M \M' = \left( \frac{V_R'}{g_s^2 l_s^8} \right)^2  \ ,
\end{equation}
which follows from the M-theory mirror symmetry relation \eqref{mir8}.
Finally, at level $1/g_s^4$, a $1_4^6$- and a $0_4^{(1,6)}$-brane are
obtained, whose masses are related to those of the KK and winding
states by \eqref{mirs7a}.

At this point a few remarks are in order about the new type IIA states that 
appear in \eqref{2a7}. The $7_2$- and $5_2^2$-brane, with mass 
proportional to $1/g_s^2$  have a conventional  dependence on the string 
coupling, but no supergravity solutions  are known for these states.
In addition, a variety of states with exotic
dependence on the string coupling, $1/g_s^3$ and $1/g_s^4$, are observed.
They arise from M-theory states with mass diverging as $1/l_p^9$ or faster. 
It is not clear what the meaning of these new states in M-theory
and type IIA string theory is. These states cannot be accommodated
in weakly coupled string theory where the most singular
behaviour is expected to be $1/g_s^2$, corresponding
to Neveu--Schwarz solitons. A higher power would imply a contribution
of a Riemann surface with Euler characteristic $\chi > 2$.
Another way to see this is by considering the  
gravitational field created by these objects, which scales as 
$\M \kappa^2_{10}$: since $\kappa^2_{10} \sim g_s^2$, states whose mass 
goes like $g_s^{n}$, $n \leq 2$ create a vanishing or at most
finite gravitational 
field in the weak coupling limit, allowing for a flat space description in the spirit
of D-branes. On the other hand, when $n>2$, the surrounding space
becomes infinitely curved at weak coupling, and these states
do not correspond to solitons anymore. In fact, the simplest of these
states, namely $6_3^1$, can be obtained by constructing an array
of Kaluza-Klein along a non-compact direction of the Taub-NUT space
\footnote{This construction first appeared in the context
of the conifold singularity in the hypermultiplet moduli space 
\cite{Ooguri:1996me}.}, and wrapping the worldvolume directions
on the string theory torus $T^6$ \cite{Blau:1997du}.
The summation of the poles in the harmonic function is
logarithmically divergent, implying that
the asymptotic space-time is logarithmically divergent as
well. This is the rule and not the exception for a pointlike
state in 3 space-time dimensions (since the Laplacian in the
two transverse coordinates has a logarithmic kernel), and
the conventional states with an asymptotically flat space-time
are simply configurations with a vanishing charge. The same
issue arises for $p$-branes in $p+3$ dimensions (or less).
We emphasize, though, that our present purpose is to
examine the consequences at the algebraic level of the presence
of the conjectured U-duality, which does require
these exotic states. The supergravity solutions describing these
states can in principle by computed using the known duality relations,
which indeed do not preserve the asymptotic flatness of the metric.

\begin{table}[H]
\hspace*{-1.7cm}
\begin{tabular}{|l||l|l|l|l|l|}
\hline 
$\begin{smallmatrix}
\irrep{248} (E_8) \supset   &SO(7,7)\\
 \cup   & \\
Sl(8) & \\ 
\end{smallmatrix} $
&  \irrep{14} ($V$)  & \irrep{64} ($S_A$)  & 
\irrep{1} + \irrep{91} ($S \oplus AS$)    &  \irrep{64} ($S_B$)  & \irrep{14} ($V'$)
\\ \hline \hline  
\irrep{8} ($m_1$) & \irrep{7} ($m_1$) & \irrep{1} ($m_s$) &  &  &  \\ 
 \irrep{28} ($m^2$) & \irrep{7} ($m^{s1}$) &  \irrep{21} ($m^2$) &  &  &  \\
 \irrep{56} ($m^5)$ &  & \irrep{35} ($m^{s4}$)  & \irrep{21} ($m^5$)  &  &  \\
 \irrep{1}+ \irrep{63} ($m^{1;7}$) &  & \irrep{7} ($m^{s;s6}$) &  \irrep{1} + \irrep{1}+ \irrep{48}  ($m^{s;7}, m^{1;s6}$) &  \irrep{7}
($m^{1;7}$)  &  \\ 
 \irrep{56} ($m^{3;8}$) &  &  & \irrep{21}  ($m^{s2;s7}$) &
 \irrep{35}
 ($m^{3;s7}$) & \\
 \irrep{28} ($m^{6;8}$) &  &  &  & \irrep{21}  ($m^{s5;s7}$) &
 \irrep{7}  
($m^{6;s7}$) \\
 \irrep{8}  ($m^{1;8;8}$) &  &  &  & \irrep{1} ($m^{s;s7;s7}$)  & 
\irrep{7} ($m^{1;s7;s7}$) \\  
\hline  
\end{tabular}
\caption[Branching of $d=8$ particle multiplet]{Branching of the
$d=8$ particle multiplet into irreps of $Sl(8)$ and $SO(7,7)$.
The entries in the table denote the irreps under the common $Sl(7)$ 
subgroup of $Sl(8)$ and $SO(7,7)$.
\label{partbranch8}}
\end{table}

\subsection{The string alias momentum multiplet \label{ssmm} }
The same analysis can be carried out for the string multiplet,
by applying a sequence of Weyl reflections on the 
highest weight $R_I/l_p^3$ describing the wrapped membrane.
After adding a multiplet of length 2 and 35 zero-weights
for $Sl(8,\Zint)$ invariance, we obtain a \irrep{3875} 
representation of $E_{8(8)}$. The precise content of
this representation is displayed in Appendix \ref{d8strm}; instead, 
we display in Table \ref{tmm} the 
more manageable result for the $d=7$ case, where
the string multiplet transforms as a \irrep{133} adjoint representation
of $E_{7(7)}$. The occurrence of the adjoint representation is 
again understood from Eq. \eqref{psi} relating the fundamental weight
$\lambda^{(1)}$ to the highest root $\psi$.

\begin{table}[H]
\begin{center}
\begin{tabular}{|c|l|l|}  
\hline
tension $\T$ mass & $Sl(7)$ irrep & charge\\ \hline  
$\frac{ R_I}{l_p^3}$ &  \irrep{7} & $n^1 $ \\ 
$ \frac{ R_I R_J R_K R_L}{l_p^6}$ &\irrep{35} & $n^4$ \\   
$\frac{ R_I^2 R_J R_K R_L R_M R_N }{l_p^9}$, 7$\frac{ V_R}{l_p^9}$ 
&\irrep{1}+\irrep{48}& 
$n^{1;6}$ \\
$ \frac{ R_I^2 R_J^2 R_K^2 R_L R_M R_N R_P}{l_p^{12}}$ & \irrep{35}&
$ n^{3;7} $\\ 
$ \frac{ R_I^2 R_J^2 R_K^2 R_L^2 R_M^2 R_N^2 R_P}{l_p^{15}}$ &
\irrep{7} & $n^{6;7}$ \\ 
\hline
\end{tabular} 
\end{center}
\caption{String/momentum multiplet \irrep{133} of $E_7$.
\label{tmm}}
\end{table}

\begin{table}[H] 
\begin{center}
\begin{tabular}{|c|c||c|c|l|}
\hline
$D$&$d$ & U-duality group & irrep & $Sl(d)$ content \\ \hline  
10&1 & 1 & \irrep{1} & \irrep{1} \\
9 &2 & $Sl(2,\Zint)$ & \irrep{2}  & \irrep{2}  \\
8 &3 & $Sl(3,\Zint)\times Sl(2,\Zint)$ & \irrep{(3,1)} & \irrep{3} \\
7 &4 & $Sl(5,\Zint)$ & \irrep{5} & \irrep{4} + \irrep{1} \\
6 &5 & $SO(5,5,\Zint)$ & \irrep{10} & \irrep{5} + \irrep{5} \\
5 &6 & $E_{6(6)}(\Zint)$ & \irrep{\bar{27}} & 
                   \irrep{6} + \irrep{15} + \irrep{6} \\
4 &7 & $E_{7(7)}(\Zint)$ & \irrep{133} & \irrep{7} + \irrep{35} + 
    \irrep{49} + \irrep{35} + \irrep{7} \\
3 &8 & $E_{8(8)}(\Zint)$ & \irrep{3875} & \irrep{8} + \irrep{70} + \ldots \\
\hline
\end{tabular}
\end{center}
\caption{String/momentum multiplets of $E_d$.
\label{tms}}
\end{table}

These states have the same interpretation as the states
in the particle multiplet, but for wrapping one
dimension less of the world-volume. In other words,
the states in the particle multiplet can be obtained
by wrapping strings on one dimension more-- except
for the Kaluza--Klein state, which is a genuine point-like
(or wave-like, rather) object. We note again that Table \ref{tmm}
is symmetric under reflection with respect to its middle line,
in agreement with 
the reality of the \irrep{133} adjoint representation of $E_7$.

The string multiplet in higher dimensions is simply obtained
by dropping the states that require too many different radii,
as displayed in Table \ref{tms};
in all cases, it corresponds to the representation dual
to the leftmost root $\alpha_1$. We note that in $d=6$ the
\irrep{\bar{27}} string multiplet is distinct from 
the \irrep{27} particle multiplet, but is related to
it by an outer automorphism of $E_6$ corresponding
to the $\Zint_2$ symmetry of its Dynkin diagram.
We also note, for later use, that in all cases 
the string multiplet representation arises in the symmetric
tensor product of two particle multiplets, \ie
$(\M\otimes_{s} \M)\otimes \T$ always contains a singlet.

Like the particle multiplet, the full string multiplet on $T^d$ 
can be easily decomposed in representations of the U-duality group 
$E_{d-1(d-1)}(\Zint)$ in one dimension
higher by using the gradation in powers of the decompactified 
radius $R_{d}$ \footnote{For $d=7$, this is 
$\irrep{133} =\irrep{27} \oplus (\irrep{78} \oplus 1) 
\oplus \irrep{\bar{27}}$.}
\begin{equation}
\label{stringdec}
\T_1^{(d)} = \T_1\vert_0  \oplus (\T_2\oplus \T'_2)\vert_1 \oplus
\T'_3\vert_2 \ ,
\end{equation}
where we have denoted the multiplets as in Eq. \eqref{hwu}
and again specified the power of $R_{d}$ in subscripts.
In particular, we note that the string multiplet on $T^d$
decomposes into a string and a membrane multiplet on $T^{d-1}$,
depending whether the state was wrapped around $x^d$.
There are also a number of additional states that disappear for $d\le 6$.

\begin{table}[H] 
\hspace*{-8mm}
\begin{tabular}{|l||l|l|l|l|l|}
\hline 
$\begin{smallmatrix}
\irrep{133} (E_7) \supset  &  SO(6,6) \\
 \cup  & \\
Sl(7)& \\ 
\end{smallmatrix}$ &  \irrep{1} ($S$)  & \irrep{32} ($S_B$)  & 
\irrep{1} + \irrep{66} ($S' \oplus AS$)    &  \irrep{32} ($S_B'$)  & 
\irrep{1} ($S''$)
\\ \hline \hline   
 \irrep{7} ($n^1$) & \irrep{1} ($n^s$) & \irrep{6} ($n^1$) &  &  &  \\ 
 \irrep{35} ($n^4$) &   &\irrep{20} ($n^{s3}$) &  
\irrep{15} ($n^4$) &  &    \\ 
 \irrep{1}+ \irrep{48} ($n^{1;6}$) &  & \irrep{6} ($n^{s;s5}$) &  
\irrep{1} + \irrep{1}+ \irrep{35} ($n^{s;6}, n^{1;s5}$) &  \irrep{6} 
($n^{1;6}$)  &  \\ 
\irrep{35} ($n^{3;7}$) &  &  & \irrep{15}  ($n^{s2;s6}$) &  
\irrep{20} ($n^{3;s6}$) &  \\ 
\irrep{7} ($n^{6;7}$) &  &   & & \irrep{6}  ($n^{s5;s6}$) & 
\irrep{1}  ($n^{6;s6}$) \\
\hline  
\end{tabular}
\caption[Branching of $d=7$ string multiplet]{Branching of the
$d=7$ string multiplet into irreps of $Sl(7)$ 
and $SO(6,6)$.
The entries in the table denote the irreps under the common $Sl(6)$ 
subgroup of $Sl(7)$ and $SO(6,6)$.
\label{stringbranch}}
\end{table}

As in the previous subsection, we give the branching of the $d=7$ string
multiplet in terms of irreps of the T-duality $SO(6,6,\Zint)$ as
well as the modular groups $Sl(7,\Zint)$ and $Sl(6,\Zint)$ of the
M-theory and string theory tori
in Table \ref{stringbranch}.

\subsection{Weyl-invariant effective action}
As in our discussion of T-duality, we would now like to write
the supergravity action \eqref{11da} in a manifestly Weyl-invariant form.
This has been carried out in Refs. \cite{Lu:1996yn,Lu:1996ge}, 
a simplified version of which will be presented here.
As in Eq. \eqref{sgrt}, we decompose the eleven-dimensional field
strength $F^{(4)}$ and metric in lower-degree forms. 
The action then takes the simple form:
\begin{multline}
\label{sgru}
S_{11-d} =\int {\rm d}^{11-d}x \sqrt{-g}~\frac{V}{l_p^9}
\left[ 
R + \sum_i \left(\frac{\partial R_i}{R_i}\right)^2 
+ \sum_{i<j} \left(\frac{R_i}{R_j}\mathcal{F}^{i(1)}_j\right)^2 
+ \sum_i \left(R_i \mathcal{F}^{(2)i} \right)^2  
\right. \\
+\left.
 \left( l_p^3 F^{(4)} \right)^2 
+ \sum_i \left(\frac{l_p^3}{R_i}F^{(3)}_i\right)^2 
+ \sum_{i<j} \left(\frac{l_p^3}{R_i R_j}F^{(2)}_{ij}\right)^2
+ \sum_{i<j<k} \left(\frac{l_p^3}{R_i R_j R_k}F^{(1)}_i\right)^2
 \right] \ , 
\end{multline}
where the first line comes from the reduction of the Einstein--Hilbert
term and the second from the kinetic term of the three-form.

In Eq. \eqref{sgru}
we again recognize in front of the one-form field strength
$F^{(1)}$ and $\mathcal{F}^{(1)}$ the positive roots 
$e_i-e_j$ and $e_i+e_j+e_k-e_0$, 
in front of the two-form field strength
$F^{(2)}$ the weights $-e_i$ of the particle multiplet,
in front of the three-form field strength
$F^{(3)}$ the weights $e_i-e_0$ of the string multiplet,
and in front of the four-form field strength
$F^{(4)}$ the weight $-e_0$ of the membrane multiplet.
However, these weights do not form complete orbits:
it is necessary to dualize the field strengths $F^{(p)}$
into lower-degree field strengths $F^{(11-d-p)}$
so as to display the Weyl symmetry. In
the $2p=11-d$ case, both the field strength and
its dual should be kept. Alternatively, all field
strengths may be doubled with their duals, and display an even larger
symmetry \cite{Cremmer:1997ct,Cremmer:1998px}.

We then obtain a manifestly Weyl-invariant action:
\begin{multline}
\label{wiu}
S_{11-d} =\int {\rm d}^{11-d}x \sqrt{-g}~\frac{V}{l_p^9}
\left[ 
R + \partial\varphi \cdot \partial\varphi 
\right.
\\
+ \sum_{\alpha\in\Phi_+}   e^{-2 \langle\varphi,\alpha\rangle} 
            \left(\mathcal{F}^{(1)}_{\alpha}\right)^2 
+ \sum_{\lambda\in\Phi_{\rm part}} e^{-2 \langle\varphi,\lambda\rangle} 
            \left(\mathcal{F}^{(2)}_\lambda\right)^2  
\\
\left. 
+ \sum_{\lambda\in\Phi_{\rm string}} e^{-2 \langle\varphi,\lambda\rangle}  
            \left( \mathcal{F}^{(3)}_\lambda         \right)^2 
+ \sum_{\lambda\in\Phi_{\rm membrane}} e^{-2 \langle\varphi,\lambda\rangle}  
            \left( \mathcal{F}^{(4)}_{\lambda}         \right)^2 
+ \dots
\right] \ , 
\end{multline}
where $\varphi=(\ln l_p^3,\ln R_1,\dots,\ln R_d)$ is the vector of 
dilatonic scalars (whose first component is non-dynamical), 
$\langle\varphi,\lambda\rangle=
x^0\ln l_p^3+ x^1\ln R_1 +\dots$ is the duality bracket \eqref{dualbu}
and $\partial\varphi\cdot\partial\varphi$ the Weyl-invariant
kinetic term ($\partial l_p^3=0$) obtained from the metric \eqref{um}.
In addition to the equations of motion from \eqref{wiu},
the duality equations $F^{(p)}=*F^{(11-d-p)}$ should also be imposed.
As in the case of T-duality, the set of positive
roots $\Phi_+$ is not invariant under Weyl reflections,
but the Peccei--Quinn scalars undergo non-linear
transformations $A ^{(0)}\rightarrow e^{-2 \langle\varphi,\alpha\rangle} 
A ^{(0)}$ that compensate for the sign change \cite{Lu:1996ge}.

\subsection{Compactification on $T^9$ and affine $\hat E_8$ 
symmetry\label{e9sec}}
As we pointed out in Subsection \ref{wgwru}, the compactification on
a nine-torus $T^9$ to two space-time dimensions gives rise
to a qualitative
change in the U-duality group: the invariant vector $\delta$ in \eqref{delta} 
corresponding to the dimensionless Newton constant becomes 
light-like in the Lorentzian metric 
$-(dx^0)^2 + (dx^1)^2 + \dots + (dx^9)^2$, so that the action
of the U-duality group generated by $S_I$ and $T$ in Eq. \eqref{sit}
cannot be restricted to its orthogonal subspace. Instead,
it generates the Weyl group of the $\hat E_8$ affine algebra,
as was shown in Ref. \cite{Elitzur:1997zn}; we shall recast 
their construction in the notation of this review, at the same
time settling several issues.

In order to see the affine symmetry $\hat E_8$ arise, we 
simply note that the Dynkin diagram of $E_9$ 
(see Table \ref{dynfigure}) is nothing
but the extended Dynkin diagram of $E_8$, where the additional
root with Coxeter label $1$ corresponds to $\alpha_8 = e_9 -e_8$.
The roots $\alpha_0, \alpha_1,\dots, \alpha_7 $ generate
the $E_8$ horizontal Lie algebra, whereas $\alpha_8$ and
$\delta = \sum_{I=1}^9 e_I - 3e_0 $ 
are the extra dimensions needed to represent
the central charge $K$ and degree $D$ generators of
the standard construction of affine Lie algebras
(see e.g. Ref. \cite{Fuchs:1992}). 
To make the identification precise,
we recall that the simple roots of an affine Lie algebra $\hat G$ can
be chosen as
\footnote{In order to keep with the standard notation, the simple roots of the
Lie algebra are now labelled by subscripts ranging from 1 to $r$, as opposed
to our notation for the simple roots of the U-duality groups $E_r$, 
which carry  labels 0 to $r-1$.}  
\begin{equation}
\label{afl}
\hat\alpha_I = (\alpha_I,0,0) \sp I = 1 \ldots r  \sp
\hat\alpha_0 = (-\psi,0,1)
\end{equation}
in the basis $(\mu,k,d)$ of the Minkovskian weight space $V_{r+2}
=\Real^r + \Real^{1,1}$ with norm $\mu ^2 + 2 k d$. Here,
$\psi$ is the highest root of $G$, $r$ is the rank of $G$,
$k$ is the affine level, and
$d$ the $L_0$ eigenvalue. 
In the case at hand,  we have
$G= E_8$ so $r=8$ and want to find
the change of basis between the roots $\alpha_I, I=0\dots 8$
and null vector $\delta$
of our formalism and the standard roots $\hat \alpha_I, I=0\dots 8$
and vectors $\gamma=(0,0,1),\kappa=(0,1,0)$. 
{}From Eq. \eqref{psi} we have,
\begin{equation}
\psi=e_1 + \dots + e_7 + 2 e_8 - 3 e_0 = \delta-\alpha_8 \ , 
\end{equation}
so that, comparing with Eq. \eqref{afl},
 we can identify $\delta$ with $\gamma=(0,0,1)$ and
\begin{subequations}
\begin{eqnarray}
\hat\alpha_I = \alpha_I\ ,I=1\dots 7\sp\\
\hat\alpha_8 = \alpha_0\sp\\
\hat\alpha_0 = \alpha_8 \ . 
\end{eqnarray}
\end{subequations}
The vector $\kappa=(0,1,0)$ can be easily calculated from
the requirements that $\kappa ^2=\kappa\cdot\hat\alpha_I=0,
I=1\dots 8$ and $\kappa\cdot\delta=1$:
\begin{equation}
\kappa= \frac{1}{2}\left(
-e_1 - \dots - e_8 + e_9 + 3 e_0 \right) = e_9 - \frac{\delta}{2}\ .
\end{equation}
The level $k$ and degree $d$ 
of any
weight vector $\lambda \in V_{10}$ can now be obtained from the products 
$\delta \cdot \lambda $ and $\delta\cdot \kappa$ 
respectively, and they both
have a simple interpretation:
\begin{equation}
k = \delta \cdot \lambda = x^1 + \dots + x^9 + 3 x^0
\end{equation} 
is simply
the length dimension of the associated monomial
$\prod R_I ^{x^I} l_p ^{3 x^0}$, and
\begin{equation}
d = \kappa \cdot \lambda = x^9 - k/2
\end{equation}
counts the power of $R_9$ appearing in the same monomial,
up to a shift $k/2$. This was expected, since the horizontal
subalgebra $E_8 \subset \hat E_8$ does not affect $R_9$ and
by definition commutes with $L_0$. $L_{-n}$ generators, on the other hand,
bring additional powers of $R_9$ and increase the degree $d$.
In particular, the $L_0$
eigenvalues are integer-spaced, as they should.

We proceed by considering the particle/flux and string/momentum multiplets
introduced in Subsections \ref{spfm} and \ref{ssmm}, with 
highest weights $\lambda^{(d-1)}=-e_9$ and $\lambda^{(1)}=e_1-e_0$
respectively (see Eq. \eqref{hwu}). The particle multiplet is therefore
a level $-1$ representation with trivial ground state
$\mu=0$ (that is, in the chiral block of the identity).
A bit of experimentation reveals the first $Sl(9)$
representations occurring in the particle multiplet:
\begin{equation}
m^{1;1;9},\ m^{1;4;9}, \ m^{2;6;9}, \ 
m^{4;7;9},\ m^{7;7;9}, \ m^{2;3;9;9}, \dots
\end{equation}
with tensions scaling from $1/l_p^{12}$ to $1/l_p^{24}$,
in addition to the representations 
already present in $d=8$,  given in Table \ref{tfm}. 
However, the full orbit is infinite.
On the other hand, the string multiplet is a level $-2$ 
representation with ground state in the \irrep{3875}
of $E_8$. 
In both cases, the representations are infinite-dimensional,
and need to be supplemented with weights of smaller
length as in the $E_7$ and $E_8$ cases. 
The instanton multiplet, on the other hand, is 
a {\it level-0} representation of $\hat E_8$, with a {\it non-singlet} 
ground state in
the adjoint of $E_8$, making it obvious that the usual
unitarity restrictions for compact affine Lie algebras  
do not apply in our case.
This concludes
our analysis of the $d=9$ case, and we now restrict ourselves to
the better understood $d\le 8$ case.

\clearpage
\section[Mass formulae on skew tori with gauge backgrounds]{Mass 
formulae on skew tori with gauge\\ backgrounds
  \label{sskew}}

We would now like to generalize the mass formulae of
the U-duality multiplets obtained so far for rectangular tori and
vanishing gauge potentials to the more general case of skew tori
and arbitrary gauge potentials, which will exhibit the full U-duality
group. This will also allow a better understanding of the
action of Borel generators on the BPS spectrum. We will concentrate
on the $d=7$ flux multiplet, but the same method applies to  the
other multiplets.

\subsection{Skew tori and $Sl(d,\Zint)$ invariance}
We have already argued that BPS states could be labelled by 
a set of tensors of integer charges describing their
various momenta and wrappings. In particular, for the
case of the $d=7$ flux multiplet, the charges
\begin{equation}  m_1, m^2, m^5, m^{1;7}
\end{equation}
describe the Kaluza--Klein momentum, membrane, fivebrane and KK6-brane
wrappings. The position of the index has been chosen in such a
way that we obtain the correct mass 
by contracting each of them with the vector of 
radii $R^I$ or inverse radii $1/R_I$. Note that for $d=7$ the tensor
$m^{1;7}$ is really a tensor $m^1$, but the extra seven indices
account for an extra factor of the volume in the tension.
Of course, a BPS state with generic charges $m$ will not be 
1/2-BPS state in general (for $d\ge 5$): some quadratic conditions
on $m$ have to be imposed, as already discussed in Subsections
\ref{bpss} and \ref{dbtd}. We shall henceforth assume these conditions
fulfilled, deferring the study of the latter to 
Subsection \ref{bpscond}.

The 1/2-BPS state mass formula for a non-diagonal metric $g_{IJ}$ can be 
straightforwardly obtained by replacing contractions
with the vector of radii by contractions with the metric,
and inserting the proper symmetry factor and power
of the Planck length on dimensional grounds:
\begin{equation}
\label{masssl}
\begin{array}{ll} 
\M^2 &= (m_1)^2 + (m^2)^2 + (m^5)^2 + (m^{1;7})^2  \\
&= m_I g^{IJ} m_J + \frac{1}{2!~l_p^6} m^{IJ} g_{IK} g_{JL} m^{KL} \\
&\quad
+\frac{1}{5!~l_p^{12}} m^{IJKLM} g_{IN} g_{JP} g_{KQ} g_{LR} g_{MS} 
m^{NPQRS} +\dots
\end{array}
\end{equation}
This formula is  invariant under $Sl(d,\Zint)$, but {\it not} yet
under the T-duality subgroup $SO(d-1,d-1,\Zint)$ of the U-duality
group. It only holds when the expectation value of the various
gauge fields on the torus vanish. 
To reinstate the dependence on the three-form $  \C_{IJK}$,  
we apply the following strategy. 
\begin{itemize}
\item Decompose the flux multiplet as a sum of T-duality irreps. 
\item Include the correct coupling to the NS two-form field $B_{ij}$
using the T-duality invariant mass formulae. 
\item Study the T-duality spectral flow $B \ra B + \Delta B$.
\item Covariantize this flow under $Sl(d,\Zint)$ into
a $\C\ra C+\Delta\C$ flow.
\item Integrate the $  \C \ra \C + \Delta \C $ flow to obtain 
the U-duality invariant mass formula.  
\end{itemize}

\subsection{T-duality decomposition and invariant mass formula}
We have already discussed the first step in Subsection \ref{exotic},
and we only need to restrict ourselves to the case $d=7$.
Table \ref{partbranch8} then truncates to its upper left-hand corner 
displayed in Table \ref{partbranch}, as can be read from the $d=7$ 
particle multiplet mass formula \eqref{masssl} 
written with $s$ and $i$ indices:
\begin{eqnarray}
\M^2 &= \left[ \frac{m_s^2}{g_s^2}+ (m_1)^2 \right]
+\left[ (m^{s1})^2 + \frac{(m^{2}) ^2}{g_s^2} \right]\nonumber\\
&+\left[ \frac{(m^{s4})^2}{g_s^2} + \frac{(m^{5})^2}{g_s^4} \right]
 +\left[ \frac{(m^{s;s6})^2}{g_s^2} + 
        \frac{m^{1;s6})^2}{g_s^4} \right]
\end{eqnarray}
corresponding to three $SO(6,6)$ irreps, 
\begin{subequations}
\begin{eqnarray}
V=&(m_1,m^{s1}) & \mbox{momentum and winding} \\
S=&(m_s,m^2,m^{s4},m^{s;s6}) & \mbox{D0-,D2-,D4-,D6-brane} \\
V'=&(m^5,m^{1;s6}) & \mbox{NS5-brane and KK5-brane} 
\end{eqnarray}
\end{subequations}

\begin{table}[h] 
\begin{center}
\begin{tabular}{|l||l|l|l|}
\hline 
$\begin{smallmatrix}
\irrep{56}(E_7)\supset &SO(6,6)\\
\cup & \\
Sl(7)& \\ 
\end{smallmatrix}$ &  \irrep{12} ($V$)  & \irrep{32} ($S_A$) &  
\irrep{12} ($V'$)
\\ \hline \hline  
 \irrep{7} ($m_1$) & \irrep{6} ($m_1$) & \irrep{1} ($m_s$) &    \\ 
 \irrep{21} ($m^2$) & \irrep{6} ($m^{s1}$) &  \irrep{15} ($m^2$) &    \\ 
 \irrep{21} ($m^5)$ &  & \irrep{15} ($m^{s4}$)  & \irrep{6} ($m^5$)    \\
 \irrep{7} ($m^{1;7}$) &  & \irrep{1} ($m^{s;s6}$) &  \irrep{6}  
($m^{1;s6}$)   
  \\ \hline 
\end{tabular}
\caption[Branching of $d=7$ particle multiplet]{Branching of 
the $d=7$ particle multiplet into irreps of $Sl(7)$ and 
$SO(6,6)$. The entries in the table denote the irreps under the common
$Sl(6)$ subgroup of $Sl(7)$ and $SO(6,6)$.
\label{partbranch}}
\end{center}
\end{table}
%
We can now use the T-duality invariant mass formulae for the T-duality
irreps that we obtained in Section \ref{tdua}. In terms
of the present charges, they schematically read (in units of $l_s$)
\begin{subequations}
\label{tduainvmass}
\begin{eqnarray}
\M_V^2&=& \left( m_i + B_{ij} m^{sj} \right) g^{ik}
\left( m_k + B_{kl} m^{sl} \right) + m^{si} g_{ij} m^{sj} \\
\M_{S_B}^2&=&\frac{1}{g_s^2}\left[
\left(m_s+m^{2}B_2 +m^{s4}B_2^2 + m^{s;s6}B_2^3 \right)^2\right.
\nonumber\\
&&\quad+ \left(m^{2}+m^{s4}B_2 + m^{s;s6}B_2^2 \right)^2\nonumber\\
&&\left.\quad+\left(m^{s4}+m^{s;s6}B_2 \right)^2
  +\left(m^{s;s6}\right)^2 \right] \\
\M_{V'}^2&=& \frac{1}{g_s^4} \left[\left(m^{5}+m^{1;s6}B_2 \right)^2 +
\left(m^{1;s6}\right)^2 \right]\ ,    
\end{eqnarray}
\end{subequations}
where we used the vector and spinor representation 
mass formulae \eqref{vectormass} and
\eqref{dparticle}. 
Adding the three contributions $\M^2_{ \{ V,S_B,V' \} }$ together, 
we now obtain the flux multiplet mass formula for vanishing values 
of the Ramond fields and arbitrary $B$-field:
\begin{subequations}
\label{massso}
\begin{eqnarray} 
\M^2 =& \left[ \frac{\tilde m_s^2}{g_s^2}+ (\tilde m_1)^2 \right]
+\left[ (\tilde m^{s1})^2 + \frac{(\tilde m^{2}) ^2}{g_s^2} \right]\\
&+\left[ \frac{(\tilde m^{s4})^2}{g_s^2} + \frac{(\tilde m^{5})^2}{g_s^4} \right]
+\left[ \frac{(\tilde m^{s;s6})^2}{g_s^2} + 
        \frac{\tilde m^{1;s6})^2}{g_s^4} \right] \ ,  
\end{eqnarray}
\end{subequations}
where the tilded charges are shifted to incorporate the
effect of the two-form as in \eqref{tduainvmass}, so that for instance
\begin{equation}
\tilde{m}_s = m_s + \frac{1}{2} B_2 m^2  + \frac{1}{8} B_2^2 m^{s4}
+ \frac{1}{48} B_2^3 m^{s;s6} 
\end{equation}
is the shift in the D0-brane charge.

\subsection{T-duality spectral flow}
In Subsections \ref{spf} and \ref{dbtd} 
we have already discussed the spectral flow $B_{ij}\rightarrow
B_{ij}+\Delta B_{ij}$ in the vectorial and spinorial representations.
We only need to rephrase this flow in terms of the present charges:
\begin{equation}
\label{tdspf}
\begin{array}{ll}
V:   & m_i \rightarrow  m_i + \Delta B_{ji} m^{sj}\ ,\quad
       m^{si} \rightarrow m^{si} \vspace{2mm} \ \\
S_B: & m_s \rightarrow m_s + \frac{1}{2}\Delta B_{ij}m^{ij} \ ,\quad
       m^{ij}\rightarrow m^{ij} + \frac{1}{2} \Delta B_{kl} m^{sklij} \ \\
     & m^{sijkl}\rightarrow m^{sijkl} + \
                \frac{1}{2} \Delta B_{mn}m^{s;smnijkl} \ ,\quad
       m^{s;s6}\rightarrow m^{s;s6} \ \vspace{2mm}\\
V' :   & m^{ijklm} \rightarrow m^{ijklm} -   \Delta B_{np} m^{n;spijklm}  \   \\
     & m^{1;s6} \rightarrow m^{1;s6} 
\ .\\
\end{array}
\end{equation}
The flow indeed acts as an automorphism on the charge lattice, and in 
particular the charges cannot be restricted to positive integers 
(except for $  m^{1;7} $).  
This fact will be of use in Subsection \ref{nahm}.

Alternatively, the above spectral flow can be recast into a system of
differential equations for the shifted charges $\tilde m$, e.g. for
the spinor representation we have
\begin{equation}
S_B:
\begin{array}{llllll}
\frac{ \partial \tilde{m}_s }{\partial B_{ij} }  &=&
\frac{1}{2} \tilde m^{ij}\ ,\quad &
 \frac{ \partial \tilde{m}^{ij} }{\partial B_{kl} }  &=&
\frac{1}{2} \tilde m^{sijkl} \ \quad\\
\frac{ \partial \tilde{m}^{sijkl} }{\partial B_{mn} }  &=&
\frac{1}{2} \tilde m^{s;sijklmn} \ , \quad
&
 \frac{ \partial \tilde{m}^{s;sijklmn} }{\partial B_{pq} }  &=& 0\ .
\end{array}
\end{equation}
This system can be integrated to yield the spinor representation 
mass formula; the constants
of integration correspond to the integer charges $m$.
The integrability of this system of differential equations follows from
the commutativity of the spectral flow.

\subsection{U-duality spectral flows \label{uduaspfl} }
The mass formula \eqref{massso}  obtained so far is 
invariant under T-duality and
holds for vanishing values of Ramond gauge backgrounds.
In order to obtain a U-duality invariant mass formula, we have
to allow expectation values of the  M-theory gauge three-form $\C_{IJK}$, 
which extends the Neveu-Schwarz two-form $  B_{ij} = \C_{sij}$; 
the expectation value
of the  Ramond one-form is already incorporated as the off-diagonal 
metric component
$\A_i=g_{si}/R_s^2 \ne 0$.  For $d\geq 6$, one should also
allow expectation values of the six-form 
$\E_{IJKLMN}$ (Poincar{\'e}-dual to $  \C_{IJK}$ in eleven dimensions). 
In string-theory language, this  
corresponds to 
the Ramond five-form $  \E_{s5}$ and the Neveu-Schwarz six-form dual to $  B_{\mu\nu}$ in 
ten dimensions
\footnote{For $d=8$, we also need to include the form $\K_{1;8}$,
which in string-theory language includes the Ramond seven-form $\K_{s;s7}$,
along with a $\K_{1;s7}$ form.}.   

In order to reinstate  the $  \C_{IJK}$ dependence in mass formula
we covariantize the
$B_{ij}=\C_{sij}$ spectral flow \eqref{tdspf} under $  Sl(d,\Zint)$, with 
the result that
\begin{equation}
\label{cf}
\begin{array}{llll}
m_I &\rightarrow& m_I &+~ \frac{1}{2} \Delta\C_{JKI}~ m^{JK}\\
m^{IJ} &\rightarrow& m^{IJ} &+~ \frac{1}{6} \Delta\C_{KLM}~ m^{KLMIJ}\\
m^{IJKLM} &\rightarrow& m^{IJKLM} &+~ \frac{1}{2}
  \Delta\C_{NPQ}~m^{N;PQIJKLM}\\
m^{1;7} &\rightarrow& m^{1;7}\ . &
\end{array}
\end{equation}
Here, however, the $\C$ spectral flow turns out to be {\it non-integrable}.
Defining 
$  \nabla^{IJK}$ as the flow induced by the shift
 $  \C_{IJK} \rightarrow \C_{IJK} + \Delta\C_{IJK}$, we have
the commutator 
\begin{equation}
\label{dcdc}
\left[ \nabla^{IJK},\ \nabla^{LMN} \right] = 20 \nabla^{IJKLMN}\ ,
\end{equation}
where 
  $  \nabla^{IJKLMN}$ is the flow induced by  the shift
$  \E_{IJKLMN} \rightarrow \E_{IJKLMN} + \Delta\E_{IJKLMN}$:
\begin{equation}
\begin{array}{llll}
m_I &\rightarrow& m_I &+ \frac{1}{5 !} \Delta\E_{JKLMNI}~ m^{JKLMN}\\
m^{IJ} &\rightarrow& m^{IJ} &+ \frac{1}{5!} \Delta\E_{KLMNPQ}~ m^{K;LMNPQIJ}\\
m^{5} &\rightarrow& m^{5} &\\
m^{1;7} &\rightarrow& m^{1;7} \ .&
\end{array}
\end{equation}

The non-integrability \eqref{dcdc} of the $\C$-flow
can be understood as a consequence of
the Chern--Simons interaction in the 11D supergravity
action Eq. \eqref{11da} \cite{Cremmer:1997ct,Cremmer:1998px}:
the equation of motion for $\C$ reads
\begin{equation}
d*F_4 + \frac{1}{2} F_4\wedge F_4 =0 \ , 
\end{equation}
so that the dual field strength of $F_4$ has a Chern--Simons
term
\begin{equation}
\label{fo}
F_7 \equiv  *F_4 = d\E_6 - \frac{1}{2} \C_3 \wedge F_4\ . 
\end{equation}
The equation of motion \eqref{fo} is invariant under
the gauge transformations 
\begin{equation}
\delta \C_3 = \Lambda_3\sp \delta\E_6 = \Lambda_6 -\frac{1}{2}
\Lambda_3\wedge \C_3\ ,
\end{equation}
for closed $\Lambda_3$ and $\Lambda_6$. Restricting to constant
shifts, this reproduces the commutation relations \eqref{dcdc}. An equivalent
statement holds in $D=3$, where the $\C_3$
and $\E_6$ shifts close on a $\K_{1;8}$ shift.

The non-integrability of the system \eqref{cf} can therefore
be evaded by combining the $\Delta \C_3$ shift
with a $\Delta \E_6$ shift
\begin{equation}  
\frac{1}{5!} \Delta \E_{IJKLMN} =  \frac{1}{12} \C_{[IJK} \Delta
\C_{LMN]}\ ,
\end{equation}  
upon which the resulting flow
\begin{equation} 
\label{intflow}
\nabla^{\prime IJK} = \nabla^{IJK} - 10 C_{KLM} \nabla^{KLMIJK}
\end{equation} 
becomes integrable
\footnote{For $d =8$ there is also a non-trivial commutator
\cite{Cremmer:1997ct}  
between the $\C_3$ and $\E_6$ flow, closing onto the $\K_{1;8}$ flow,
which induces further shifts.}.
The extra shift is 
invisible in the type IIA picture for
zero Ramond potentials since it does not contribute to the T-duality
spectral flow. 
We  emphasize again that these extra terms are generated as a
consequence of the integrability of the flow, which we take as a guiding
principle for reconstructing the invariant mass formula.
The explicit form of the resulting flow equations that follow
from \eqref{intflow} is then given by \cite{Obers:1997kk}
\begin{subequations}
\label{fleq}
\begin{eqnarray}
\nabla^{\prime JKL} \tilde m_I  &  =  & \frac{1}{2}  \tilde m^{JK} \delta_I^L\\
\nabla^{\prime KLM} \tilde m^{IJ} & = & \frac{1}{6} \tilde m^{KLMIJ}\\
\nabla^{\prime NPQ}  \tilde m^{IJKLM} & = &  \frac{1}{2} \tilde m^{N;PQIJKLM}\\
\nabla^{\prime RST} \tilde m^{I;JKLMNPQ} & = & 0\ ,
\end{eqnarray}
\end{subequations}
\begin{subequations}
\begin{eqnarray}
\nabla^{JKLMNP} \tilde m_I  &  =  & \frac{1}{5!}  \tilde m^{JKLMN} \delta_I^P\\
\nabla^{KLMNPQ} \tilde m^{IJ} & = & \frac{1}{5!} \tilde m^{K;LMNPQIJ}  \\
\nabla^{NPQRST}  \tilde m^{IJKLM} & = &  0 \\
\nabla^{RSTUVW} \tilde m^{I;JKLMNPQ} & = & 0\ ,
\end{eqnarray}
\end{subequations}
which now can be integrated, as will be shown in Subsection \ref{partmass}.  

\subsection{A digression on Iwasawa decomposition}
In order to understand the non-commutativity of
the spectral flow from another perspective, it is worthwhile
coming back to a simpler example of a non-compact group, namely
the prototypical $G(\Real)=Sl(n,\Real)$ group. The Iwasawa
decomposition \eqref{iwa} then takes the form
\begin{equation}
\label{iwa2}
g= k\cdot a\cdot n \in K \cdot A \cdot N\ ,
\end{equation}
where $K=SO(n,\Real)$ is the maximal compact subgroup of $G(\Real)$,
$A$ is the Abelian group of diagonal
matrices with determinant 1 
and $N$ is the nilpotent group of upper triangular
matrices. The factor $k$ is absorbed in the coset $G(\Real)/K$, and
the coset space is really parametrized by $A \cdot N$. 

Now
the subgroup of $G(\Zint)$ leaving $A$ invariant is nothing
but the Weyl group $\mathcal{S}_n$ of permutations of entries
of $A$,  whereas that leaving $N$ invariant is the Borel
group of integer-valued upper triangular matrices with 1's
on the diagonal. The latter is graded by the distance away from
the diagonal, in the sense that 
\begin{equation}
[ B_p , B_{p'} ] \subset B_{p+p'}\ ,
\end{equation}
where $B_p$ is the subset of upper triangular matrices with 1's
on the diagonal and other non zero entries on the $p$-th diagonal
only. In particular, $B_p$ is a non-compact Abelian subgroup
when $p>n/2$. 

Returning to the case at hand, we see that $\nabla^{3}$,
$\nabla^{6}$ (and $\nabla^{1;8}$ in the $d=8$ case)
are analogous to the $B_1$, $B_2$ (and $B_3$) Borel generators
of $Sl(3)$ (or $Sl(4)$). More precisely, they correspond to
the grading of the root lattice of $E_d$ with
respect to the simple root $\alpha_0$ extending the 
$Sl(d,\Zint)$ Lorentz subgroup to the full $E_{d(d)}(\Zint)$ subgroup,
or in other words the grading of the adjoint representation
in powers of $l_p^3$. This can be seen from Table \ref{tfm}
for $d=8$ since, in this case, the particle multiplet
happens to be in the adjoint representation 
\irrep{248} of $E_8$. For $d<8$, this can also be seen
from the Coxeter label $a_0$ of $\alpha_0$ in
Table \ref{dynfigure}, \ie the $\alpha_0$ component
of the highest root of $E_d$: the degree $p$ of all the positive
roots then runs from $0$ (corresponding to the $g_{IJ}$ Borel
generators) to $a_0$, with intermediate values $1$ for
the $\C_{3}$ flow, $2$ for $\E_{6}$ and $3$ for
$\K_{1;8}$. 

We finally note that,  
in the notation of Eq. \eqref{iwa2}, 
the mass formula we are seeking 
takes the form, 
\begin{equation}
\mathcal{M}^2= m^t\  R^t(a\cdot n) R(a\cdot n) \ m\ ,
\end{equation}
where $m$ is the vector of integer charges transforming in
the appropriate linear representation $R$ of $E_{d(d)}(\Real)$.

\subsection{Particle multiplet and U-duality invariant mass formula 
\label{partmass} }

The flow \eqref{fleq} can be integrated to  
obtain the $E_{7(7)}(\Zint)$-invariant
mass formula for the particle multiplet of M-theory
compactified on a torus $T^7$ with arbitrary shape
and gauge background. The result is:
\begin{equation}
\label{mass} 
\M^2  = \left( \tilde m_1 \right)^2 +
         \frac{1}{2!\ l_p^6} \left( \tilde m^{2} \right)^2 +
         \frac{1}{5!\ l_p^{12}} \left( \tilde m^{5} \right)^2 +
         \frac{1}{7!\ l_p^{18}}  \left( \tilde m^{1;7} \right)^2\ ,
\end{equation}
where the shifted charges depend on the gauge potentials as
\begin{subequations}
\begin{align}
\label{chshft}
\begin{split}
\tilde m_I   &= m_I  + \frac{1}{2} \C_{JKI} m^{JK}  +
\left( \frac{1}{4!} \C_{JKL} \C_{MNI} + \frac{1}{5!} \E_{JKLMNI}
       \right) m^{JKLMN} \\
 &+ \left( \frac{1}{3! 4!}  \C_{JKL} \C_{MNP} \C_{QRI}
+ \frac{1}{2 \cdot 5!} \C_{JKL} \E_{MNPQRI} \right) m^{J;KLMNPQR} 
\end{split}\\
\begin{split}
\tilde m^{IJ}  &= m^{IJ} + \frac{1}{3!} \C_{KLM} m^{KLMIJ} \\
&+ \left( \frac{1}{4!} \C_{KLM} \C_{NPQ} + \frac{1}{5!} \E_{KLMNPQ}  \right)
m^{K;LMNPQIJ}
\end{split}\\
\tilde m^{IJKLM}   &= m^{IJKLM}  + \frac{1}{2} \C_{NPQ}  m^{N;PQIJKLM}  \\
\tilde m^{I;JKLMNPQ}   &= m^{I;JKLMNPQ}\ .
\end{align}
\end{subequations}
The shifts induced by the expectation values of $\C_3$ and $\E_6$
give an explicit parametrization of the upper
triangular\footnote{$\mathcal{V}$ is actually upper triangular in
blocks, because we did not decompose the metric $g_{IJ}$ in
a product of upper triangular vielbeins.}
vielbein $\mathcal{V}$
in terms of the physical compactification parameters (see Eq. \eqref{zvq}).
The mass formula \eqref{mass}
is now invariant under T-duality, besides 
the manifest $Sl(d,\Zint)$ symmetry.

 As an illustration,
we can look at the shift in T-duality vector charge $  m^{s1}$ implied by 
the above equation:
\begin{multline}
\tilde m^{s1} + \A_1 \tilde{m}^2 =m^{s1}\\ 
              + \A_1 m^{2} 
              +\left( \C_3 + \A_1 B_2 \right) m^{s4} 
              +\left( \E_{s5} + \C_3 B_2 + \A_1 B_2 B_2 \right)
              m^{s;s6}\\
             +\left(  \A_1 \C_3  \right) m^{5} 
 +\left(  \E_6 + \C_3^3 +  \A_1 \E_{s5} + \A_1 B_2 \C_3 \right) m^{1;s6} \ .  
\end{multline}
The second line precisely involves the tensor product of the 
{\it charge} spinor representation $S$ with the spinor 
representation made up by the Ramond moduli.  
In fact, to see that the set $( \A_1$,$\C_3 + \A_1 B_2$,
$\E_{s5} + \C_3 B_2 + \A_1 B_2 B_2)$  
transforms as a spinor, one may simply note that
it is precisely the combination that appears in the expansion in powers
of $F$ of the T-duality invariant D-brane coupling 
$  \int e^{B+l_s^2 F} \wedge  \Ra$. Formula \eqref{mass}
reduces to the $d=5$ result of Ref. \cite{Dijkgraaf:1997hk}
for vanishing expectation values of the gauge backgrounds
(see also \cite{Sugawara:1997xh}).

\subsection[String multiplet and U-duality invariant tension formula]{String multiplet and U-duality invariant tension formula}
Exactly the same analysis can be done for the
momentum multiplet. We give here the result for $d=6$. 
The contributing charges $n^1, n^4, n^{1;6}$
decompose into $SO(6,6)$ T-duality multiplets 
\begin{equation}  I=(n^s) \ ,\quad
   S'=(n^1,n^{s3},n^{s;s5}) \ ,\quad
V=(n^4,n^{1;s5}) \ , 
\end{equation}
and we obtain the 
$E_{6(6)}(\Zint)$-invariant tension formula for   
the $d=6$ string multiplet:  
\begin{equation}
\label{tension}
\T^2 =    \left[ \frac{1}{l_p^6} \left( \tilde n^1 \right)^2 +
         \frac{1}{l_p^{12}} \left( \tilde n^{4} \right)^2 +
         \frac{1}{l_p^{18}} \left( \tilde n^{1;6} \right)^2 \right] \ ,   
\end{equation}
where the shifted charges are 
\begin{equation}
\label{cshft}
\begin{array}{lllllll}
\tilde n^1    &=& n^1 &+& \C_3 n^4 &+& 
                 \left(  \C_3 \C_3 + \E_6 \right) n^{1;6}  \\
\tilde n^4    &=& n^4 &+& \C_3 n^{1;6} 
                &&  \\
\tilde n^{1;6}&=& n^{1;6} &&  && \\
\end{array}
\end{equation}
The combinatorial factors and explicit index contractions are easily reinstated
in this equation by comparison with \eqref{chshft}.  
This yields the parametrization of the vielbein $\mathcal{V}$ 
of Eq. \eqref{zvq} in the representation appropriate to the
string multiplet.

\subsection{Application to $R^4$ couplings\label{rfour}} 
As an illustration of the result \eqref{tension}, we display the
$d\le 5$ string multiplet invariant tension formula. 
Because of antisymmetry, only the charges $n^1$ and $  n^4$ contribute, so
that the tension of the string multiplet is given by
\begin{multline}
\label{tso5}
\T^2 =  \frac{1}{l_p^6} \left( n^I + \frac{1}{3!} n^{IJKL} C_{JKL} \right)
g_{IM} \left( n^M + \frac{1}{3!} n^{MNPQ} C_{NPQ} \right)   \\
+ \frac{1}{4!~l_p^{12}} n^{IJKL} g_{IM} g_{JN} g_{KP} g_{LQ}
  n^{MNPQ} \ . 
\end{multline}
This is precisely the U-duality invariant quantity that was obtained in
the study of instanton corrections to $R^4$ corrections in type II
theories in Ref. \cite{Kiritsis:1997em}, where it was conjectured 
that the coupling for $d=5$ is given by the $SO(5,5,\Zint)$ Eisenstein series 
\begin{equation}
\label{thres}
A= \frac{V}{l_p^9} \hat{\sum}_{n^I,n^{IJKL}} \left[\T^2\right]^{-3/2} =
\frac{2\pi V}{l_p^9} \int_0^{\infty} \frac{{\rm d} t}{t^{5/2}} 
\hat{\sum}_{n^I,n^{IJKL}} e^{-\pi \T^2 /t} \ ,
\end{equation}
where $\T$ is given by the tension formula \eqref{tso5}, and $V$
denotes the volume of the M-theory torus $T^d$. 
As will become clear in the next Subsection (see Eq. \eqref{so55cond}),
the sum has to be restricted to integers such that
$n^{[I} n^{JKLM]}=0$, in order to pick up the contribution
of half-BPS states only. The generalization of this construction
to compactifications of M-theory to lower dimensions was addressed
in Ref. \cite{OP:progress}.

Under Poisson resummation on the charge $n^s$, the U-duality invariant
function \eqref{thres} 
exhibits a sum of instanton effects of order $e^{-1/g_s}$,
corresponding to the D0-branes (with charge $n^1$) and
D2-branes (with charge $  n^{s3}$), but there is also
a contribution of the extra charge $n^4$ superficially 
of order $e^{-1/g_s^2}$.
The NS5-brane does not yield any instanton on $T^4$,
so these
effects seem rather mysterious. On the other hand, we may
interpret Eq. \eqref{thres} as a sum of loops
from all perturbative and non-perturbative strings. The occurrence
of the NS5-brane of the string multiplet is then no longer surprising.
This soliton loop interpretation should, however, be
taken with care, since in any case we have not succeeded yet in 
recovering the one-loop $R^4$ coupling from the 
$SO(5,5,\Zint)$ Eisenstein series.

\subsection{Half-BPS conditions and Quarter-BPS states\label{bpscond}} 
The U-duality mass formulae \eqref{mass} and \eqref{tension} that we have
obtained only hold for 1/2-BPS states, and require particular
conditions on the various integer charges. These conditions can be 
obtained from a precise analysis of the BPS eigenvalue equation,
as in Subsection \ref{bpss}, or from a sequence of U-dualities
from the perturbative level-matching condition 
$\| m \|^2=0$ in Eq. \eqref{vectormass}. 
In analogy to the latter condition, they should be quadratic in the integer
charges, be moduli-independent, and constitute a
representation of the U-duality group $E_{d(d)}(\Zint)$, appearing
in the symmetric tensor product of two charge
multiplets. 

We have already noticed in Subsection \ref{ssmm} that 
the string multiplet
always appears in the symmetric product of two particle 
multiplets, and indeed all the computations in Appendix
\ref{bpsmass} point to the fact 
\footnote{The naive inclusion of the KK6-brane as an extra
 $\Gamma_{IJKLMN}  Z^{0IJKLMN}$ term does not seem, however, to yield a 
U-duality invariant mass formula by this method.} 
that {\it the 1/2-BPS condition 
on the particle multiplet is the string multiplet constructed
out of the particle charges}. This has also been observed
in Ref. \cite{Ferrara:1997ci}, where it was shown that for 
$d=7$ the 1/2-BPS conditions on the \irrep{56} particle multiplet
were transforming in a \irrep{133} adjoint
representation of $E_7$, which is the corresponding string multiplet. 

In order to extract the
precise conditions, it is convenient to consider the branching
under the ST-duality group: 
\begin{eqnarray}
E_{7(7)}&\supset& SO(6,6)\times Sl(2)\\
 \irrep{56}&=& (\irrep{12},\irrep{2})+(\irrep{32},\irrep{1})\nonumber\\
 \irrep{133}&=& (\irrep{1},\irrep{3}) + (\irrep{\bar{32}},\irrep{2})
+(\irrep{66},\irrep{1}),\nonumber
\end{eqnarray}
where the \irrep{32} correspond to the D-brane charges 
$m_s,m^2,m^{s4},m^{s;s6}$ and
the two \irrep{12}'s to the Kaluza--Klein and winding charges 
$m_1,m^{1s}$ and
the NS5-brane/KK5-brane charges $m^5,m^{1;s6}$ respectively
(see also Tables \ref{partbranch} and \ref{stringbranch}).   
The \irrep{133} in the symmetric tensor product $\irrep{56}\otimes_s
\irrep{56}$ of
two particle multiplets is therefore
\begin{equation}
\label{tensprod}
(\irrep{1}\in \irrep{12}\otimes_s \irrep{12},\irrep{3}) 
+ (\irrep{\bar{32}}\in \irrep{12}\otimes \irrep{32},\irrep{2}) + (
\irrep{66}\in
\irrep{32}\otimes_s \irrep{32}+\irrep{12}\wedge \irrep{12},\irrep{1})
\ . 
\end{equation}

So as to work out the tensor products in Eq. \eqref{tensprod},
it is advisable to consider the further branching
\begin{eqnarray}
SO(6,6)&\supset&Sl(6)\times O(1,1)\\
 \irrep{12}&=&\irrep{6}_{1}+\irrep{\bar{6}}_{-1}\nonumber\\
 \irrep{32}&=& \irrep{1}_3 + \irrep{\bar{15}}_1 + \irrep{15}_{-1} 
       + \irrep{1}_{-3} \nonumber\\
 \irrep{66}&=& \irrep{15}_2 + \irrep{1}_0 + \irrep{35}_{0} +
        \irrep{\bar{15}}_{-2} \ . \nonumber
\end{eqnarray}
The decomposition of the $\irrep{133}$ conditions in terms of the 
various $Sl(6)\subset SO(6,6)$ charges is therefore 
\begin{subequations}
\label{tcond}
\begin{eqnarray}
\irrep{1}_{2}&:& k^s \equiv m_1 m^{s1} \\
\label{spc1}
\irrep{32}_{1}&:&
\left\{ \begin{array}{lcl}
        k^1& \equiv &m_1 m^2 + m_s m^{s1} \\
        k^{3s}& \equiv &m_1 m^{s4} + m^{s1} m^2 \\
        k^{s;s5}& \equiv &m_1 m^{s;s6} + m^{s1} m^{s4} 
        \end{array}
\right. \\
\irrep{66}_0&:& 
\left\{ \begin{array}{lcl}
         k^4   &\equiv & m_s m^{s4} + m^2 m^2 + m_1 m^5 \\
         k^{1;s5}&\equiv & m^2 m^{s4} + m_s m^{s;s6} + m^{s1} m^5 + m_1 m^{1;s6} \\
         k^{s2;s6}&\equiv & m^2 m^{s;s6} + m^{s4} m^{s4} + m^{s1} m^{1;s6}
        \end{array}
\right. \\
\irrep{1}_0&:& k^{s;6}\equiv m^{s1} m^5 + m_s m^{s;s6} \\
\label{spc2}
\irrep{32}_{-1}&:& 
\left\{ \begin{array}{lcl}
         k^{1;6}&\equiv &m^5 m^2 + m^{1;s6} m_s \\ 
         k^{3;s6}&\equiv &m^5 m^{s4} + m^{1;s6} m^2 \\
         k^{s5;s6}&\equiv &m^5 m^{s;s6} + m^{1;s6} m^{s4} 
        \end{array}
\right. \\
\irrep{1}_{-2}&:& k^{6;s6}\equiv m^5 m^{1;s6}
\end{eqnarray}
\end{subequations}
where the subindex denotes the $SO(1,1)\subset Sl(2)$ charge,
and the contractions are the obvious ones.
In particular, for D-brane charges only, the condition $\irrep{66}_0$
reduces to the one introduced in Subsection \ref{dbtd}. The condition
$\irrep{1}_{2}$ is the familiar perturbative level-matching
condition, whereas $\irrep{1}_{-2}$ is the analogous condition
on NS5--KK5 bound states. The other conditions mix different
T-duality multiplets. For example, the spinor constraints 
\eqref{spc1} and \eqref{spc2} are composed of products of D-brane
charges with either KK- and winding charges or NS5- and KK5-brane
charges.  

As suggested by the index structure of the conditions $k$ in \eqref{tcond}, 
the constraints combine in a string or momentum multiplet
as\footnote{One could have alternatively derived these conditions
from the branching $E_7\supset Sl(7)$, but the one we used
is more constrained and more convenient.}
\begin{subequations}
\begin{eqnarray}
k^1 &=& m_1 m^2\\
k^4 &=& m_1 m^5 + m^2 m^2\\
k^{1;6} &=& m_1 m^{1;7} + m^2 m^5\\
k^{3;7} &=& m^2 m^{1;7} + m^5 m^5\\
k^{6;7} &=& m^5 m^{1;7}
\end{eqnarray}
\end{subequations}
If these composite charges do not vanish, the state
is at most 1/4-BPS, in which case its mass formula is given by
\begin{equation}
\label{14mass}
\M^2 = \mathcal{M}^2_0(m) + \sqrt{ \left[ \T(k) \right]^2 } \ ,  
\end{equation}
where $\M_0(m)$ and $\T(k)$ are given by the half-BPS
mass and tension formulae \eqref{mass} and \eqref{tension}. 

Noting from Eq. \eqref{partdec} that the string multiplet $\T_1$
appears in the decompactification of the particle multiplet $\M$,
we can obtain the half-BPS condition on the
string multiplet by allowing non-zero $m^{s1}, m^{s4},
m^{1;s6}$ charges only, where $s$ denotes a fixed direction on
the torus:
\begin{subequations}
\begin{eqnarray}
k^{s;s5} &=& m^{s1} m^{s4}\\
k^{s2;s6} &=& m^{s1} m^{1;s6} + m^{s4} m^{s4}\\
k^{s5;s6} &=& m^{s4} m^{1;s6}
\end{eqnarray}
\end{subequations}
and identifying these charges with the $n^1,n^4,n^{1;6}$ charges
of the string multiplet in one dimension lower.
We therefore obtain a multiplet of
half-BPS conditions
\begin{subequations}
\begin{eqnarray}
k^{5} &=& n^{1} n^{4} \label{so55cond}\\
k^{2;6} &=& n^{1} n^{1;6} + n^{4} n^{4}\\
k^{5;6} &=& n^{4} n^{1;6}
\end{eqnarray}
\end{subequations}
This is easily seen to transform as a $\T'_3$ multiplet, as can
also be inferred from the decomposition \eqref{stringdec} at level 2
of the
string multiplet under decompactification. For $d=6$, this is
a $\irrep{\bar{27}}$ quadratic condition on the $\irrep{\bar{27}}$ 
string multiplet of $E_6$, whereas for $d=5$ only the first
condition remains, giving a singlet condition on the $\irrep{10}$
multiplet of $SO(5,5)$. For $d<5$, a BPS string state is automatically
1/2-BPS, while for $d=7$ the $\T'_3$ condition transform as a
\irrep{1539} of $E_7$. The tension of a 1/4-BPS string can also be obtained
by decompactifying one direction in Eq. \eqref{14mass}, and has
an analogous structure
\begin{equation}
\label{14tension}
\T^2 = \mathcal{T}^2_0(n) + \sqrt{ \left[ \T_3' (k) \right]^2 } 
\end{equation}
where $\T_0(n)$ and $\T_3'(k)$ are given by the half-BPS
tension \eqref{tension} and the half-BPS 3-brane tension,
which can be worked out easily.

For $d\ge 6$ (resp $d\ge 5$),
there still remain conditions to be imposed on the particle multiplet
(resp. string multiplet)
in order for
the state to be 1/4-BPS and not 1/8. In the $d=7$ case,
it  should be required that the \irrep{56} in the third symmetric
tensor power of the \irrep{56} particle charges 
vanishes \cite{Ferrara:1997ci}. For $d=6$, this reduces
to the statement that the singlet in $\irrep{27}^3$ should
vanish. This condition is empty for $d\le 5$.
We shall however not investigate the 1/8-BPS case any
further, and refer to Appendix \ref{ns5kkw} for the 1/8-BPS
mass formula of a NS5--KK-winding bound
state in $d=6$ ($D=5$). In contrast to 1/2-BPS states, 1/4-BPS and 1/8-BPS
states in general have a non-trivial degeneracy and therefore entropy,
which still has to be a U-duality invariant quantity depending on the
charges $m$ \cite{Horowitz:1996ac,Kallosh:1996uy,Cvetic:1996zq,
Dijkgraaf:1997cv,Andrianopoli:1997hb}. 
This allows non-trivial checks on U-duality and
predictions on BPS bound states, which we shall only mention
here \cite{Vafa:1996zh,Vafa:1996bm,Sen:1996hb}.

\clearpage
\section{Matrix gauge theory \label{smgt} }


The definition of M-theory
as the strong-coupling limit of type IIA string theory
and the finite energy extension of the eleven-dimensional SUGRA
does not allow the systematic computation of   
S-matrix elements, since type IIA theory is only defined
through its perturbative expansion and 11D SUGRA is severely
non-renormalizable. In Ref.\ \cite{Banks:1997vh}, 
Banks, Fischler, Susskind and
Shenker (BFSS) formulated 
a proposal for a non-perturbative definition of
M-theory, in which M-theory in the infinite momentum frame (IMF) 
with IMF momentum $P=N/R$, 
is related to the  supersymmetric quantum mechanics 
\footnote{This model was first introduced
in Ref. \cite{Claudson:1985th,Flume:1985mn,Baake:1985ie}.} 
of $N\times N$ 
Hermitian matrices in the large-$N$ limit, the same as the one
describing the interactions of $N$ D0-branes induced by fluctuations
of open strings. 
Despite the powerful constraints of supersymmetry,
it is still a formidable problem to solve this 
quantum mechanics in the large-$N$ limit. 

As was argued by Susskind 
\cite{Susskind:1997cw}, 
sense can however be made of the finite-$N$
Matrix gauge theory, as describing the Discrete Light-Cone 
Quantization (DLCQ) of M-theory, that is quantization on a light-like
circle. This stronger conjecture has been
further motivated in Ref.\ \cite{Seiberg:1997ad},  
relating through an infinite Lorentz boost
the compactification of M-theory on a light-like circle to 
compactification on a vanishing space-like circle, \ie to type IIA
string theory in the presence of D0-branes. This argument gives
a general prescription for compactification of M-theory
(see also Sen's argument Ref. \cite{Sen:1997we}), and we shall 
briefly go through it in this Section.

Upon toroidal compactification on $T^d$, the extra degrees
of freedom brought in by the wrapping modes of the open strings
extend this quantum mechanics to a quantum field theory, namely
a $U(N)$ Yang--Mills theory with 16 supersymmetries on the T-dual torus
$\tilde{T}^d$ in the large-$N$ limit 
\cite{Ganor:1997zk,Taylor:1997ik}. This prescription is consistent
up to $d\le 3$, but breaks down for compactification on 
higher-dimensional tori, owing to the ill-definition of SYM theory at
short distances. Several  proposals have been made 
as to how to supplement the SYM theory with additional degrees of freedom
while still avoiding the coupling to gravity, which will be briefly
discussed in this section. Besides their relevance for M-theory
compactification, these theories are also interesting theories
in their own right, as non-trivial interacting field theories in higher
dimensions.

Our aim is to provide the background to discuss
in Section \ref{usymmat}  the implications of U-duality for the
Matrix gauge theory describing toroidal compactification of
M-theory. The relation between the M-theory
compactification moduli, including gauge backgrounds, and
Matrix gauge-theory parameters will be obtained, as well as the
spectrum of excitations that Matrix gauge theory should exhibit
in order to describe compactified M-theory. This will leave open
the issue of what is the correct Matrix gauge theory reproducing
these features.

\subsection{Discrete Light-Cone Quantization} 

The finite-$N$ conjecture of Ref.\ \cite{Susskind:1997cw} 
is formulated in the framework of the DLCQ, the essentials of which we
review first. 
In field theory, it is customary to use equal-time ($t=x^0$) quantization,
which breaks Poincar{\'e} invariance, but preserves invariance under
the kinematical generators
consisting of spatial rotations and translations. However, an alternative
quantization procedure exists, in which  the theory is quantized with
respect to the proper time $x^+ = (x^0 + x^1)/\sqrt{2}$, which is 
referred to as light-cone quantization. In this case, 
the transverse  translations $P^i$ and rotations $L^{ij}$,
as well as the longitudinal momentum $P^+$ and the boosts $L^{-i}$,
$L^{+-}$
do not depend on the dynamics, while the generator
$P^-$ generates the translations in  the $x^+$ direction and plays the role
of the Hamiltonian. 
The  usual dispersion relation $H =\sqrt{P^i P_i + \M^2}$  in equal-time 
quantization, is replaced in the light-cone quantization by 
\begin{equation}
\label{lcd}
P^- = \frac{P^i P_i +\M^2}{2 P^+} \ ,  
\end{equation}  
exhibiting Galilean invariance on the transverse space.
Particles, with positive energy $P^- > 0$, 
necessarily have positive longitudinal momentum $P^+$,
while antiparticles will have negative $P^+$. The vacuum of $P^-$ is
hence reduced to the Fock-space state $| 0 \rangle$, and the negative-norm
ghost states are
decoupled as well. This simplification of the theory is at the expense
of instantaneous non-local interactions due to the $P^+=0$ pole in
\eqref{lcd}. 

{\it Discrete} light-cone quantization proceeds by compactifying
the longitudinal direction $x^-$ on a circle of radius $R_l$: 
\begin{equation} 
x_- \simeq  x_- + 2 \pi R_l \ .  
\end{equation} 
This results into a quantization of the longitudinal momentum
of any particle $i$ according to
\begin{equation}
P^+_i = \frac{n_i}{R_l}\ .
\end{equation}
Because the total momentum is conserved, the Hilbert space decomposes
into finite-dimensional superselection sectors labelled by 
$N=\sum n_i$.
Note that the finite dimension does {\it not} require imposing
any ultraviolet cut-off on the eigenvalues $n_i$, but follows
from the condition $n_i>0$. 

It is important 
to note that, because the $x^-$ direction is a
light-like direction, the length $R_l$ of the radius is not invariant,
but can be modified by a Lorentz boost $L^{+-}$, 
\begin{subequations}
\label{lcb}
\begin{equation}
\left(  \begin{array}{c} x^0 \\ x^1 \end{array} \right)
 \ra 
\left(  
\begin{array}{cc} \cosh \beta  & - \sinh \beta  \\ - \sinh \beta & \cosh \beta 
 \end{array} 
\right)
\left(  \begin{array}{c} x^0 \\ x^1 \end{array} \right)
\ , 
\end{equation} 
\end{subequations}
which amounts to 
\begin{equation}
 R_l \ra e^\beta R_l \sp P^- \ra e^\beta P^- \sp
P^+ \ra e^{-\beta} P^+ \ .  
\end{equation} 
This implies that the Hamiltonian $P^+$ depends on the radius $R_l$
through an over-all factor
\begin{equation}
P^+ = R_l H_N \ , 
\label{psc}
\end{equation}  
so that the mass $\M^2 = 2 P^+ P^- $ is independent of $R_l$.  

\subsection{Why is Matrix theory correct ?\label{deriv} }  
Following Ref. \cite{Seiberg:1997ad}, 
we will now derive the Hamiltonian $H_N$ describing
the DLCQ of M-theory, and obtain the BFSS Matrix-theory conjecture.
The basic idea is to consider the 
compactification on the light-like circle as Lorentz-equivalent to
a limit of a compactification on a space-like circle.
Acting with a boost \eqref{lcb} on an ordinary space-like circle, 
we find 
\begin{equation}
\left(  
\begin{array}{cc} \cosh \beta  & - \sinh \beta  \\ - \sinh \beta & \cosh \beta 
 \end{array} 
\right)
\left(  \begin{array}{c} 0 \\ R_s  \end{array} \right)
= \frac{R_l}{\sqrt{2}}  
\left(  \begin{array}{c} -1 + e^{-2 \beta} \\ 1 + e^{2 \beta} 
 \end{array} \right)
\ra
 \frac{1}{\sqrt{2}}  
\left(  \begin{array}{c} -R_l \\ R_l  
 \end{array} \right)
\end{equation} 
where $R_s=R_l e^{-\beta}$. Sending $\beta\rightarrow\infty$ while 
keeping $R_l$ finite, we see that the light-like circle is
Lorentz-equivalent to a space-like circle of radius $R_s
\rightarrow 0$. 

In order to keep the energy finite, which 
from Eq. \eqref{psc} and on dimensional ground scales as
$R_l/l_p^2$, we should also rescale the Planck length
(and any other length) as $l_{p,s}  = e^{-\beta/2} l_p$.
Altogether, M-theory with Planck length $l_p$ 
on the light-like circle of radius
$R_l$ in the momentum $P^+ = \frac{N}{R_l}$ sector
is equivalent to M-theory with Planck length $l_{p,s}$
on the space-like circle of radius $R_s$ 
in the momentum $P=\frac{N}{R_s}$ sector, with
\begin{equation} 
\label{rrl}
R_s = R_l e^{-\beta} \sp l_{p,s}  = e^{-\beta/2} l_p     
\end{equation} 
in the limit $\beta\rightarrow\infty$. Eliminating $\beta$,
we obtain the following scaling limit:
\begin{equation}
\label{scl}
R_s \ra 0 \sp M = \frac{R_s}{l^2_{p,s}} = \frac{R_l}{l_p^2} = {\rm fixed}    
\ . 
\end{equation} 
Following Ref. \cite{Seiberg:1997ad}, we shall denote the latter theory
as $\tilde M$ theory.

Since the space-like circle $R_s$ shrinks to zero in $l_{p,s}$ units,
this relates the DLCQ of M-theory to weakly coupled type IIA string
theory in the presence of $N$ D0-branes carrying the momentum
along the vanishing compact dimension. 
Using Eq. \eqref{matching}, the scaling limit becomes
\begin{equation} 
g_s = (R_s M)^{3/4} 
\sp
\a' = l_s^2 = \frac{R_s^{1/2}}{M^{3/2}}
\sp R_s \ra 0
\sp M = \mbox{fixed}\ .
\end{equation} 
In particular, $g_s$ and $\a'$ go to zero, so that the bulk degrees
of freedom decouple, and only the leading-order Yang--Mills
interactions between D0-branes remain.
This validates the BFSS conjecture, up to the possible ambiguities
in the light-like limit $\beta\ra\infty$ 
\cite{Hellerman:1997yu,Bilal:1998ys}. Several difficulties
have also been shown to arise for compactification on curved
manifolds \cite{Douglas:1997pj,Douglas:1997uy}, 
but since we are only concerned with
toroidal compactifications, we will ignore these issues.

\subsection{Compactification and Matrix gauge theory}
For toroidal compactifications of M-theory, we consider the 
same scaling limit as in \eqref{scl}, and keep the torus
size constant in Planck length units, that is 
\begin{equation} 
\label{rsc}
R_I = r_I \left( \frac{R_s}{M} \right)^{1/2}   
  \sp r_I = \frac{R_I}{l_{p,s}} = {\rm fixed}  \ .  
\end{equation} 
However, comparing \eqref{scl} and \eqref{rsc}, we find
that the size of the torus goes to zero in the
scaling limit. To avoid this it is convenient to consider the theory on
the T-dual torus $\tilde{T}^d$, 
obtained by a maximal T-duality in all $d$ directions. From 
Eq. \eqref{tdua2a2b},
this has the effect that,    
\begin{equation}   
\label{2tt}
\mbox{IIA with}\;N\;\mbox{D0-branes}
\ra
\left\{
\begin{array}{ll}
\mbox{IIA with}\;N\;\mbox{D}d\mbox{-branes} & d={\rm even} \\
\mbox{IIB with}\;N\;\mbox{D}d\mbox{-branes} & d={\rm odd} \\
\end{array}
\right. 
\end{equation} 
Using the maximal T-duality transformation $\prod_{I=1}^d T_I$, with  
$T_I$ given in \eqref{tdua2a2b}, the type II parameters then become 
\begin{subequations}
\begin{equation}   
\label{ttp}
g_s = \frac{(R_s M)^{(3-d)/4}}{\prod r_I }  
\sp
\a' = l_s^2 = \frac{R_s^{1/2}}{M^{3/2}}
\sp
\tilde{R}_I = \frac{1}{r_I M} \ ,  
\end{equation} 
\begin{equation}
   R_s \ra 0 \sp M = \frac{R_s}{l_{p,s}^2} = {\rm fixed}
\sp  r_I = \frac{R_I}{l_{p,s}}  = {\rm fixed} \ ,  
\end{equation} 
\end{subequations}
so that, in particular,  the size of the dual torus is fixed 
in the scaling limit. We will sometimes refer to the type II theory in
this T-dual picture as the $\tilde{\rm II}$-theory.  

The behaviour of the string coupling in the scaling limit is now different
according to the dimension of the torus:
\begin{equation}   
g_s \ra 
\left\{
\begin{array}{ll}
0 & d < 3 \\ 
{\rm finite} & d = 3 \\ 
\infty & d > 3 \\ 
\end{array}
\right. 
\end{equation} 
In particular for $d<3$ we still have weakly coupled type IIA or IIB 
string theory in the presence of $N$ D$d$-branes, so that M-theory 
is described by the SYM theory with 16 supercharges living on the 
world-volume of the $N$ D$d$-branes.
The gauge coupling constant of this Matrix gauge theory
and the radii $s_I$ of the torus on which the D-branes are wrapped
read
\begin{equation}
\label{ymr}
g_{\rm YM}^2 = g_s l_s^{d-3} = \frac{M^{3-d}}{V_r} \sp V_r \equiv \prod_I r_I
\sp s_I = \tilde{R}_I = \frac{1}{r_I M} \, 
\end{equation} 
showing, in particular, that $g_{YM}^2$ is finite in the scaling limit. 

The special case of Matrix theory on a circle $(d=1)$ yields 
(after an S-duality transforming
the background D1-strings into  fundamental strings) Matrix string theory
\cite{Motl:1997th,Dijkgraaf:1997vv,Dijkgraaf:1997ku}, 
in which an identification
between the large-$N$
limit of two-dimensional N=8 supersymmetric YM theory and type IIA string
theory is established. We will not further discuss this topic here,
and refer to Ref. \cite{Fischler:1997kp,Sethi:1997sw} for the next
case $d=2$ and its relation to type IIB string theory.
Moving on to the case $d=3$, the same conclusion as in the $d <3$ case
continues to hold,  since although the string
coupling is finite, the string length goes to zero
so that loop corrections are suppressed in the 
$\a' \ra 0$ limit. Consequently, the $d=3$ Matrix gauge theory is N=4
supersymmetric Yang Mills theory.  

For $d >3$, however, the coupling $g_s$ blows up, and the weakly
coupled string description of the D-branes is no longer valid.
This coincides with the fact that the Yang--Mills theory
becomes non-renormalizable and strongly coupled in the UV.
Hence, in order to define a consistent quantum theory,  one needs 
to supplement the theory with additional degrees of freedom. 
In the following we briefly review the proposals for
$d=4$ and $d=5$, and show the complication
that arises for $d=6$. These proposals follow from the
above prescription, using further duality symmetries, which will
be examined in more detail in Section \ref{usymmat}. Other decoupling
limits have been considered in \cite{Hull:1998zp}.

\subsection{Matrix gauge theory on $ T^4$}

In the case $d=4$, it follows from \eqref{2tt} that the 
effective theory is 4+1 
SYM coming from the type IIA D4-brane world-volume theory. In the scaling
limit the type IIA theory becomes strongly coupled and using the
correspondence between strongly coupled IIA theory and M-theory 
a new eleventh dimension is generated,
which plays the role of a fifth  space
dimension in the gauge theory
\cite{Rozali:1997cb,Berkooz:1997cq,Berkooz:1997wq}.
Using Eqs. \eqref{msr} and \eqref{ttp},
the radius and 11D Planck length are
\begin{equation}   
\tilde{R} = g_s l_s = \frac{1}{M V_r}
\sp
\tilde{l}_p = g_s^{1/3} l_s = R_s^{1/6} M^{-5/6} V_r^{-1/3} \ .   
\end{equation} 
Moreover, comparing with \eqref{ymr} we find that
the radius $\tilde{R}$  is in fact equal
to the YM coupling constant
\begin{equation}
\label{ym4}
\tilde{R} = g_{\rm YM}^2 \ .  
\end{equation} 

Hence, in the scaling limit \eqref{scl}, the Planck length $\tilde{l}_p$ 
goes to zero so that the bulk degrees of
freedom decouple, while
the radius $\tilde{R}$ remains finite. The  
$N$ type IIA D4-branes become  $N$  M5-branes wrapped around the
extra radius $  \tilde{R}$, and 
M-theory on $  T^4 \times S^1$ is  then
described by the (2,0) world-volume theory of $N$ M5-branes, wrapped
on $T^4$ and the extra radius $\tilde{R}$, related to 
the Yang--Mills coupling constant by  Eq. \eqref{ym4}. The proper
formulation of this theory is still unclear, but Matrix light-cone
descriptions have been proposed in Refs. 
\cite{Brodie:1996ti,Seiberg:1997td,Losev:1997hx,Argurio:1997cu,Aharony:1998th,
Kachru:1998rk,Ganor:1998jx}
and the low-energy formulation studied in Refs. \cite{Ganor:1998ve,
Grojean:1998zt}. In particular, 
at energies of order $1/g_{\rm YM}^2$ the Kaluza--Klein states along the extra
circle come into play. They can be identified as instantons of 4D SYM
lifted as particles in the (4+1)-dimensional gauge theory. 
Additional evidence for this conjecture that follows
from the U-duality symmetry will be discussed in Section
\ref{usymmat}.

\subsection{Matrix gauge theory on $T^5$\label{mt5}}

In the case $d=5$, we have $N$ type IIB D5-branes at strong string
coupling, so that it is useful to perform an S-duality that maps
the D5-branes to NS5-branes. Using Eqs. \eqref{bsdua}) and
\eqref{ttp}, we find that 
the string coupling and length become 
\begin{equation}   
\hat{g}_s= \frac{1}{g_s} = (R_s M)^{1/2} V_r 
\sp 
\hat{l}_s^2 = g_s l_s^2 =   \frac{1}{M^2 V_r }    
 \sp
\hat{R}_I =\tilde{R}_I    \ .  
\end{equation} 
Moreover, comparing with Eq. \eqref{ymr}, we find that
the string tension is related to the gauge coupling constant by
\begin{equation}
\label{ym5}
\hat{l}_s^2 = g_{\rm YM}^2 \ .   
\end{equation} 

The string coupling $\hat{g}_s$ goes
to zero in the scaling limit, so that the bulk modes are decoupled from those
localized on the NS5-branes. However, the
string theory on the NS5-branes is still non-trivial, and has
a finite string tension in the scaling limit \cite{Berkooz:1997cq}. 
As a consequence, we find that M-theory on $T^5 \times S^1$ is
described by a theory of non-critical strings
propagating on the NS5-brane world-volume
with a tension related
to the gauge coupling by Eq. \eqref{ym5}. The proper formulation
of this theory is still unclear, but light-cone Matrix formulations
have been proposed \cite{Seiberg:1997td,Sethi:1997zz}.
The string
can be identified with a 4D Yang--Mills instanton
lifted to 1+5 dimensions.
This description is close but not identical to the
proposal in Refs. \cite{Dijkgraaf:1997cv,Dijkgraaf:1997hk}
according to which 
the (1/4-) BPS sector of M-theory should be described by the 
(1/2-) BPS excitations of the M5-brane, 
whose dynamics would be described
by a (ground-state) non-critical ``micro-string'' theory on its 
six-dimensional world-volume.
In particular, the theory on the type IIB NS5-brane is non-chiral,
whereas that on the M5-brane is chiral. We refer the reader to
the work of \cite{Dijkgraaf:1997ku} for a discussion of these two
approaches.


\subsection{Matrix gauge theory on $T^6$\label{han}} 

Finally, we discuss the problems that arise for $d=6$, in which case
we have $N$ type IIA D6-branes at strong coupling. As in the $d=4$
case, an eleventh dimension opens up, and we find M-theory
compactified on a circle of radius $\tilde R$ with
\begin{equation} 
\tilde{R} = g_s l_s = \frac{1}{R_s^{1/2} M^{3/2} V_r  }  
\sp 
\tilde{l}_p = g_s^{1/3} l_s =   \frac{1}{M V_r^{1/3} }  \ .  
\end{equation} 
The $N$ D6-branes actually correspond to $N$ coinciding Kaluza--Klein monopoles
with Taub--NUT direction along the eleventh direction, and as
$\tilde R\ra \infty$, the monopoles shrink to zero size and reduce
to an $A_N$ singularity in the eleven-dimensional metric.
It was suggested in Ref. \cite{Hanany:1997xc} that the bulk dynamics still 
decouples from the (6+1)-dimensional world-volume, and that
the latter can be described in the IMF by the m(atrix) quantum
mechanics of $N_1$ D0-branes inside $N$ ten-dimensional Kaluza--Klein
monopole, in the large-$N_1$ limit. This is very reminiscent to 
the BFSS description of M-theory, but the quantum mechanics is now 
a matrix model with eight supersymmetries and corresponds to the
Coulomb phase of the quiver gauge theory in 0+1 dimensions associated
to the Dynkin diagram $A_N$ \cite{Douglas:1996sw}. 
In other words, this is a sigma model
with vector multiplets in the adjoint representation of $[U(N_1)]^{\otimes N}$
and hypermultiplets in bifundamental representations $(N_1,\bar N_1)$
of $U(N_1)_k \times U(N_1)_{k+1}$ for $k=1\dots N$, with
$U(N_1)_k$ denoting the $k$-th copy of $U(N_1)$ and $U(N_1)_1$
identified with $U(N_1)_N$; this model is restricted to its
Coulomb phase, where the hypers have no expectation value.
In the low-energy limit, it is expected to reduce to
SYM in 1+6 dimensions, with gauge coupling
\begin{equation} 
\label{ym6}
\tilde{l}_p^3 = g_{\rm YM}^2  \ .  
\end{equation} 
Other approaches have been proposed in Refs.
\cite{Brunner:1997jx,Ganor:1997yu}. We shall come back to the $d=6$
case in Section \ref{usymmat} when we display 
the BPS states in terms of the low-energy SYM theory.

\subsection{Dictionary between M-theory and Matrix gauge theory}

We finally give the dictionary that allows us to go from
M-theory on $T^d \times S^1$ (with $S^1$ a light-like circle) in the
sector $P^+ = N/R_l$ and
Matrix gauge theory on $\tilde T^d$.
This can be obtained by solving \eqref{ttp} and \eqref{ymr}, for
the parameters $(s_I,N, g_{\rm YM})  $  
of the $  U(N)$ Matrix gauge theory in terms
of the parameters $ (R_I,R_l ,l_p)$ of  
 M-theory compactification on $  T^d \times S^1$:
\begin{subequations}
\label{dic}
\begin{equation} 
\label{duar}
   s_I = \frac{l_p^3 }{R_l R_I}  
\end{equation}  
\begin{equation} 
   g_{\rm YM}^2 =\frac{l_p^{3(d-2)} }{R_l^{d-3} \prod R_I} \ .  
\end{equation} 
\end{subequations}
For completeness we also give the inverse relations
\begin{equation} 
R_I  = \frac{1}{s_I} \left( \frac{R_l V_s}{g^2_{\rm YM}} \right)^{1/2} 
\sp l_p^3 = \left( \frac{R_l^3 V_s}{g^2_{\rm YM}} \right)^{1/2} 
\sp P^+ = \frac{N}{R_l} \ , 
\end{equation} 
where we have defined $V_s = \prod_I s_I$ as the volume of the dual torus
on which the Matrix gauge theory lives. 

\subsection{Comparison of M-theory and Matrix gauge theory SUSY \label{mvsmg} }
In order to describe the M-theory BPS states from the point of
view of the gauge theory, we need to understand how the space-time
supersymmetry translates to the brane world-volume. This is in complete
analogy with the perturbative string in the Ramond--Neveu--Schwarz
formalism, in which space-time supersymmetry emerges from 
world-sheet supersymmetry (see \cite{Green:1987sp}, Section 5.2),
and the case of the M5-brane has been thoroughly discussed in
Ref.\ \cite{Dijkgraaf:1997hk}. We will abstract their argument
and discuss the case of a general 1/2-BPS brane, whether D,
M, KK or otherwise, referring to that work for computational details.

In the presence of a $p$-brane, the breaking of the 11D $N=1$
space-time supersymmetry is only spontaneous. The unbroken SUSY 
charges generate a superalgebra on the world-volume of the
brane, whereas the broken ones generate fermionic zero modes.
The fixing of the reparametrization invariance on the world-volume
is most easily done in the light-cone gauge.
The 32-component supercharge $Q_\alpha$ then decomposes as 
a\footnote{or a sum of, depending on the parity of $p$.}
(spinor,spinor,spinor) of the unbroken Lorentz group
$SO(1,1)\times SO(p-1) \times SO(10-p)$. The algebra is
graded by the eigenvalue $\pm 1/2$ of the generator of
$SO(1,1)$, so that the {\it unbroken} generators $Q^{+}_{a\alpha}$ 
have charge $+1/2$ and the broken ones $Q^{-}_{a\alpha}$ charge
$-1/2$, where $a$ is the spinorial index of the  $SO(10-p)$ R-symmetry
and $\alpha$ the spinorial index of the $SO(p-1)$ Lorentz world-volume
symmetry. The anticommutation relations then take the form
\begin{subequations}
\begin{eqnarray}
\label{broksusy}
\{ Q^{+}, Q^{+}\} &=& H + P + Z^{++} \\
\{ Q^{+}, Q^{-}\} &=& p + Z^{0}  \\
\{ Q^{-}, Q^{-}\} &=& Z^{--} \ . 
\end{eqnarray}
\end{subequations}
In this expression, $H$ and $P$ are the world-volume Hamiltonian
and momentum, $Z^{++}$, $Z^{0}$, $Z^{--}$
 some possible central charges and  
$p$ is the transverse momentum.
A contraction of the central
charges with the appropriate Gamma matrices is also assumed. 
In the following, we
absorb the momentum in the charges
$Z^{++}$, and set $p=0$ by considering a particle at rest
in the transverse directions. 

The central charges 
$Z^0$ and $Z^{\pm\pm}$ are
simply a renaming of the $Z^{MN}$, $Z^{MNPQR}$ central charges
of the 11D superalgebra \eqref{susy1}. As their indices show, 
$Z^{0}$ is a singlet of $SO(1,1)$, whereas $Z^{++}$ and
$Z^{--}$ combine in a vector of $SO(1,1)$; $Z^{0}$ is therefore
identified with the $Z^{IJ}$, $Z^{IJKLM}$ charges, whereas
$Z^{\pm\pm}$ correspond to the $Z^{1I}$, $Z^{1IJKLM}$
charges, where as usual $I,J,\dots$, are directions on the
torus and $1$ is the space-time direction combined with the time direction
on the light cone. In other words, $Z^{0}$ is identified
with the {\it particle} charges, whereas $Z^{\pm\pm}$
correspond to the {\it string} charges. In order for the
superalgebra \eqref{broksusy} to reproduce the space-time
superalgebra \eqref{susy1} {\it with particle charges only},
we therefore need to impose $Z^{\pm\pm}=0$ on the
physical states. This is the analogue of the $L_0=\bar L_0$
level-matching condition.

The broken generators $Q^{-}$ and the central charges
$Z^0$ are given by the fermionic and bosonic zero modes only.
On the other hand, the unbroken generators as well
as the central charges $Z^{\pm\pm}$ have a non-zero-mode contribution:
\begin{equation}
Q^{+}=Q^{+}_0 + \hat Q^{+}\sp
Z^{\pm\pm}=Z_0^{\pm\pm}+\hat Z^{\pm\pm}\sp
H=H_0 + \hat H \ . 
\end{equation}
The zero-mode part of the generators 
$Q^{+}_0$ is built out of the bosonic and fermionic
zero modes $Z^0$ and $Q^{-}$, and anticommutes with the
oscillator part $\hat Q^{+}$. It generates the same algebra
as in Eq. \eqref{broksusy}, while the oscillator parts generate
the same algebra on their own and anticommute with the
zero-mode broken generators $Q^{-}=Q^{-}_0$. 
The level-matching
conditions $Z^{\pm\pm}=0$ are achieved through a cancellation
of the zero-mode part, quadratic in the particle charges 
$Z^0$, and the oscillator parts.

Let us now consider the Hamiltonian $H$. Because
of supersymmetry, both $H_0$ and $\hat H$ are positive
operators and for given zero modes $Z^0$, the
supersymmetric ground state is given by the condition $\hat H=0$,
or $\hat Q^+ |0\rangle = 0$. This state is therefore annihilated
by all the $\hat Q^+$ supersymmetries, that is half the space-time
supersymmetries, and must have vanishing $\hat Z^{\pm\pm}$ charge,
that is from the level-matching conditions
$Z_0^{\pm\pm} = (Z^0)^2 = 0$. This condition is, in less
detail, the 1/2-BPS condition $k=0$ with $k$
defined as in Eq. \eqref{bps2c}.
The energy of this state is given by the zero-mode part 
$H_0=\{Q^+_0,Q^+_0\}$ quadratic in the {\it particle}
charges $Z^0$. This is equivalent to the mass formula 
Eq. \eqref{bps2} for 1/2-BPS states in space-time.

On the other hand, BPS states preserving 1/4 of the space-time 
supersymmetry are only annihilated by {\it half} 
the world-volume supercharges $\hat Q^+$, and their energy
is shifted by the non-zero-mode contribution $\hat H$.
The latter is quadratic in the non-zero-mode part of the
{\it string} charges $\hat Z^{\pm\pm}$
$=-Z_0^{\pm\pm} = -(Z^0)^2$, therefore {\it quartic} in
the particle charge. This is precisely what was found in Eq.
\eqref{bps4}:
\begin{equation}
E = \mathcal{M}^2(Z) + \sqrt{ \left[ \T(K) \right]^2 } \ .  
\end{equation}
This equation has a simple interpretation: the quadratic
term corresponds to the 1/2-BPS bound state between the
heavy mass $\mathcal{M}_p$ $p$-brane and the mass $\mathcal{M}$
particle, with binding energy
\begin{equation}
\label{nzbs}
E = \sqrt{ \mathcal{M}_p^2 + \mathcal{M}^2 }- \mathcal{M}_p
\simeq \frac{\M^2}{ \M_p} \ ,  
\end{equation}
whereas the second corresponds to a 1/4-BPS bound state between the
$p$-brane and the mass $\M=R \T$ of the string with tension $\T$
wrapped on the circle $R$:
\begin{equation}
\label{zbs}
E = \left(\mathcal{M}_p+ \mathcal{M} \right)- \mathcal{M}_p
= R \T \ .  
\end{equation}
There is therefore a complete identity between
i) the space-time supersymmetry algebra and particle spectrum
{\it in the absence of the $p$-brane}, ii) the $p$-brane world-volume 
gauge theory and iii) the bound states of the $p$-brane with other
particles. This also holds at the level of space-time field configurations,
which can be seen as configurations on the world-volume
\cite{Callan:1997kz,Bergshoeff:1998bh}. 

\subsection{SYM masses from M-theory masses \label{symmass} }

We shall now explicit the correspondence of the previous subsection
in the D-brane case, relevant for
Matrix gauge theory, and relate the energies in the Yang--Mills 
theory to the masses in space-time. This has been discussed in
particular in Refs. \cite{Susskind:1996uh,Ganor:1997zk,Fischler:1997kp,Gopakumar:1997py}.
Based on the last interpretation
as bound states of the $N$ background D-branes with other particles,
we identify $R=R_l$ and $\M_p= P_+=N/R_l$, where $R_l$ is the radius
of the light-like direction, and fix the normalization of the 
Yang--Mills energies as
\begin{equation}
E_{\rm YM} = 
\frac{R_l}{N} \mathcal{M}^2(Z) + R_l \sqrt{ \left[ \T(K) \right]^2 } \ .  
\end{equation}
We then proceed by using the dictionary \eqref{dic} to obtain the Yang--Mills
energy of the BPS states we discussed previously.

We now apply these considerations to 
the highest-weight states of the
two U-duality multiplets of Subsections  
\ref{spfm} and \ref{ssmm}. 
The highest-weight
state of the particle multiplet is a    
Kaluza--Klein excitation on the $I$-th direction,
which becomes, after the maximal T-duality, 
an NS-winding state bound to the background D$d$-brane.
Hence, it is a bound state with non-zero binding energy,
and using Eq. \eqref{nzbs} we find
\begin{equation}
\label{efs}
E_{\rm YM} = \frac{(1/R_I)^2}{N/R_l}=
 \frac{g_{\rm YM}^2 s_I^2}{N V_s}      \ ,  
\end{equation} 
where in the second step we used the dictionary \eqref{dic} to translate to
Matrix gauge theory variables. This is the energy of a state in the
gauge theory carrying electric
flux in the $I$-th direction.  For this reason, the particle multiplet 
is also called the {\it flux multiplet}.

Next we turn to the highest-weight state of the string multiplet, 
wrapped on the light-like direction $R_l$.
The highest weight is a membrane wrapped on $R_I$ and $R_l$,  
which becomes, after the  
maximal T-duality,  a Kaluza--Klein state bound to the background 
D$d$-brane. These two states form a bound state at threshold and according
to \eqref{zbs} we have  
\begin{equation}
 \label{sts}
 E_{\rm YM}  = \frac{R_l R_I}{l_p^3} 
= \frac{1}{s_I} \ ,  
\end{equation} 
where Eq. \eqref{dic} was used again in the second step. This is the energy of 
a massless particle with momentum along the $I$-th direction
in the gauge theory, so that we may  alternatively 
call the string multiplet the {\it momentum multiplet} from the point of
view of Matrix gauge theory.  

This translation can be carried out for all other members
of the U-duality multiplets, and since U-duality preserves the 
supersymmetry properties of the bound state, one finds
the following general relation between SYM masses and M-theory masses:  
\begin{subequations}
\label{mar}
\begin{eqnarray}
\label{pfm}
\mbox{particle/flux multiplet} &\co& E_{\rm YM} = \frac{R_l}{N} \M  \ , \\
\label{smm}
\mbox{string/momentum multiplet} &\co&  E_{\rm YM} =  R_l \T_1 \ .  
\end{eqnarray} 
\end{subequations}  
In Subsection \ref{udmmg}, we will explicitly see for the cases $d=3,4,5$
that indeed all non-zero binding energy and threshold bound states
appear in the particle/flux and string/momentum
multiplets respectively.  
Finally, we remark that the 
equalities in the two equations \eqref{efs} and \eqref{sts} can be solved to 
yield the dictionary \eqref{dic}, so that the comparison of these 
two types of energy quanta gives a convenient short-cut to \eqref{dic}.

\clearpage
\section{U-duality symmetry of Matrix gauge theory \label{usymmat}}

If any of the previously discussed Matrix gauge theories
purports to describe compactified Matrix gauge theory, it
should certainly exhibit U-duality invariance. In this section,
we wish to investigate the implications of U-duality on 
the Matrix gauge theory at the algebraic level, irrespective
of its precise realization.

To this end we use the dictionary \eqref{dic} between compactified M-theory
and Matrix gauge theory. We first
recast the Weyl transformations of the U-duality
group (see Subsection \ref{ssweyl}) in the gauge-theory language and interpret
them as generalized electric--magnetic dualities of the gauge theory. 
Then, we
translate the U-duality multiplets of Subsections \ref{spfm} and \ref{ssmm}   
in Matrix gauge theory
and discuss the interpretation of the states. Finally, we use the results
of Subsection \ref{uduaspfl} to discuss the realization of the
full U-duality group in Matrix gauge theory
and in particular the couplings induced by non-vanishing gauge potentials. 

At the end of this section a more speculative aspect of finite-$N$ matrix
gauge theory is discussed. By promoting the 
rank $N$ to an ordinary charge, we show 
the existence of an $E_{d+1(d+1)}(\Zint)$ action on the spectrum of BPS states.
In this way, we find that the conjectured extended U-duality symmetry of
matrix theory on $T^d$ in DLCQ has an
implementation as action of $E_{d+1(d+1)}(\Zint)$
on the BPS spectrum, as demanded by
eleven-dimensional Lorentz invariance.

\subsection{Weyl transformations in Matrix gauge theory} 

The discussion of Matrix gauge theory from M-theory in Section \ref{smgt} 
has been restricted to rectangular tori with vanishing gauge
potentials, 
so that we first focus on 
the transformations in the Weyl subgroup of the U-duality
group
\footnote{We restrict to the case $d \geq 3$; the case $d=1$ has trivial
Weyl group, while for the case $d=2$ there is only the permutation symmetry
$\mathcal{S}_2$.}  
\begin{equation}
\mathcal{W} (E_{d(d)} (\Zint)) = \Zint_2 \bowtie {\mathcal S}_d \ . 
\end{equation} 
The permutation group $S_d$ that interchanges the radii $R_I$ of the M-theory
torus obviously still permutes the radii $s_I$ of the Matrix gauge
theory T-dual torus.
On the other hand, the generalized T-duality $T_{IJK}$ in \eqref{mthz2},
using the  dictionary \eqref{dic}, translates into the
following transformation of the Matrix gauge theory parameters:
\begin{equation}
\label{sdua}
S_{IJK} :\;\;\;
\left\{
\begin{array}{lcll} 
g_{\rm YM}^2  & \ra & \frac{g_{\rm YM}^{2(d-4)}}{W^{d-5}} & 
W \equiv \prod_{a \neq I,J,K} s_a \\  
s_{\alpha}   & \ra & s_{\alpha}  &  \alpha = I,J,K \\
s_a  & \ra & \frac{g_{\rm YM}^2 }{W} s_a &  a \neq I,J,K \\
 \end{array}
\right. 
\end{equation} 

For $d=3$ the transformation \eqref{sdua}
 is precisely the (Weyl subgroup of) S-duality symmetry of 
$N=4$ SYM in 3+1 dimensions \cite{Susskind:1996uh,Ganor:1997zk}:
\begin{equation}
  g_{\rm YM}^2 \ra 1/g_{\rm YM}^2 \ , 
\end{equation} 
obtained for zero theta angle. 
The transformation \eqref{sdua} generalizes this symmetry to the
case $d>3$, by acting as S-duality in the (3+1)-dimensional
theory obtained by reducing the Matrix gauge theory in $d+1$ dimensions
to the directions $I,J,K$ and the time only \cite{Elitzur:1997zn}. Indeed, 
the coupling constant for the effective (3+1)-dimensional
gauge theory reads
\begin{equation}
\frac{1}{g_{\rm eff}^2}= \frac{W}{g_{\rm YM}^2} \ ,  
\end{equation} 
and the transformation \eqref{sdua} becomes
\begin{equation}
(g_{\rm eff}^2, s_\alpha, s_a ) \ra 
(1/g_{\rm eff}^2, s_\alpha, g_{\rm eff}^2 s_a ) \ .  
\end{equation} 
To summarize, we see that from the point of view of the 
Matrix gauge theory the U-dualities are accounted for by 
the modular group of the torus on which the gauge theory
lives (yielding the $Sl(d,\Zint)$ subgroup) 
as well as by generalized electric--magnetic dualities
(implementing the T-dualities of type IIA string)
\footnote{From the point of view of type IIB theory, it can be
shown that the latter also account for the restoration
of the transverse Lorentz invariance \cite{Sethi:1997sw}.}.

We now discuss in more detail the $d=4,5,6$ cases, in order
to give more support to the proposals discussed in Section
\ref{smgt}. Explicitly, one obtains
\begin{subequations} 
\begin{equation}
\label{sdua4}
d=4: \;\;S_{IJK} \;\;\;\; 
\left\{
\begin{array}{lcll} 
g_{\rm YM}^2  & \leftrightarrow & s_a  & a \neq I,J,K \\  
s_{\alpha}   & \ra & s_{\alpha}  &  \alpha = I,J,K \\
 \end{array}
\right. 
\end{equation} 
\begin{equation}
\label{sdua5}
d=5: \;\;S_{IJK} \;\;\;\; 
\left\{
\begin{array}{lcll} 
g_{\rm YM}^2  & \rightarrow & g_{\rm YM}^2   &  \\  
s_{a}   & \ra & \frac{g_{\rm YM}^2}{s_{b}}  &  a ,b  \neq I,J,K \\
s_{\alpha}   & \ra & s_{\alpha}  &  \alpha = I,J,K \\
 \end{array}
\right. 
\end{equation} 
\begin{equation}
\label{sdua6}
d=6: \;\;S_{IJK} \;\;\;\; 
\left\{
\begin{array}{lcll} 
g_{\rm YM}^2  & \rightarrow & \frac{g_{\rm YM}^4}{s_a s_b s_c}    & a,b,c \neq I,J,K \\ 
s_{a}   & \ra & \frac{g_{\rm YM}^2}{s_{b} s_c }  &   \\
s_{\alpha}   & \ra & s_{\alpha}  &  \alpha = I,J,K \\
 \end{array}
\right. 
\end{equation} 
\end{subequations} 
For $d=4$ we see that \eqref{sdua4} induces a permutation of the YM coupling
constant with the radii, in accordance with the interpretation \eqref{ym4}
of the YM coupling constant as an extra radius. 
For $d=5$,  Eq. \eqref{sdua5} takes the form of
a T-duality symmetry \eqref{tid} of the non-critical
string theory living on the type IIB NS5-brane world-volume
with the YM coupling related to the string length
as in \eqref{ym5}. Finally, for $d=6$, we see by comparing \eqref{sdua6} with 
the U-duality transformation in \eqref{mthz2} that we recover the symmetry 
transformation $T_{IJK}$ in M-theory with the YM coupling constant related to the
Planck length by \eqref{ym6}. 

\begin{table}[h] 
\begin{center}
\begin{tabular}{|r|r||l|l|}
\hline $D$ & $d$  & U-duality & origin  \\ \hline
8 &3 & $  Sl(3,\Zint)\times Sl(2,\Zint)$ & S-duality $\times$ symmetry of $T^3$ \\
7 & 4 & $  Sl(5,\Zint)$ & symmetry of $T^5$ of M5-brane  \\
6 & 5 & $  SO(5,5,\Zint)$ & T-duality symmetry on NS5-brane\\
$D\leq 5$&$d \geq 6 $ & $E_{d(d)}(\Zint) $ & unclear \\ \hline    
\end{tabular}
\end{center}
\caption{Interpretation of U-duality in Matrix gauge theory.
\label{tum}} 
\end{table}

At this point, it is also instructive to recall 
the full U-duality groups for toroidal compactifications of M-theory,
as summarized
in Table \ref{tud}, and discuss their interpretation in view of the
Matrix gauge 
theories for $d=3,4,5$ (see Table \ref{tum}). 
For $d=3$, the $Sl(3,\Zint) \times Sl(2,\Zint)$ U-duality symmetry
is the product of the (full) S-duality and the reparametrization group of
the three-torus. For $d=4$, the $Sl(5,\Zint)$ symmetry is the modular
group of the five-torus, corroborating the interpretation of this
case as  the (2,0) theory on the M5-brane \cite{Rozali:1997cb}.
Finally, for $d=5$ the $SO(5,5,\Zint)$ symmetry should be interpreted as the
T-duality symmetry of the string theory living on the NS5-brane
\cite{Dijkgraaf:1997hk,Berkooz:1997cq}.  
The $E_{6(6)}(\Zint)$ symmetry is by no means obvious in the 
IMF description discussed in Subsection \ref{han}, but this
is expected since part of it are Lorentz transformations broken
by the IMF quantization. 
The interpretation of the exceptional groups $E_{d(d)} (\Zint)$, $d=7,8$ is 
not obvious either, since
 a consistent quantum description for these cases is lacking as well.

In Subsections \ref{invmass3}--\ref{invmass5}, 
the precise identification of the full U-duality groups for $d=3,4,5$ 
will be discussed in further detail.
 Note also that as we are considering M-theory compactified on
a torus times a light-like circle, it has been conjectured that the
$E_{d(d)} (\Zint)$ U-duality symmetry should be extended to $E_{d+1(d+1)}(\Zint)$,
as a consequence of Lorentz invariance. This extended U-duality
symmetry will be discussed in Subsection \ref{uduaext}.

Finally, we can translate the U-duality invariant Newton constant
\eqref{uduainv} in the
Matrix gauge theory language. The most convenient form is obtained by
writing  
\begin{equation}
\label{udsmg}
I_d = \frac{V_s^{d-5}}{g_{\rm YM}^{2(d-3)}}= \left( \frac{V_R}{l_p^9} \right)^2 R_l^{9-d}  \ ,  
\end{equation} 
which depends on the invariant $D$-dimensional Planck length and the 
radius of the light-like circle, invariant under the $E_{d(d)} (\Zint)$
transformations acting on the transverse space. 
Again, in agreement with the Matrix gauge theory
descriptions, we see that for $d=3$
the invariant $I_3= 1/V_s^{2}$ is related to the volume $V_s$ of the 
three-torus;
for $d=4$ the invariant $I_4 = 1/(V_s g_{\rm YM}^2)$ is related to
the total volume 
of the five-torus, constructed from the four-torus 
and the extra radius $\tilde{R} = g_{\rm YM}^2$; 
for $d=5$ the invariant $I_5 = 1/g_{\rm YM}^4$ is related to
the finite string tension
$T = 1/g_{\rm YM}^2$ of the string theory. Finally, note also that for $d=6$ 
the U-duality invariant $I_6 = V_s/g_{\rm YM}^6$ is
related to the 5-D Planck length, when using $l_p^3 = g_{\rm YM}^2$.   

\subsection{U-duality multiplets of Matrix gauge theory \label{udmmg} }

We now turn to the translation of the U-duality multiplets of Subsections
 \ref{spfm} and \ref{ssmm}  
in the Matrix gauge theory picture. To this end we use the 
dictionary \eqref{dic} and the mass relations in Eq. \eqref{mar}. 
Equivalently,
one may start with the highest-weight states corresponding to
electric flux \eqref{efs} and momentum states \eqref{sts}
in 
the Matrix gauge theory and subsequently act with the transformations
\eqref{sdua} of the Weyl subgroup. Of course these two methods lead to the same
result, which are summarized in Tables \ref{tfmg} and \ref{tmmg}, 
for the particle/flux and
string/momentum multiplet respectively. As a compromise between
explicitness and complexity, we have chosen
to write down
the content for $d=7$ in the first case, and for $d=6$ in the latter case.  
The tables 
list the mass $\M$ in M-theory variables, the corresponding gauge theory
energy $E_{\rm YM}$ and their associated charges, obtained
from the M-theory charges by raising lower indices or lowering 
upper indices. 

\begin{table}[h] 
\begin{center} 
\begin{tabular}{|c||c|l|} \hline 
$\M $   & $E_{\rm YM}$  & charge  \\ \hline
$\frac{1}{R_I}$   &  $\frac{g^2_{\rm YM} s_I^2 }{N V_s}$    &  $m^1$  \\
$\frac{R_I R_J}{l_p^3}$ &  
$\frac{V_s}{ N g^2_{\rm YM} (s_I s_J)^2 }$    &  $m_2 $  \\
$    \frac{R_I R_J R_K R_L R_M}{l_p^6}$ & 
$\frac{V_s^3}{ N g^6_{\rm YM} (s_I s_J s_K s_L s_M  )^2 }$    &  $m_5 $  \\
$\frac{R_I; R_J R_K R_L R_M R_N R_P R_Q }{l_p^9}$ & 
$\frac{V_s^5}{ N g^{10}_{\rm YM} (s_I; s_J s_K s_L s_M s_N s_P s_Q  )^2 }$  &  
$m_{1;7}$   \\
\hline
\end{tabular}
\end{center}
\caption[Flux multiplet for Matrix gauge theory on
$T^7$]{Flux multiplet (\irrep{56} of $E_7$) for Matrix gauge theory
on $T^7$. 
\label{tfmg}}
\end{table}

\begin{table}[h] 
\begin{center}
\begin{tabular}{|c||c|l|} \hline 
$\M $   & $E_{\rm YM}$  & charge  \\ \hline
$\frac{R_l R_I }{l_p^3}$   &  $\frac{1}{s_I}$    &  $n_1$  \\
$\frac{R_l R_I R_J R_K R_L }{l_p^6}$ &  
$\frac{V_s}{ g^2_{\rm YM} s_I s_J s_K s_L  }$    &  $n_4  $  \\
$    \frac{R_l R_I; R_J R_K R_L R_M R_N R_P}{l_p^9}$ & 
$\frac{V_s^2}{  g^4_{\rm YM} s_I; s_J s_K s_L s_M s_N s_P  }$    &  $n_{1;6}   $  \\
$    \frac{R_l R_I R_J R_K; R_L R_M R_N R_P R_Q R_R R_S}{l_p^{12}}$ & 
$\frac{V_s^3}{  g^6_{\rm YM} 
 s_I s_J s_K; s_L s_M s_N s_P s_Q s_R s_S}$    &  $n_{3;7}   $  \\
$    \frac{R_l R_I R_J R_K R_L R_M R_N ; R_P R_Q R_R R_S R_T R_U R_V }{l_p^{15}}$ & 
$\frac{V_s^4}{  g^8_{\rm YM} 
 s_I s_J s_K s_L s_M s_N;  s_P s_Q s_R s_S s_T s_U s_V }$    &  $n_{6;7}   $  \\
\hline
\end{tabular}
\end{center} 
\caption[Momentum multiplet for Matrix gauge theory on $T^7$.]
{Momentum multiplet (\irrep{133} of $E_7$) for Matrix gauge theory
on $T^7$.
\label{tmmg}}  
\end{table}

In Table \ref{tfmg}, the first entry corresponds to a state with electric flux
 in the $  I$-th direction, while the second one 
carries magnetic flux in the $I,J$ direction. The first entry in Table 
\ref{tmmg} is a KK state of the gauge theory, while the second one 
is a YM instanton in 3+1 dimensions,  
lifted to $  d+1$ dimensions.
For $d \geq 5$, new states appear. 
As a further illustration,
we take a closer look 
at the special cases $d=3,4,5,6$, which can be obtained from the tables
by omitting those states that have too many compactified dimensions.
The Tables \ref{tfm3}--\ref{tfm6} list  
the content 
of each of the two multiplets for these 
cases \cite{Elitzur:1997zn,Obers:1997kk,Ryang:1998sz}, 
including the M-theory mass,
the YM energy, the multiplicity of each type of state and 
its interpretation both in the Matrix gauge theory and as a bound
state with the $N$ background type $\tilde{\rm II}$ D$d$-branes. 
For $d=4,5$ we have also added a column
giving the bound-state interpretation
in the M5- and NS5-brane theories
respectively.   

\begin{table}[ht] 
\begin{center} 
\begin{tabular}{|c||c|l|l||l|} \hline  
$\M$ & $E_{\rm YM}$ & \#   & YM state  & b.s. of $N$ D3 \\ \hline \hline
{\large $\frac{1}{R_I}$} &  
{\large $  \frac {g^2_{\rm YM} s_I^2 }{N V_s}$} &  3  & electric flux
& NS-w \\   
{\large $\frac{R_I R_J}{l_p^3}$} &  
 {\large $   \frac {V_s}{N g^2_{\rm YM} (s_I s_J)^2 }$} &  3  & magnetic flux & D1\\
\hline \hline  
{\large $\frac{R_l R_I}{l_p^3}$ } &
{\large $   \frac{1}{s_I} $}  &  3  & momentum  & KK \\     
\hline  
\end{tabular} 
\end{center}
\caption[Flux and momentum multiplet for $d=3$]
{Flux and momentum multiplet for $d=3$: \irrep{(3,2)} and \irrep{(3,1)}
of $Sl(3) \times Sl(2)$. 
\label{tfm3}}
\end{table} 

\begin{table}[h] 
\begin{center} 
\begin{tabular}{|c||c|l|l||l||l|} \hline  
$\M$ & $E_{\rm YM}$ & \#   & YM  state  & b.s. of $N$ D4 & b.s. of $N$ M5
\\ 
\hline \hline
{\large $\frac{1}{R_I}$} &  
 {\large $   \frac {g^2_{\rm YM} s_I^2 }{N V_s}$} &  4  & electric flux & NS-w 
& \multirow{2}{2cm}{M2}\\  
{\large $\frac{R_I R_J}{l_p^3}$} &  
{\large $   \frac {V_s}{N g^2_{\rm YM} (s_I s_J)^2 }$} &  6  & magnetic flux & D2
 & \\ 
\hline \hline  
{\large $\frac{R_l R_I}{l_p^3}$ } &
{\large $   \frac{1}{s_I} $}  &  4  & momentum   &  KK & 
\multirow{2}{2cm}{KK}\\   
{\large $\frac{R_l V_R}{l_p^6}$} &
 {\large $   \frac{1}{g^2_{\rm YM}} $}  &  1  & YM particle  & D0 &  \\     
\hline  
\end{tabular} 
\end{center}
\caption[Flux and momentum multiplet for $d=4$]
{Flux and momentum multiplet for $d=4$: \irrep{10} and
  \irrep{5}  of $Sl(5)$.
\label{tfm4}}
\end{table} 

\begin{table}[h]  
\begin{center} 
\begin{tabular}{|c||c|l|l||l||l| }\hline  
$\M$ & $E_{\rm YM}$ & \#   & YM state  & b.s. of $N$ D5 & b.s. of $N$ NS5
\\ 
\hline \hline
{\large $\frac{1}{R_I}$} &  
{\large $   \frac {g^2_{\rm YM} s_I^2 }{N V_s}$} &  5  & electric flux & NS-w
 & D1  \\   
{\large $\frac{R_I R_J}{l_p^3}$} &  
{\large $   \frac {V_s}{N g^2_{\rm YM} (s_I s_J)^2 }$} &  10  & magnetic flux & D3 &
D3\\  
{\large $\frac{V_R}{l_p^6}$} &  
{\large $   \frac {V_s}{N g^6_{\rm YM}}$} &  1  & new sector & NS5 & D5 \\  
\hline \hline  
{\large $\frac{R_l R_I}{l_p^3}$ } &
{\large $   \frac{1}{s_I} $}  &  5  & momentum  & KK   & KK \\   
{\large $\frac{R_l V_R}{R_I l_p^6}$} &
{\large $   \frac{s_I}{g^2_{\rm YM}} $}  &  5  & YM string   & D1    & NS-w  \\   
\hline  
\end{tabular} 
\end{center}
\caption[Flux and momentum multiplet for $d=5$]{Flux and momentum  multiplet for $d=5$: \irrep{16} and 
\irrep{10} of $SO(5,5)$.
\label{tfm5}}
\end{table}  

\begin{table}[H] 
\begin{center}
\begin{tabular}{|c||c|l|l||l|} \hline  
$\M$ & $E_{\rm YM}$ & \#   & YM state  & b.s. of $N$ D6 \\ \hline \hline
{\large $\frac{1}{R_I}$} &  
{\large $   \frac {g^2_{\rm YM} s_I^2 }{N V_s}$} &  6  & electric flux & NS-w
\\    
{\large $\frac{R_I R_J}{l_p^3}$} &  
{\large $   \frac {V_s}{N g^2_{\rm YM} (s_I s_J)^2 }$} &  15  & magnetic flux & D4\\  
{\large $\frac{V_R}{R_I l_p^6}$} &  
{\large $   \frac {V_s s_I^2}{N g^6_{\rm YM}}$} &  6  & new sector & KK5 \\ \hline  
\hline 
{\large $\frac{R_l R_I}{l_p^3}$ } &
{\large $   \frac{1}{s_I} $}  & 6  & momentum   & KK \\   
{\large $\frac{R_l V_R}{R_I R_J l_p^6}$} &
{\large $   \frac{s_I s_J}{g^2_{\rm YM}} $}  &  15  & YM membrane      & D2  \\     
{\large $\frac{R_l R_I V_R}{ l_p^9}$} &
{\large $   \frac{V_s }{g^4_{\rm YM} s_I} $}  &  6 & new sector     & NS5   \\    
\hline  
\end{tabular}
\caption[Flux and momentum multiplet for $d=6$]{Flux and momentum multiplet for $d=6$: \irrep{27} and \irrep{\bar{27}} of $E_6$.
\label{tfm6} }
\end{center}
\end{table}

A few comments on these tables are in order.    
\begin{itemize}
\item A number of states in the Matrix gauge theory
have a uniformly valid interpretation as bound states
with the background D$d$-branes, namely, for the flux multiplet,
\begin{subequations}
\label{flbs}
\begin{equation}
\mbox{electric flux} = {\rm D}d\mbox{--NS-winding} \;\,\mbox{bound state}
\end{equation}
\begin{equation}
\mbox{magnetic flux} = {\rm D}d\mbox{--D}(d-2) \;\,\mbox{bound state}
\end{equation}
\end{subequations}
and for the momentum multiplet,
\begin{subequations}
\label{mobs}
\begin{equation}
\mbox{KK momentum} = {\rm D}d\mbox{--KK}\;\,\mbox{bound state}
\end{equation}
\begin{equation}
\mbox{YM state} = {\rm D}d\mbox{--D}(d-4) \;\,\mbox{bound state}
\end{equation}
\end{subequations} 
where the YM state denote the 4D Yang--Mills instanton lifted
to $d+1$ dimensions. The correspondences in Eq. \eqref{flbs}
and \eqref{mobs} were noted in Refs. \cite{Papadopoulos:1997ca,
Douglas:1995bn,Witten:1996im}.
\item 
In the $d=3$ case, only perturbative states are observed in Table \ref{tfm3}. 
\item
For $d=4$ one non-perturbative state occurs in Table \ref{tfm4}, 
which corresponds 
precisely to momentum along the dynamically generated fifth direction,
\ie to a Yang--Mills instanton lifted to 4+1 dimensions
\cite{Rozali:1997cb}. 
{}From the M5-brane point of view, 
the flux multiplet describes the M2-brane excitations, 
while the momentum multiplet comprises the KK states, as indicated in
the last column. 
\item
For the case $d=5$ in Table \ref{tfm5}, we focus on the last column obtained by
S-duality from the D5-brane picture of the $\tilde{\rm II}$ theory.
The YM string in the momentum 
multiplet arises in this case from the wound strings on the NS5-brane.
The wrapped transverse fivebrane on $T^5$ appears as a 
bound state of D5-branes with the background NS5-branes,
with non-zero binding energy (since it is related by
electric--magnetic duality to the D1--NS1 bound state). 
It corresponds to a new sector in the Matrix gauge theory
Hilbert space, with energy scaling as $1/g^6_{\rm YM}$.
This state does not correspond to any known configuration
of the 1+5 gauge theory, but may be understood as a magnetic
flux along one ordinary dimension together with the dynamically
generated dimension in a 1+4 gauge theory obtained by
reducing the original one on a circle \cite{Halyo:1997dp}.

\end{itemize}
For the $d=6$ case, we see from 
Table \ref{tfm6} that all BPS states of type IIA
theory on $T^6$ are involved in the bound states of the flux and
momentum multiplet, except for the D6--D0 ``bound state''. It has been 
argued  that the latter forms a non-supersymmetric
resonance with the unconventional mass relation \cite{Dhar:1998ip,
Taylor:1997ay}: 
\begin{equation}
\label{dp6}
\M = \left(\M_{\rm D6}^{2/3} + \M_{\rm D0}^{2/3}\right)^{3/2} \ .
\end{equation}
As a consequence we expect to find a state in the gauge theory with
energy
\begin{equation}
E_{\rm YM} = 
\left(\M_{N{\rm D6}}^{2/3} + \M_{\rm D0}^{2/3}\right)^{3/2} 
- \M_{N{\rm D6}}
\simeq  
\M_{\rm D0}^{2/3} \M_{N{\rm D6}}^{1/3} \ .  
\end{equation} 
Using the corresponding D-brane masses and the relation $g_{\rm YM}^2 = g_s l_s^3$
we then obtain
\begin{equation}
\label{d06}
E_{\rm YM} = N^{1/3} \frac{V_s^{1/3} }{g_{\rm YM}^2}   = N^{1/3} I_6^{1/3} \ .  
\end{equation} 
In the last step we have expressed the mass in terms of the
U-duality invariant \eqref{udsmg}, explicitly showing that this extra
state transforms  as a singlet
under the U-duality group $E_{6(6)}(\Zint)$. 
Since the D0-brane is mapped onto a D6-brane under the maximal T-duality,
the space-time interpretation of this
extra U-duality multiplet follows from the M-theory origin of the D6-brane,
\ie the state is KK6-brane with the TN direction along 
the light-like direction. The corresponding data of this extra singlet
are summarized in Table \ref{exm6}. 
The $d=7$ case is discussed in 
Appendix \ref{mgt7}, and exhibits a 
 number of similar states (with M-theory masses depending on
multiple factors of $R_l$)  as the extra singlet in $d=6$.  

\begin{table}[h] 
\begin{center}
\begin{tabular}{|c||c|l|l||l|} \hline  
$\M$ & $E_{\rm YM}$ & \#   & YM state  & b.s. of $N$ D6 \\ \hline \hline
{\large $\frac{R_l^2 V_R}{l_p^9}$} &  
{\large $   \frac{N^{1/3} V_s^{1/3} }{g_{\rm YM}^2 }$} &  1  & new sector
 & D0  
\\    
\hline  
\end{tabular}
\caption[Additional multiplet for $d=6$]{Additional multiplet for $d=6$: \irrep{1} 
 of $E_6$.
\label{exm6} }
\end{center}
\end{table}

\subsection{Gauge backgrounds in Matrix gauge theory}

Our discussion of the Matrix gauge theory U-duality symmetries and
mass formulae has so far been restricted to the rectangular-torus case,
with zero expectation values for the M-theory gauge potentials.
However, gauge backgrounds in M-theory yield moduli, and
should have a counterpart as  couplings in the Matrix gauge theory. 

As a simple example, consider first M-theory
on $T^3$, in which case we can switch on an expectation value
for the component $\C_{123}$ of the three-form. Together with
the volume $V$ of $T^3$, it forms a complex scalar
\begin{equation}
\tau=\C_{123}+i~\frac{V}{l_p^3} \ , 
\end{equation}
which transforms as a modular parameter
under the subgroup $Sl(2,\Zint)$ of the U-duality 
group $Sl(3,\Zint)\times Sl(2,\Zint)$
\cite{Green:1997di}. On the other hand, according to Eq. \eqref{duar}
the volume is identified in the Matrix gauge theory
with $1/g^2_{\rm YM}$, which together
with the theta angle forms a complex scalar
\begin{equation}
\label{theta}
S= \frac{\theta}{2\pi} + i\frac{4\pi}{g_{\rm YM}^2} \ , 
\end{equation} 
transforming as a modular parameter under the electric--magnetic
duality group $Sl(2,\Zint)$. One should therefore identify
$\C_{123}$ with $\theta$, or in other words the three-form background
induces a topological coupling $\int F\wedge F$ on the D3-brane
world-volume.

This can be derived more generally for any $d$ by making use
of Seiberg's argument and the well-known coupling of
Ramond gauge fields to the D-brane world-volume. 
Details of the derivation can be found in
Ref. \cite{Obers:1997kk} and we only quote the result
which is that the 
expectation value of the three-form induces the following  
topological coupling in Matrix gauge theory:
\begin{equation}
\label{topc}
S_{\C}  = \C^{IJK}~\int {\rm d } t \int_{\tilde{T}^d} 
F_{0I} F_{JK} \ .  
\end{equation} 
This coupling reduces to the $\theta$ term \eqref{theta} for $d=3$ and
was conjectured in Ref. \cite{Berenstein:1997cp}. As we now show, 
the coupling \eqref{topc} can also be inferred from the U-duality invariant
mass formulae.  

To see this, we first translate the general U-duality invariant mass formulae
\eqref{mass} into the gauge theory language using \eqref{dic} 
and \eqref{mar}, 
restricting to  $  d\leq 6 $ for simplicity:   
\begin{multline}
\label{gtmass}
E_{\rm YM} =\frac{g_{\rm YM}^2}{N V_s}
\left[  \left( \tilde m^1 \right)^2 +
         \left(\frac{V_s}{g_{\rm YM}^2}\right)^2 \left( \tilde m_{2} \right)^2   
 +       \left(\frac{V_s}{g_{\rm YM}^2}\right)^4 \left( \tilde m_{5} \right)^2 
\right]
\\
+ 
 \sqrt{  \left( \tilde n_1 \right)^2 +
         \left(\frac{V_s}{g_{\rm YM}^2}\right)^2 \left( \tilde n_{4} \right)^2 +
         \left(\frac{V_s}{g_{\rm YM}^2}\right)^4 \left( \tilde n_{1;6} \right)^2 }
\end{multline}
in which we have added the flux multiplet and momentum multiplet
together, as was argued in Subsection \ref{mvsmg}. 
Index contractions are performed with the dual metric 
$ \tilde g_{IJ}= g^{IJ} l_p^6/R_l^2$, and 
upper (lower) indices in the M-theory picture have become  
lower (upper) indices in the 
Matrix gauge theory picture. We also recall that
$V_s$ is the volume of the dual torus $\tilde{T}^d$ on which the Matrix
gauge theory lives.  
The expression of shifted charges is then given by
\begin{subequations}
\begin{eqnarray}
\tilde m^1  &=&  m^1 +  \C^3 m_2 + \left(  \C^3 \C^3 + \E^6\right) m_{5} \\  
\tilde m_2  &=& m_2 + \C^3 m_5 \\
\tilde m_5  &=& m_5
\end{eqnarray} 
\begin{eqnarray}
\tilde n_1    &=& n_1 + \C^3 n_4 + \left( \C^3 \C^3 +\E^6 \right) n_{1;6} \\  
\tilde n_4    &=& n_4 + \C^3 n_{1;6} \\
\tilde n_{1;6}    &=&  n_{1;6} \ .  
\end{eqnarray} 
\end{subequations} 
As we will see below, the 
linear shift in $\C^3$ is in agreement
with the coupling obtained in Eq. \eqref{topc}. 
As a preview, the interpretation of the $\C^3$ coupling in the various Matrix
gauge theories is summarized in Table \ref{tgp}.
We will discuss these formulae in further detail for $d=3,4,5$ below.  
There is as yet no derivation of the coupling of the  $\E_6$  gauge
potential to the Matrix gauge theory.

\begin{table}[h] 
\begin{center}
\begin{tabular}{|c|c||c|l|} \hline 
$D$ & $d$ &   $\C_{IJK}$  &  interpretation \\ \hline \hline 
8 &3 & 1 & $\theta$-parameter \\ \hline  
7 &4 & 4 & off-diagonal component $\A_I$ of $T^5$-metric \\ \hline  
6 &5 & 10 & $B_{IJ}$-background field of string theory on 5-brane  \\ \hline  
$D\leq 5$&$d \geq 6 $ & 
$ \binom{d}{3}  $  
 & unclear  \\ \hline   
\end{tabular} 
\end{center}
\caption{Matrix gauge theory interpretation of three-form potential.
\label{tgp}}
\end{table}

\subsection{$Sl(3,\Zint) \times Sl(2,\Zint)$-invariant 
mass formula for N=4 SYM\\ in 3+1 dimensions \label{invmass3} }
As a first case, we consider the mass formula \eqref{gtmass}
for $d=3$,
\begin{equation}
\label{mass3}
E_{\rm YM}=
\frac{g_{\rm YM}^2}{N V_s} \left( m^I + \frac{1}{2} \C^{IJK} m_{JK} \right)
\tg_{IL} \left( m^L + \frac{1}{2} \C^{LMN} m_{MN} \right)
\end{equation} 
$$
+\frac{V_s}{N g_{\rm YM}^2} \left( m_{IJ} \tg^{IK} \tg^{JL} m_{KL} \right) 
+\sqrt{  n_I \tg^{IJ} n_J } \ .  
$$
This includes the energy of the electric flux
 $  m^I$ (\ie the momentum conjugate to $  \int F_{0I}$)
and the   magnetic flux $  m_{IJ}=\int F_{IJ}$ 
in the diagonal Abelian subgroup of $  U(N)$, together with  
the energy of a massless excitation with  quantized momentum
$  n_I$. 
We observe that the effect of
the M-theory background value of the three-form $\C_3$ is to shift the electric
flux $m^I$,  which is a manifestation of the Witten effect 
and indicates that the coupling 
of $  \C_3$ to gauge theory occurs through the topological term \eqref{topc}. 
Indeed, the only
effect of such a coupling is to shift 
the  momentum conjugate to $  \partial_0 A_I$ by a quantity
$  \C^{IJK} \int F_{JK}$. 

Moreover, introducing the dual magnetic charge 
$m_*^I = \frac{1}{2} \epsilon^{IJK} m_{JK}$ and setting $\C^{IJK} = \theta
\epsilon^{IJK}$, the mass formula \eqref{mass3} can be written in the 
alternative 
form
\begin{equation}
\label{sl2mani}
\begin{split} 
E_{\rm YM}=
\frac{1}{N V_s} (   
 g_{\rm YM}^2 (m^I + \theta m_*^I ) \tg_{IJ} (m^J + \theta m_*^J ) &  +
 \frac{1}{g_{\rm YM}^2} m_*^I \tg_{IJ}  m_*^J  ) \\  
 & +\sqrt{  n_I \tg^{IJ} n_J } \ ,  
\end{split}
\end{equation} 
which manifestly exhibits the $Sl(2,\Zint)$ S-duality symmetry 
as well as the $Sl(3,\Zint)$ modular group of the three-torus.

\subsection{$Sl(5,\Zint)$-invariant mass formula for (2,0) theory on 
the M5-brane\label{invmass4} }

Moving on to the case $d=4$, an extra momentum charge $n_4$  appears
in \eqref{gtmass},
which corresponds to the
momentum along the dynamically generated 5th dimension.
After some algebra, the total mass \eqref{gtmass} can be rewritten in a 
manifestly 
U-duality ($  Sl(5,\Zint)$)-invariant form:  
\begin{equation}
\label{gtmass4}
  E_{\rm YM} = \frac{1}{N V_5} m^{AB} \tg_{AC} \tg_{BD} m^{CD} + 
\sqrt{ n_A \tg^{AB} n_B } 
\end{equation} 
where  $  A,B,\dots = 1 \ldots  5$ and
 $  V_5=V_s g_{\rm YM}^2$ is the  volume of the five-dimensional torus. 
Here, the two-form and vector charges $m^{AB}$, $n_A$ on the five-torus
are related to the original set on the four-torus by
\begin{subequations}
\begin{eqnarray}
m^{I5}&=&m^I\ ,\qquad m^{IJ}=\frac{1}{2} \epsilon^{IJKL} m_{KL}\\ 
n_I&=&n_I\ ,\qquad n_5=\frac{1}{4 !} \epsilon^{IJKL} n_{IJKL}
\ \qquad I,J,\dots=1,\dots,4
\end{eqnarray} 
\end{subequations} 
where the charge $m^{AB}$ 
is the quantized flux (in the diagonal Abelian group)
conjugate to the two-form gauge
field that lives on the (5+1)-dimensional world-volume,
and $n_A$ is simply the momentum along the direction $A$. 
The gauge potential $\C_{IJK}$ combines 
with the gauge coupling and the $T^4$ metric to make
the metric on $T^5$:
\begin{subequations}
\label{ftm}
\begin{equation}
{\rm d}s_5^2 = \tilde{R}^2 ( {\rm d}x^5 + \A^I {\rm d}x_I)^2 +   
{\rm d}s_4^2 
\end{equation}
\begin{equation} 
\label{cf4}
\tilde{R} =g_{\rm YM}^2 \sp \A_I = \frac{1}{3!} \epsilon_{IJKL} \C^{JKL} \ .  
\end{equation}
\end{subequations}
In particular, it is seen that the three-form potential plays the role
of the off-diagonal component of the five-dimensional metric relevant to
the M5-brane. 

As a check, we recall that the bosonic part of the M5-brane action 
can be written in a non-covariant form by solving the self-duality condition 
after singling out a special (fifth) space-like direction and
integrating the resulting equations of motion
\cite{schwarz:1997a,pasti:1997gx,aganagic:1997zq}. 
In particular, it contains the coupling
\begin{subequations}
\label{m5c} 
\begin{equation}
\mathcal{L} = -\frac{1}{4} \epsilon_{\mu \nu \lambda \rho \sigma}
\frac{G^{5 \lambda}}{G^{55}} \tilde{H}^{\mu \nu}   \tilde{H}^{\rho \sigma}   
\end{equation} 
\begin{equation}
\tilde{H}^{\mu \nu} = \frac{1}{6} \epsilon^{\mu \nu \rho \lambda \sigma}
H_{\rho \lambda \sigma} \sp
\mu, \nu = 0 \ldots 4 \ ,  
\end{equation} 
\end{subequations}
which precisely reproduces, 
upon the identifications in \eqref{ftm}, the
topological coupling \eqref{topc} in the effective (4+1)-dimensional SYM
theory, where the field strength $F_{\m \n}$ is identified with the
dual field strength $\tilde{H}_{\m \n}$.  
Finally, we note that 
$E_{\rm YM}$ in \eqref{gtmass4}
depends on the volume of $T^5$ through an over-all factor 
$V_5^{-1/5}$, in agreement with the scale invariance of 
the conjectured (5+1)-dimensional (2,0) theory.

\subsection[$SO(5,5,\Zint)$-invariant mass formula for 
non-critical string theory on\\ the NS5-brane]
{$SO(5,5,\Zint)$-invariant mass formula for 
non-critical string\\ theory on the NS5-brane \label{invmass5} } 
Finally, we consider the case $d=5$, for which according to
 the reasoning
in  Subsection \ref{mt5} the Matrix
gauge theory should correspond to a
non-critical string theory on the type IIB NS5-brane with 
vanishing string coupling $\hat{g}_s$ \footnote{$\hat{g}_s$
cancels out in the following formulae.} and finite
string tension $ \hat{l}_s^2 = g_{\rm YM}^2$.  
After some algebra, the mass formula \eqref{gtmass}
can be rewritten in the manifestly U-duality ($SO(5,5)$) invariant
form 
\begin{equation}
\label{mass5}
  E_{\rm YM} = \frac{1}{N \M_{{\rm NS}5}}  \M^2 (\mbox{D1,D3,D5}) 
 + \sqrt{ \M^2 (\mbox{KK,F1} )} \ ,  
\end{equation} 
where
$\M_{\rm NS5} = \frac{V_s}{\hat{g}_s^2 \hat{l}_s^6}$ is the mass of the
background NS5-brane. 

The second part of \eqref{mass5} involves the 
momentum ($n_1$) and winding ($n^1$, dual to $n_4$) 
excitations of the strings living on 
the world-volume, which form the vector representation \irrep{10}
of the $SO(5,5)$ T-duality group. The corresponding invariant mass
\begin{equation}
\M^2 (\mbox{KK,F1}) = (n_1 +B_2 n^1)^2 + \frac{1}{\hat l_s^2} (n^1)^2    
\end{equation} 
directly follows from the second part of \eqref{gtmass},  
using the identification 
\begin{equation}
\label{cf5}
 B_{IJ} = \frac{1}{3!} \epsilon_{IJKLM} \C^{KLM}
\end{equation} 
for the background antisymmetric tensor field in terms of the components of
the three-form gauge potential on the five-torus.  

The first term in \eqref{mass5}
 involves the D-brane excitations arising from the
charges $(m^1,m_2,m_5)$ that can be dualized into $(m^1,m^3,m^5)$.
It exhibits the correct invariant mass  
Eq. \eqref{dstring} for a spinor representation
of $SO(5,5)$:
\begin{subequations}
\begin{equation}
\M^2 (\mbox{D1,D3,D5}) = 
\left( \frac{\tilde m^1}{\hat g_s \hat l_s^2} \right)^2  
+ \left( \frac{\tilde m^3}{\hat g_s \hat l_s^4} \right)^2  
+ \left( \frac{ m^5}{\hat g_s \hat l_s^6} \right)^2  
\end{equation} 
\begin{equation}
  \tilde m^1 = m^1 + B_2 m^3 + B_2^2 m^5
\sp
\tilde m^3 = m^3 + B_2 m^5
\end{equation} 
\end{subequations}
where we again used the identification \eqref{cf5}.  

As a further check, let us note that the Green--Schwarz term 
$\int {\rm d}^6 x B \wedge F \wedge F $
in the effective action of the six-dimensional string theory,
correctly gives the topological term \eqref{topc} after using 
the relation \eqref{cf5}
between the background string theory $B$-field and the vacuum expectation
values of the M-theory three-form. 
 


\subsection{Extended U-duality symmetry and Lorentz invariance \label{uduaext}} 

M(atrix) theory still lacks a proof of eleven-dimensional Lorentz
covariance to shorten its name to M-theory. In the original
conjecture, this feature was credited to the large-$N$ 
infinite-momentum limit. The much stronger Discrete Light
Cone (DLC) conjecture, if correct, allows
Lorentz invariance to be checked at finite $N$ -- or rather 
at finite $N$'s, since the non-manifest Lorentz generators mix
distinct $N$ superselection sectors. In particular, M(atrix) theory
on $T^d$ in the DLC should exhibit a U-duality $E_{d+1(d+1)}(\Zint)$,
if it is assumed that U-duality is unaffected by light-like
compactifications \cite{Hull:1997jc,Blau:1997du,Obers:1997kk}. 
In this section, we show that an action of $E_{d+1}(\Zint)$ on the 
M-theory BPS spectrum can be defined when we include the light-like
circle $S^1$ on an equal footing with the space-like torus $T^d$.

\begin{table}[h]
\hspace*{-1.7cm}
\begin{tabular}{|c|l||c|l||l||l|}  
\hline
particle multiplet &charge & string multiplet & charge & missing
charges& ext. part.\\ \hline\hline
$\frac{1}{R_I}$ & $m_1(\irrep{7})$ & & & $N(\irrep{1})$ & $M_1(\irrep{8})$
\\  \hline
$\frac{R_I R_J}{l_p^3}$ & $m^2(\irrep{21}) $ & $\frac{R_l R_J}{l_p^3}$ &
$n^1(\irrep{7})$ & & $M^2(\irrep{28})$\\ \hline
$\frac{R_I R_J R_K R_L R_M}{l_p^6}$ & $m^5(\irrep{21})$ &
$\frac{R_l R_I R_J R_K R_L}{l_p^6}$ & $n^4(\irrep{35})$ & & $M^5(\irrep{56})$
\\  \hline
$\frac{R_I^2 R_J R_K R_L R_M R_N R_P}{l_p^9}$ & $m^{1;7}(\irrep{7})$ &
$\frac{R_l R_I^2 R_K R_L R_M R_N R_P}{l_p^9}$ & $n^{1;6}(\irrep{49})$ &
$N^6(\irrep{7}), N^7(\irrep{1})$& $M^{1;7}(\irrep{64})$\\ \hline
&&$ \frac{ R_l R_I^2 R_J^2 R_K^2 R_L R_M R_N R_P}{l_p^{12}}$ &$
n^{3;7}(\irrep{35}) $& $N^{2;7}(\irrep{21})$& $M^{3;8}(\irrep{56})$\\ \hline
&&$ \frac{ R_l R_I^2 R_J^2 R_K^2 R_L^2 R_M^2 R_N^2 R_P}{l_p^{15}}$ &
$n^{6;7}(\irrep{7})$ & $N^{5;7}(\irrep{21})$& $M^{6;8}(\irrep{28})$\\ \hline
&&&& $N^{1;7;7}(\irrep{7}),N^{7;7}(\irrep{1})$& $M^{1;8;8}(\irrep{8})$\\ 
\hline
\end{tabular} 
\caption[Particle multiplet and wrapped string multiplet for $d=7$.]
{Particle multiplet and string multiplet wrapped on $R_l$ for
  $d=7$. Together with the rank multiplets, they form
  the $d=8$ particle multiplet.
\label{tfmm}}
\end{table}

In the presence of an extra (light-like) compact direction of radius
$R_l$, the states from the string multiplet in Table \ref{tmm}
can be wrapped to yield extra particles in the spectrum that join
the already existing states from the particle multiplet 
in Table \ref{tfm}. We have summarized in Table \ref{tfmm}
the various particles obtained in the case $d=7$. It clearly
appears that altogether, the $d=7$ particle and string multiplets
build a particle multiplet of the $d=8$ U-duality group, whose
charges $M$ are obtained from the particle $m$ and string $n$
charges through the relations
\begin{equation}
\begin{array}{lp{2cm}l}
m_1 = M_1  & &
m^{1;7} = M^{1;7} \sp n^{1;6}=M^{1;l6} \\
m^2 = M^2\sp n^1 = M^{l1} & &
m^{3,8} = M^{3;8} \sp n^{3,7}=M^{3;l7} \\
m^5 = M^5\sp n^4 = M^{l4} & &
m^{6,8} = M^{6,8} \sp n^{6,7}=M^{6;l7} \\
\end{array}
\end{equation}
where we have denoted the light-cone direction by an index
$l$ \footnote{As
usual, the same relations hold for $d<7$ by dropping the
tensors with too many antisymmetric indices.}.
This is not quite correct, however, since in particular there is no candidate
for the $M_l$ state, which would correspond to a Kaluza--Klein
excitation along the light-like direction. Obviously, this missing
charge is nothing but the rank of the gauge group
\begin{equation}
N=M_l\ ,
\end{equation} which indeed
denotes the momentum along the light-like direction, and should
therefore be considered as a charge on the same footing as the others.
labelling the vacuum of some M(eta) theory on which
the eleven-dimensional Lorentz group is represented.
This charge has to be invariant under the U-duality group
$E_{d(d)}(\Zint)$, but it gets mixed with other charges under
$E_{d+1(d+1)}(\Zint)$.

While $N$ is the only missing charge for $d\leq 5$, there is still, 
for $d\geq 6$, an extra missing U-duality singlet 
\begin{equation}
\label{exc6} 
N^{6} = M^{l;l6}
\end{equation}
which can be interpreted as the D6--D0 bound state of Eq. \eqref{d06}.
For $d=7$, one needs even more extra charges, namely
\begin{equation}
\label{rank}
N^{2;7}\equiv M^{l2;l7}\sp N^{6}\equiv M^{l;l6}\sp N^{5;7}\equiv M^{l5;l7}\sp
N^{1;7;7}\equiv M^{1;l7;l7} \ , 
\end{equation}
which form the \irrep{56} of $E_7$, isomorphic to
the particle multiplet of
Table \ref{partbranch}, as well as the two singlets   
\begin{equation}
\label{exc7} N^{7}=M^{l;7}\sp N^{7;7}=M^{l;l7;l7}\ , 
\end{equation}
for which Table \ref{tra7} gives the bound-state interpretation as well.  
These extra charges along with $N$ were referred to in 
\cite{Obers:1997kk} as the
{\it rank multiplet}.  
The results are summarized in the Table \ref{trm}, which lists, for all 
$d$'s, 
the dimensions of the particle and string multiplet, as well as the
rank multiplets that are needed to complete the first two into the
particle multiplet of the $d+1$ case.   

We note that the above discussion follows immediately from the decomposition
\eqref{partdec} of the particle multiplet of $E_{d(d)}$ into the particle
and string multiplet of $E_{d-1(d-1)}$ plus extra irreps for $d\geq 6$. 
In particular, the extra
representations that appear are nothing but the extra charges forming the
rank multiplet. If we omit the light-like direction, we indeed see
an extra $\T_1'\vert_1$ for $d=6$; for $d=7$ we have the extra 
representations
$\T_1'\vert_1$, $\T_2'\vert_2$ and $(\T_1')^{2}\vert_3$, whose subscripts are  in precise correspondence 
with the number of times the light-like direction appears in the
charges of \eqref{exc6} and \eqref{rank},\eqref{exc7}.

\begin{table}[h] 
\begin{center}  
\begin{tabular}{|c|c||c||c|c|c||c|}
\hline
& & U-duality &   Flux   &  Mom.   &  
Rank &  Total  \\ 
$D$&$d$ & $E_{d(d)}(\Zint)$ &    $\{m\}$  &    $\{n\}$ &  
$ \{ N \}$ &  $ \{ M\}$  \\ \hline \hline  
10&1 & 1  & \irrep{1} & \irrep{1}  & \irrep{1} & \irrep{3}  \\ \hline 
9 &2 & $Sl(2)$  & \irrep{3} & \irrep{2} & \irrep{1} & \irrep{6} \\ \hline  
8 &3 & $Sl(3) \times Sl(2)$   & \irrep{(3,2)}  & \irrep{(3,1)}& \irrep{1} & \irrep{10} 
\\  \hline  
7 &4 & $Sl(5)$ & \irrep{10} & \irrep{5}  & \irrep{1} & \irrep{16} \\ \hline  
6 &5 & $SO(5,5)$ &  \irrep{16} & \irrep{10} & \irrep{1} &  \irrep{27} \\ \hline  
5 &6 & $E_6 $   & \irrep{27}   & \irrep{\bar{27}}   & \irrep{1}+\irrep{1} & 
\irrep{56} \\ \hline  
4 &7 & $E_7 $ & \irrep{56}  & \irrep{133}& \irrep{56} + \irrep{1} & \irrep{248} \\ 
 & &   &  & & +\irrep{1}+\irrep{1} &  \\ \hline 
3 &8 & $E_8 $  &  \irrep{248} & \irrep{3875} & $\infty$ &  $\infty$  \\ \hline 
\end{tabular}
\end{center} 
\caption{Flux, momentum and rank multiplets.  
\label{trm}} 
\end{table} 
\subsection{Nahm-type duality and interpretation of rank\label{nahm}} 

To see the physical significance of the U-duality enhancement, we discuss
the  extra generators in $E_{d+1(d+1)}(\Zint)$.
First there is the
{\it Weyl} generator, 
exchanging the
light-cone direction with a chosen direction $I$ on $T^d$: 
\begin{equation}
R_l \leftrightarrow R_I \ .
\end{equation} 
The action of this Weyl transformation leaves
the other $R_J$'s and $l_p$ invariant.
In particular, Newton's constant in $11-(d+1)$ dimensions
\begin{equation}
\frac{1}{\kappa^2} = \frac{ V_R R_{l} }{l_p^9}
= R_{l}^{(d-7)/2} \frac{ V_s^{(d-5)/2}}{ g^{d-3} }
\end{equation}
is invariant under U-duality. 
In terms of Matrix gauge theory, this means
\begin{equation}
g_{\rm YM}^2 \rightarrow \left(\frac{R_l}{R_I}\right)^{d-4}g_{\rm YM}^2 \sp
   s_I \rightarrow s_I \sp s_{J\ne I} \rightarrow
              \left(\frac{R_l}{R_I}\right) s_J \ .
\end{equation}
Note that the transformed parameters depend on the original ones
{\it and on} $R_l$. On the other hand, the only dependence
of the gauge theory on $R_l$
should be through a multiplicative factor in the Hamiltonian,
since $R_l$ can be rescaled by a Lorentz boost (see Eq. \eqref{lcb}).
This leaves open the question of how the M(eta) theory itself depends
on $R_l$.

The action on the charges follows
from the exchange of the $I$ and $l$ indices, so that restricting to
$d=6$ for simplicity,   we have  
\begin{equation}
\begin{array}{l}
 N \leftrightarrow m_I\\
   n^{1} \leftrightarrow m^{I1}  \\ 
   n^{4} \leftrightarrow m^{I4}  \\
   n^{1;6} \leftrightarrow m^{1;I6} \ . \\
\end{array}
\end{equation} 
In particular, the rank $N$ of the gauge group is exchanged with the
electric flux $m_I$, whereas the momenta are exchanged with
magnetic fluxes. This is reminiscent of Nahm duality,
relating (at the classical level) a $U(N)$ gauge theory
on $T^2$ with background flux $m$ to a $U(m)$ gauge theory on
the dual torus with background flux $N$
\cite{Nahm:1982}. In the context of higher-dimensional Yang-Mills
theories, this symmetry was first observed at the level of
the multiplicities of the BPS spectrum of SYM in $1+3$ dimensions
\cite{Hacquebord:1997nq}, and extended in the context of
Matrix theory on $T^d$ in Refs. \cite{Hull:1998kb,Blau:1997du,
Obers:1997kk,deMelloKoch:1998hx}. Non-commutative geometry
may provide the correct framework for this duality
\cite{Nekrasov:1998ss}.

The other generator is the Borel generator,
\begin{equation}
\C_{lJK} \rightarrow \C_{lJK} \ ,  
+ \Delta\C_{lJK}
\end{equation} 
which is obtained from the usual $E_{d(d)}(\Zint)$ shifts by conjugation
under Nahm-type duality. It is therefore not an independent generator,
but still gives a spectral flow on the BPS spectrum 
\begin{equation}
\begin{array}{lcl}
N       &\rightarrow& N   + \Delta \C_{l2}~ m^2\\
m_1     &\rightarrow& m_1 + \Delta \C_{l2}~ n^{1}\\
m^2     &\rightarrow& m^2 + \Delta \C_{l2}~ n^4\\
m^5     &\rightarrow& m^5 + \Delta \C_{l2}~ n^{1;6} \ . \\
\end{array}
\end{equation} 
In particular, this implies that states with negative $N$
need to be incorporated in the M(eta) theory if it is
to be $E_{d+1(d+1)}(\Zint)$-invariant. This is somewhat
surprising since the DLC quantization selects $N>0$,
and it seems to require a revision both of the interpretation
of $N$ as the rank of a gauge theory and of the relation
between $N$ and the light-cone momentum $P^+$ 
\footnote{See Ref. \cite{Rey:1997gc} for a discussion of DLCQ with
negative light-cone momentum.}.  

Finally, let us comment in some more generality on the occurrence of
this extended U-duality group. At least at low energies, the Matrix 
gauge theory describing the DLCQ of M-theory compactified
on $T^d$ is nothing but the gauge theory on the $N$ D$d$-brane
wrapped on $T^d$.
The latter is certainly invariant under the T-duality 
$SO(d,d,\Zint)$, and not only $SO(d-1,d-1,\Zint)\bowtie Sl(d)$.
Its spectrum of excitations, or equivalently bound states,
is therefore invariant under $SO(d,d,\Zint)$, and very
plausibly under the extended duality group $E_{d+1(d+1)}(\Zint)$.
On the other hand, we have expanded the bound-state mass
in the limit where the $N$ D$d$-branes are much heavier than
their bound partners, whereas T-duality can exchange the
D$d$-branes with some of their excitations. 
$SO(d,d,\Zint)$ is therefore explicitly broken,
and $E_{d+1(d+1)}(\Zint)$ is broken to $E_{d(d)}(\Zint)$.
The invariance of the mass spectrum can be restored
by using the full non-commutative Born--Infeld dynamics instead of
its small $\alpha'$ Yang--Mills limit
\cite{Connes:1998cr}. While 
not relevant for M(atrix) theory anymore, interesting insights
can certainly be obtained by studying these truly
stringy gauge theories.

\subsection*{Acknowledgements}

The material presented in this review grew from lectures given
by the authors on various occasions during the 1997--98 academic 
year, including CERN-TH Workshop
and journal club, Nordita and the NBI, the
Tours meeting of the Working Group 
on integrable systems and string theory, Ecole Normale Sup{\'e}rieure
and CEA SPhT seminars, Amsterdam Summer Workshop on String Theory and
Black Holes, 
Hamburg Workshop on Conformal Field Theory of D-branes,
Corfu Summer Institute on Elementary Particle Physics.   
We are very grateful to the organizers
for invitation and support.
It is our pleasure to thank E. Rabinovici for an enjoyable collaboration
and early participation to this work.
We are grateful to C. Bachas, I. Bars, D. Bernard, J. de Boer, E. Cremmer, 
E. Eyras, K. F{\"o}rger, A. Giveon, F. Hacquebord, M. Halpern,
C. Hofman,  K. Hori, R. Iengo, B. Julia, E. Kiritsis, M. Krogh, D. Olive,
S. Ramgoolam, C. Schweigert, K. Stelle, W. Taylor, P. di Vecchia,
E. Verlinde and G. Zwart, for useful
remarks or discussions. This work is supported in part by the EEC
under the TMR contracts ERBFMRX-CT96-0045 and ERBFMRX-CT96-0090.

\subsection*{Note added in proof~: \\Boundaries of M-theory moduli space}
\addcontentsline{toc}{section}{Note added in proof~:
  Boundaries of M-theory moduli space}

As we discussed in Section 2, M-theory arises in the strong coupling 
regime of type II string theory, and reduces at low energy to 11D
supergravity. It is important to determine what portion of the 
M-theory moduli space are covered by these weakly coupled
descriptions, and thus what room is left for truly M-theoretic
dynamics. The techniques we developed in Section 4 allow us to
easily answer this question, first addressed in
Ref. \cite{Witten:1995ex} for compactifications down to $D\ge 4$,
and recently for $D=2$ in Ref. \cite{Banks:1998vs}. We first
consider the case $D>2$, and consider an asymptotic direction
in the moduli space, represented by an arbitrarily large 
weight vector $\lambda$ in the
weight space $V_{d+1}$, see Section 4.6. Modulo U-duality, 
$\lambda$ can be chosen in the fundamental Weyl chamber
$\lambda\cdot\alpha>0$ for all positive roots $\alpha$. This
corresponds to choosing
\begin{equation}
\label{note1}
R_1<R_2<\dots <R_d\ ,\qquad R_1 R_2 R_3>l_p^3\ ,
\end{equation}
where the inequalities are understood to be large inequalities,
in order to have a maximal degeneration in the moduli space 
\cite{Witten:1995ex}.
The 11D supergravity description is valid provided all radii are
larger than the Planck length, \ie
\begin{equation}
\label{note2}
11D~SUGRA~:\qquad l_p<R_1\ .
\end{equation}
On the other hand, when the radius $R_1$ is much smaller than $l_p$, we can 
have a type IIA description with weak coupling $g_s^2=(R_1/l_p)^3$, 
provided all radii are larger than the string length
$l_s^2=l_p^3/R_1$~:
\begin{equation}
\label{note3}
IIA~:~\qquad R_1< l_p\ , \qquad R_1 R_2^2 > l_p^3\ .
\end{equation}
If this is not the case, then we may instead try a type IIB
description with weak coupling $g_s=R_1/R_2$, same string 
length $l_s^2=l_p^3/R_1$ and 10-th radius $R_B=l_p^3/(R_1 R_2)$.
The IIB radii $R_B$ and $R_3,\dots, R_d$ 
are larger than the string length provided
$R_1 R_2^2 <l_p^3$ and $R_1 R_3^2>l_p^3$, and it is not difficult
to see that, using Eq. (\ref{note1}),
the first implies $R_1<l_p$, and the second is automatically
satisfied. The type IIB description thus hold in the region
\begin{equation}
\label{note4}
IIB~:~\qquad R_1< l_p\ , \qquad R_1 R_2^2 < l_p^3\ .
\end{equation}
The weakly coupled 11D supergravity, type IIA and type IIB
descriptions therefore cover, up to U-duality, the entire
asymptotic moduli space of M-theory on $T^d$, $d>2$. Of course,
these descriptions fail when any of the large inequalities above
become approximate equalities, hence the need for a more fundamental
definition of M-theory. On the contrary, when $D\le 2$, there are
asymptotic sectors of the moduli space where no perturbative
description is possible. Indeed, the weight
space $V_{d+1}$ is now intrinsically Minkovskian, and the light-cone
$\lambda\cdot\lambda=-(x^0)^2+\sum (x^i)^2=0$ 
separates the moduli space into three 
sectors that can never be related to each other by U-duality.
For instance, the 11D supergravity region \eqref{note1} has
$x^i>x^0/3$ for all $i$, so that $\lambda\cdot\lambda>0$.
It therefore sits in the interior of the future light-cone if $x^0>0$,
or past light-cone if $x^0<0$ (one may choose $x^0=0$ by working in 
$l_p$ units). In fact, the time-like region can be shown to 
have a weakly coupled 11D supergravity, type IIA or type IIB
description, whereas the spacelike region can be argued to
be cosmologically forbidden by the holographic principle \cite{Banks:1998vs}. 

\vskip 1cm
\centerline{\Large \bf Appendices}
\vskip .7cm

\appendix

\section{BPS mass formulae \label{bpsmass}}
In this Appendix, we analyse the BPS eigenvalue equation \eqref{bps}
for various choices of non-vanishing central charges. This gives
a check on the mass formulae obtained on the basis of duality, and
yields the conditions on the charges for a state to preserve
a given fraction of supersymmetry.

\subsection{Gamma Matrix theory \label{gamma} }
In order to maintain manifest eleven-dimensional Lorentz invariance, we use
the 11D Clifford algebra $\left[ \Gamma_M, \Gamma_N \right]=
2\eta_{MN}$, with signature $(-,+,\dots)$, even after
compactification. The matrices $\Gamma_M$ are then $32\times 32$
real symmetric
except for the charge conjugation matrix $C=\Gamma_0$, which is 
real antisymmetric. All products of Gamma matrices are traceless
except for
\begin{equation}
\Gamma_0 \Gamma_1 \dots \Gamma_9 \Gamma_s = 1 \ , 
\end{equation}
where we denote by $s$ the eleventh direction. 
We define $\Gamma_{MN\dots}=\Gamma_M\Gamma_N\dots$
if the $p$ indices $M,N,\dots$ are distinct, zero otherwise, 
and abbreviate it as $\Gamma_{(p)}$. We have
\begin{equation}
(\Gamma_{(p)})^2 = (-1)^{\left[ \frac{p}{2} \right]}\sp
(\Gamma_0 \Gamma_{(p)})^2 = (-1)^{\left[ \frac{p-1}{2} \right]} \ , 
\end{equation}
where the $p$ indices are non-zero and the square brackets
denote the integer part. Furthermore,
\begin{subequations}
\begin{eqnarray}
\Gamma_{(p)}\Gamma_{(q)}+(-)^{pq}\Gamma_{(q)}\Gamma_{(p)}
&=&\sum^\infty_{\substack{k=0\\p+q-4k\geq|p-q|}} \Gamma_{(p+q-4k)}\label{g1}\\
\Gamma_{(p)}\Gamma_{(q)}-(-)^{pq}\Gamma_{(q)}\Gamma_{(p)}
&=&\sum^\infty_{\substack{k=0\\p+q-2-4k\geq|p-q|}} \Gamma_{(p+q-2-4k)}\label{g2}
 \ , 
\end{eqnarray}
\end{subequations}
with no restrictions on the $p+q$ indices.
On the right-hand side of Eq. \eqref{g1} (resp. \eqref{g2}),
a contraction between the first 
$2k$ (resp. $2k+1$) indices of $\Gamma_{(p)}$ and the first $2k$
(resp. $2k+1$)
indices of $\Gamma_{(p)}$ is implied. In particular,
\begin{equation}
[ \Gamma_{(2)}, \Gamma_{(p)} ] = \Gamma_{(p)} \ , 
\end{equation}
since $\Gamma_{(2)}$ generates Lorentz rotations.

\subsection{A general configuration of KK-M2-M5 on $T^5$}
Here we consider M-theory compactified on $T^5$, and allow
for non-vanishing central charges 
$Z_I$,
$Z_{IJ}$,
$Z_{IJKLM}$, where the indices $I,J,\dots$ are internal indices
on $T^5$. We therefore look for solutions to the eigenvalue
equation
\begin{eqnarray}
\Gamma \epsilon &=& \M \epsilon \\
\Gamma & \equiv &  Z_I \Gamma^{0I} + Z^{IJ} \Gamma_{0IJ} 
+ Z^{IJKLM} \Gamma_{0IJKLM} \ . \nonumber
\end{eqnarray}
Squaring this equation, we obtain
\begin{multline}
Z_I Z_{J} \{\Gamma_I,\Gamma_J\} - Z^{IJ} Z_{KL} \{ \Gamma_{IJ},\Gamma_{KL} \}
+ Z^{IJKLM} Z^{NPQRS} \{\Gamma_{IJKLM}, \Gamma_{NPQRS}\} \\
+ 2 Z_I Z^{JK} \left[ \Gamma^{I},\Gamma_{JK} \right] 
+ 2 Z_I Z^{JKLMN} \left\{ \Gamma^{I},\Gamma_{JKLMN} \right\}  \\
- 2 Z^{IJ} Z^{KLMNP} \left[ \Gamma_{IJ},\Gamma_{KLMNP} \right] 
\circeq \M^2 \ , 
\end{multline}
where the symbol $\circeq$ denotes the equality when acting on
$\epsilon$. 
Using the identities \eqref{g1},\eqref{g2}, this reduces to
\begin{equation}
\label{sy1}
(Z_I)^2 + (Z^{IJ})^2+ (Z^{IJKLM})^2 +
+ Z_J Z^{IJ} \Gamma_I 
+ \left( Z_M Z^{MIJKL} +Z^{IJ} Z^{KL} \right) \Gamma_{IJKL}
\circeq  1
\end{equation}
A 1/2-BPS state is obtained under the conditions
\begin{subequations}
\begin{eqnarray}
k^I & \equiv & Z_J Z^{IJ} =0  \\
k^{IJKL} & \equiv & Z_M Z^{MIJKL} + Z^{IJ} Z^{KL} = 0 \ ,   
\end{eqnarray}
\end{subequations}
which indeed form a string multiplet (\irrep{10}) of $E_5 =SO(5,5)$,
and has a mass given by
\begin{equation}
\M_0^2 = (Z_I)^2 + (Z^{IJ})^2+ (Z^{IJKLM})^2 \ .  
\end{equation}
If the conditions are not satisfied, we can define
$k_I=\epsilon_{IJKLM}k^{IJKL}/4!$, $\Gamma_6=\Gamma_{12345}$
and rewrite Eq. \eqref{sy1} as
\begin{equation}
\label{sy1b}
k^I \Gamma_I 
+ k_I \Gamma_6 \Gamma ^{I} \circeq  \M^2 - \M_0^2 \ . 
\end{equation}
Note that the $SO(5,5)$ vector $(k_I,k^I)$ is null:
$k_I k^I = 0$.
Squaring again yields the 1/4-BPS state mass formula 
\begin{equation}
\M^2 = (Z_I)^2 + (Z^{IJ})^2+ (Z^{IJKLM})^2 +\sqrt{ (k^I)^2 + (k_I)^2} \ . 
\end{equation}
This result can be straightforwardly made invariant under the full
U-duality group by including the couplings to the gauge potentials 
through the lower charges as
found for the particle and string multiplet in \eqref{mass} and
\eqref{tension}.

\subsection{A general configuration of D0,D2,D4-branes on $T^5$ \label{d0d2d4} }
We now consider the D-brane sector of M-theory on $T^6$, that
is a general configuration of D0,D2,D4-branes. The
eigenvalue equation becomes
\begin{subequations}
\begin{eqnarray}
\Gamma \epsilon = \M \epsilon \\
\Gamma \equiv  Z \Gamma_{0s} + Z^{ij} \Gamma_{0ij} 
+ Z^{ijkl} \Gamma_{0ijkls}\ ,  
\end{eqnarray}
\end{subequations}
where $Z$, $Z^{ij}$, $Z^{ijkl}$ denote the D0,D2,D4-brane
charges respectively, and $i,j,\dots$ run from 1 to 5.
Squaring this equation, we obtain
\begin{multline}
\label{s5}
2 Z^2 + Z^{ij} Z^{kl} \{\Gamma_{ij},\Gamma_{kl}\} 
+ Z^{ijkl} Z^{mnpq} \{ \Gamma_{ijkl} \Gamma_{mnpq}\} \\
+4 Z\ Z^{ijkl} \Gamma_{ijkl}
+ 2 Z^{ij} Z^{klmn} \left[ \Gamma_{ij},\Gamma_{klmn} \right] \Gamma_s 
\circeq  \M^2 \ . 
\end{multline}
Using identities \eqref{g1},\eqref{g2}, this becomes
\begin{equation}
\label{s6}
Z^2 + (Z^{ij})^2+ (Z^{ijkl})^2 + 
k^{ijkl} \Gamma_{ijkl} + (k')^{ijkl} \Gamma_{ijkls} \circeq  \M^2 \ , 
\end{equation}
where we defined 
\begin{subequations}
\begin{eqnarray}
k^{ijkl} \equiv Z^{[ij}Z^{kl]} + Z\ Z^{ijkl}\\
k^{'ijkl} \equiv Z^{m[i} Z^{jkl]m} \ . 
\end{eqnarray}
\end{subequations}
The second combination can be rewritten on $T^5$ as
a form $k^{i;jklmn}=Z^{i[j}Z^{klmn]}$. 
Then, $k^4$ and $k^{1;5}$ can be dualized
into a \irrep{10} null vector $(k_i,k^i)$ of the T-duality
group $SO(5,5)$. A state with $k=k'=0$ is 1/2-BPS
with mass
\begin{equation}
\M_0^2 = (Z)^2 + (Z^{ij})^2 + (Z^{ijkl})^2 \ . 
\end{equation}
If these conditions are not met, we can rewrite Eq. \eqref{s6} as
\begin{equation}
k_i \Gamma^{i}\Gamma_6 + k'_i \Gamma^{i}\Gamma_6\Gamma_s
\circeq \M^2 - \M_0^2 \ , 
\end{equation}
implying a mass formula
\begin{equation}
\M^2 =  (Z)^2 + (Z^{ij})^2 + (Z^{ijkl})^2 + 2
\sqrt{ (k^i)^2  + (k_i)^2}
\end{equation}
or, in terms of the natural undualized charges, 
\begin{equation}
\M^2 = (Z)^2 + (Z^{ij})^2 + (Z^{ijkl})^2  + 2
\sqrt{ (k^{ijkl})^2  + (k^{i;jklmn})^2} \ . 
\end{equation}

\subsection{A general configuration of KK--w--NS5 on $T^5$\label{ns5kkw}}
Finally, we consider the Neveu--Schwarz sector of the
theory considered in the Appendix \ref{d0d2d4}, namely
the bound states of NS5-branes, winding and Kaluza--Klein
states. The eigenvalue equation then reads
\begin{equation}
\label{ss1}
\left( z_i \Gamma ^{0i} + z^{i} \Gamma_{0s i} 
+ z^{ijklm} \Gamma_{0ijklm} \right) \epsilon = \M \epsilon \ . 
\end{equation}
Taking the square gives
\begin{equation}
\label{ss2}
z^2 + (z^{i})^2+ (z_{i})^2 + 2 z z^i \Gamma_{6i}
+2 z z^i  \Gamma_{6s i} -2 \Gamma_{s}z^i z_i \circeq  \M^2 \ ,  
\end{equation}
so the 1/2-BPS conditions appear to be
\begin{equation}
z~z^i = z~z_i = z^i z_i =0 \ . 
\end{equation}
This agrees with the vanishing of the entropy $z z^i z_i$ and its
first derivatives, as obtained in Ref. \cite{Ferrara:1997ci}.
We can go further and find the complete 1/8-BPS mass formula:
multiply Eq. \eqref{ss1} by  $z\Gamma_{06}$:
\begin{equation}
-z~z_i \Gamma_{6i} -z~z^{i} \Gamma_{6s i} 
- z^2 \epsilon \circeq  z \M \Gamma_{06} 
\end{equation}
and combine  with Eq. \eqref{ss2} to obtain:
\begin{equation}
\left(-z^2 + (z^{i})^2+ (z_{i})^2 + 2 z \M \Gamma_{60}  
-2 z^i z_i \Gamma_{s}
\right) \circeq \M^2 \ .  
\end{equation}
Now $\Gamma_s$ and $\Gamma_{06}$ commute, are traceless
and square to 1, so this is a second-order equation:
\begin{equation}
-z^2 + (z^{i})^2+ (z_{i})^2 \pm 2 z \M 
\pm 2 z^i z_i  \circeq \M^2 \ ,  
\end{equation}
with solutions
\begin{equation}
\M = \pm z \pm \sqrt{ (z_i \pm z^i)^2 }
\end{equation}
or, equivalently: 
\begin{equation}
\M^2 = z^2 + (z_i)^2 + (z^i)^2 + 2 |z_i z^i| 
+ 2|z|\sqrt{ (z_i)^2 + (z^i)^2 + 2 |z_i z^i|} \ . 
\end{equation}
This reduces to the usual mass formula for perturbative string
states ($z=0$) and for KK--NS5 or w--NS5 bound states.
For momentum and winding charges along a single
direction, this reduces to $\M=\pm z \pm z_1 \pm z^1$,
in agreement with the identification of central charges
in Ref. \cite{Ferrara:1997ci}. The U-duality invariant
generalization of this mass formula is however unclear.

\section{The $d=8$ string/momentum multiplet \label{d8strm} }

For completeness, we give  
in Table \ref{string8} the content of the string/momentum
multiplet for $d=8$ in the \irrep{3875} of $E_{8(8)}$.
It comprises the 2160 states in the Weyl orbit
of the highest weight $R_i/l_p^3$ of length 4, together
with 7 copies of the 240 weights of length 2 with tension
\begin{equation}
\T = \frac{V_R}{l_p^9} \times (d=8\;\,\mbox{particle multiplet}) \ ,  
\end{equation} 
as well as 35 zero weights with tension 
\begin{equation}
\T = \left( \frac{V_R}{l_p^9} \right)^2  \ .  
\end{equation} 
As in $d=7$, the resulting multiplet exhibits a mirror symmetry, which
relates each state with tension $R^{3a-2}/l_p^{3a} $, $a= 1 \ldots  6 $
to another state with tension $R^{34-3a}/l_p^{3(12-a)}$ 
through the relation  
\begin{equation}
\label{mirs8}
\M \M' = \left( \frac{V_R}{l_p^9} \right)^4 \ ,    
\end{equation}  
where $V_R$ is the volume of the eight-torus.
For this reason, Table \ref{string8} only gives the explicit form of 
the tensions for the lower half $a=1\ldots 5$ and the self-mirror part
$a=6$. The second
column gives the $Sl(8)$ irreps at each level graded by $1/l_p^{3a}$, 
while the last column lists the corresponding charges.  
Here the notation is as
follows: a semicolon denotes an ordinary tensor product as before (so in general
contains more than one $Sl(8)$ irrep); two superscripts $(p;q)$ grouped within
parentheses and separated by a semicolon denote the irrep, 
whose Young tableau is
formed by juxtaposition of a column with $p$ rows and one with $q$ rows.

\begin{table}[h]
\begin{center}
\begin{tabular}{|c|l|l|}  
\hline
mass & $Sl(8)$ irrep & charge\\ \hline  
$\frac{ R_I}{l_p^3}$ &  \irrep{8} & $n^1 $ \\ 
$ \frac{ R_I R_J R_K R_L}{l_p^6}$ & \irrep{70} & $n^4$ \\   
$\frac{ R_I^2 R_J R_K R_L R_M R_N }{l_p^9}$, 7$\frac{ V_R  }{R_I l_p^9}$ 
& $\irrep{8} + \irrep{216}$  & 
$n^{1;6}$ \\
$ \frac{ R_I^2 R_J^2 R_K^2 R_L R_M R_N R_P}{l_p^{12}}$ &  & \\  
$ \frac{ V_R R_I^2 }{l_p^{12}}$,
7$ \frac{ V_R R_I R_J }{l_p^{12}}$ & $\irrep{28} + \irrep{36} + \irrep{420}$  
  & $ n^{3;7}, n^{(1;1);8} $ \\ 
$ \frac{ R_I^2 R_J^2 R_K^2 R_L^2 R_M^2 R_N^2 R_P}{l_p^{15}}$ &  & \\  
$ \frac{V_R  R_I^2 R_J R_K R_L }{l_p^{15}}$,
7$ \frac{V_R  R_I R_J R_K R_L R_M }{l_p^{15}}$ &\irrep{56} + \irrep{168} +
\irrep{404} 
&  $n^{(6;7)}$, $n^{1;4;8}$ \\  
$ \frac{V_R  R_I^2 R_J^2 R_K R_L R_M R_N}{l_p^{18}}$,
7$ \frac{V_R  R_I^2 R_J R_K R_L R_M R_N R_P }{l_p^{18}}$,
35$\frac{V_R^2}{l_p^{18}} $
 & $\irrep{1} + \irrep{63} + \irrep{720}$ &  $n^{(1;7);8}$, $n^{2;6;8}$ \\  
\hline
\end{tabular} 
\end{center}
\caption{String/momentum multiplet {\bf 3875} of $E_8$. 
\label{string8}}
 \end{table}

As an aid to the reader, we give the charges of the dual states 
at level $l_p^{3(12-a)}$:
\begin{equation}
a=1: \;\; n^{7;8;8;8}  \sp \;\;\; 
a=2: \;\; n^{4;8;8;8} \sp \;\;\; 
a=3: \;\; n^{2;7;8;8} 
\end{equation}
$$
a=4: \;\; n^{1;5;8;8},  n^{(7;7);8;8} \sp \;\;\; 
a=5 : \;\; n^{(1;2);8;8},  n^{4;7;8;8} 
$$

Finally, we display the decomposition of the $d=8$ string multiplet 
under the T-duality subgroup group $SO(7,7,\Zint)$. 
Here, we may again restrict to those states with (type IIA) tensions
$ M \sim 1/g_s^a$, $a=0 \ldots 4$, for each of which there is a
dual state with tension $\M '$ related to it by
\begin{equation}
\label{mirs8a}
\M \M ' = \left( \frac{V'_R}{g_s^2 l_s^8} \right)^4 \ ,  
\end{equation}
where $V'_R$ stands for the seven-dimensional type IIA torus. The type IIA
mirror symmetry \eqref{mirs8a}  easily follows from \eqref{mirs8} using
the M-theory/type IIA connection in \eqref{msr}.   
The results are summarized in Table \ref{partbranch8s}. 

\begin{table}[h] 
\begin{center} 
\begin{tabular}{|l||l|l|l|l|l|l|l|}
\hline
$\begin{smallmatrix}
\irrep{3875}(E_8)\supset &SO(7,7)\\
\cup &\\
Sl(8)&\\ 
\end{smallmatrix}$ &  {1}   & {64}   & 378 & 896 & 1197  & 896 & \ldots \\  
 \hline \hline  
 {8}    & {1} &  {7} &     &     &     &  &\\ 
 {70}   &   & {35} &  {35} &     &     &  & \\ 
 224  &   & {21} & 154 & 49  &     &  & \\
 484  &   &  {1} & 154 & 294 &  {35} &  & \\
 728  &   &    &  {35} & 292 & 294 & {7}  & \\
 847  &   &    &     & 154 & 539 & 154  & \\
 728  &   &    &     &   {7} & 294 & 292  & \ldots  \\
 484  &   &    &     &     &  {35} & 294  & \ldots \\ 
 \vdots  &    &    &     &     &   & $\vdots $  & $\ddots$   \\
\hline
\end{tabular}
\end{center} 
\caption[Branching of $d=8$ string multiplet]{Branching of
the $d=8$ string multiplet into representations of $Sl(8)$ and 
$SO(7,7)$.
The entries in the table denote the common $Sl(7)$ reps. The full table
can be reconstructed using mirror symmetry in the point with $Sl(7)$ 
representation 539.  
\label{partbranch8s}}
\end{table}

Here the type IIA states in the $a$-th column, have a tension proportional
to $1/g_s^{a-1}$. The first column is the singlet irrep formed by the
fundamental string.
The second column is the spinor irrep consisting of D$p$=0-,2-,4-,6-branes,
with one unwrapped world-volume direction. The third column can be decomposed
into the $SO(7,7)$ irreps 
$378 = \irrep{1} \oplus 3 \times \irrep{91} \oplus  \irrep{104} $, 
and contains, together with
NS5 and KK5 with one unwrapped direction,  many non-standard
states with tension $\sim 1/g_s^2$. The fourth column contains
the representation $896 =
\irrep{14} \otimes \irrep{64}$ 
formed by tensor product of vector and spinor representation, and
   has states with 
tension $\sim 1/g_s^3$. The fifth  column consists of 1197 states
with tension $\sim 1/g_s^4$. The set of duals of these states includes  
states with tension up to $1/g_s^8$, all of which are at present
far from understood. 

We note that the string state with tension $\frac{V_R R_I^2 }{l_p^{12}}$
is presumably related to the conjectured M9-brane 
\cite{Bergshoeff:1996ui,Bergshoeff:1998bs}, which should more
properly be called M8-brane.
In fact, for $d=9$ there will be a corresponding particle with mass
$\frac{V_R R_I^2 }{l_p^{12}}$, where 
$V_R$ is now the volume of the nine-torus. Taking $R_I =R_s = l_s g_s$,  
this reduces to the mass of the type IIA D8-brane, while taking $R_s$ in one of
the other world-volume directions gives an 8-brane with exotic mass
$\frac{V_R' R_i^2}{l_s^{11} g_s^3}$. Vertical reduction, on the other hand,
would give a type IIA 9-brane.

\section{Matrix gauge theory on $T^7$ \label{mgt7} }

In this appendix we discuss in some detail the Matrix gauge theory on
$T^7$, performing the analysis of Subsection \ref{udmmg} for the case $d=7$.

For our discussion, it will be useful to first consider
the type IIB states obtained from the set of type IIA states 
in \eqref{2a7}, by performing a 
maximal T-duality on the seven-torus. Using \eqref{tid}, we find the following 
T-duality multiplets for type IIB string theory compactified on $T^7$ 
\begin{subequations}
\label{2b7}
\begin{eqnarray}
V: & 
\frac{R_i}{l_s^2},  
\frac{1}{R_i} \\ 
S_B: & \frac{1}{g_s} \left( 
  \frac{V_R}{l_s^8},  
  \frac{R_iR_j R_k R_l R_m}{l_s^6}, 
\frac{R_iR_jR_k}{l_s^4}, 
\frac{R_i}{l_s^2} 
\right) \\  
S+ AS : & \frac{1}{g_s^2} \left( 
  \frac{V_R R_i R_j }{l_s^{10} },  
   8 \frac{V_R }{l_s^8}, 
  \frac{R_i^2 R_j R_k R_l R_m R_n}{l_s^8},   
\frac{R_iR_j R_k R_l R_m }{l_s^6}
\right) \\ 
S_A : & \frac{V_R}{g_s^3 l_s^8}  \left( 
  \frac{R_i R_j R_k R_l R_m R_n }{l_s^6},  
  \frac{R_i R_j R_k R_l}{l_s^4}, 
\frac{R_i R_j}{l_s^2}, 
  1  \right) \\ 
V' : & \left( \frac{V_R}{g_s^2 l_s^8} \right)^2  \left( 
R_i,  \frac{l_s^2}{R_i}
\right) \ .  
\end{eqnarray}
\end{subequations}
Here we have given the states in each multiplet in the order in which
they are obtained from the corresponding type IIA states. At the levels
$1/g_s^a$, with $a$ even, we obtain the same set of states as in type IIA. 
At odd level, however, the spinor representations are interchanged,
so that at level $1/g_s$ we obtain the 
odd D$p$ branes, while
at level $1/g_s^3$ we find the set of $p_3^{7-p}$ branes with 
$p=1,3,5,7$. 

We also give the S-duality structure of the type IIB states \eqref{2b7}. 
Using \eqref{bsdua}, the following list of
S-duality singlets (appearing at each level) is found:  
\begin{equation}
\mbox{KK} \sp   
\mbox{D3} \sp  
7_{2} \sp  
\mbox{KK5} \sp  
3_{3}^4 \sp
0_{4}^{(1,6)} \ .  
\end{equation}
The remaining states pair up into S-duality doublets
\begin{equation} 
\mbox{F1--D1} \sp
\mbox{D5--NS5} \sp
\mbox{D7--}7_{3}  \sp
5_{2}^2\mbox{--}5_{3}^2   \sp
1_3^6\mbox{--}1_{4}^6 \ .  
\end{equation}   

The M-theory mass,
gauge-theory energy and bound-state interpretation of the 
flux and momentum multiplets is given in Table \ref{tfm7}.  

\begin{table}[h] 
\begin{center}
\begin{tabular}{|c||c|l|l||l|} \hline  
$\M$ & $E_{\rm YM}$ & \#   & YM  state  & b.s. of $N$ D7 \\ \hline \hline
{\large $\frac{1}{R_I}$} &  
{\large $   \frac {g^2_{\rm YM} s_I^2 }{N V_s}$} &  7  & electric flux & NS--w
\\    
{\large $\frac{R_I R_J}{l_p^3}$} &  
{\large $   \frac {V_s}{N g^2_{\rm YM} (s_I s_J)^2 }$} &  21  & magnetic flux & D5\\  
{\large $\frac{V_R}{R_I R_J l_p^6}$} &  
{\large $   \frac {V_s (s_I s_J)^2}{N g^6_{\rm YM}}$} &  21  & new sector & 
$5_2^2$  \\ 
{\large $\frac{V_R R_I }{ l_p^9}$} &  
{\large $   \frac {V_s^3 }{N g^{10}_{\rm YM} s_I^2}$} &  7  & new sector & 
$1_3^6$  \\ \hline  
\hline 
{\large $\frac{R_l R_I}{l_p^3}$ } &
{\large $   \frac{1}{s_I} $}  & 7  & KK momentum   & KK \\   
{\large $\frac{R_l V_R}{R_I R_J R_K l_p^6}$} &
{\large $   \frac{s_I s_Js _K}{g^2_{\rm YM}} $}  &  35  & YM threebrane    & D3  \\     
{\large $\frac{R_l R_I V_R}{ R_J l_p^9}$} &
{\large $   \frac{V_s s_J  }{g^4_{\rm YM} s_I} $}  &  49 & new sector     
& KK5, 7${}_{2}$    \\    
{\large $\frac{R_l R_I R_J R_K  V_R}{  l_p^{12} }$} &
{\large $   \frac{V_s^2  }{g^6_{\rm YM} s_I s_J s_K} $}  &  35 & new sector     
& $3_{3}^4$    \\    
{\large $\frac{R_l  V_R^2}{R_I  l_p^{15} }$} &
{\large $   \frac{V_s^2 s_I  }{g^8_{\rm YM} } $}  &  7 & new sector     
& $0_{4}^{(1,6)}$    \\    
\hline  
\end{tabular}
\caption[Flux and momentum multiplet for $d=7$]
{Flux and momentum multiplet for $d=7$: \irrep{56} and 
\irrep{133} of $E_7$.
\label{tfm7} }
\end{center}
\end{table}  

Comparing the states in the last column of this table 
with the total set of 1/2 BPS 
states \eqref{2b7} for type IIB on $T^7$, we note that there is 
a large number  of states that do not appear. 
In analogy with the extra D6--D0 multiplet (a singlet) that appeared
for $d=6$, we can construct in this case an extra multiplet that contains
the D7--D1 bound state, for which we conjecture (by T-duality)
 the same bound-state mass 
formula as in Eq. \eqref{dp6}, so that 
\begin{equation}
\label{d7d1}
E_{\rm YM} = \M_{D1}^{2/3} \M_{ND7}^{1/3}  = N^{1/3}  
\frac{V_s^{1/3} s_I^{2/3} }{g_{\rm YM}^2} \ ,  
\end{equation}
where we used $g_{\rm YM}^2 = g_s l_s^4$. 
The relevant data of the corresponding U-duality multiplet, which forms
the \irrep{56} of $E_{7(7)}(\Zint)$, 
is given in Table \ref{tra7}. The easiest way to obtain this
table, starting with the gauge-theory mass \eqref{d7d1} obtained for the D7--D1 
bound
state, is by noticing that this state is, up to a multiplicative U-duality
invariant factor $I_7^{1/3}$ (see Eq. \eqref{udsmg}) and up to a power of 1/3, 
exactly analogous to the flux
multiplet of Table \ref{tfm7}. Note that the 56 states are precisely
the S-dual states of those involved in the flux multiplet bound states.
The bound states relevant to the momentum multiplet, on the other hand,
involve S-duality singlets.  

 Besides the D7 itself, this leaves two more possible
states left in the type IIB, which can form a bound state with the D7, 
namely the two 7-branes with mass
\begin{equation}
\label{2sin} \frac{V_s}{g_s^2 l_s^8} \sp    
\frac{V_s}{g_s^3 l_s^8} \ .  
\end{equation} 
For the first one, we know already from the momentum multiplet that the
mass relation is 
\begin{equation}
\label{72b}
E_{\rm YM} = \M = \frac{V_s}{g_{\rm YM}^4} = I_7^{1/2}
\end{equation}
and hence a U-duality singlet. 
For the second state in \eqref{2sin}, we deduce the mass relation by the
requirement that the bound-state energy be such that 
\begin{equation}
E_{\rm YM} = [\M_{N{\rm D}7}^a + \M^a]^{1/a} -  
\M_{N{\rm D}7} \simeq \M^a  
\M_{N{\rm D}7}^{1-a}   
\end{equation}
i) can be written in gauge-theory variables, and ii) is a U-duality
singlet. Either of these requirements yields $a=1/2$, and we are left
with  a gauge-theory state with energy  
\begin{equation}
\label{73b}
E_{\rm YM} = 
\M_{7_3}^{1/2}   
\M_{N{\rm D}7}^{1/2}  = N^{1/2} \frac{V_s}{g_{\rm YM}^4}   
= N^{1/2} I_7^{1/2}  \ .  
\end{equation}
The singlets in Eqs. \eqref{72b} and \eqref{73b} 
are also given in Table \ref{tra7}. 

\begin{table}[h] 
\begin{center}
\begin{tabular}{|c||c|l|l||l|} \hline  
$\M$ & $E_{\rm YM}$ & \#   & YM  state  & b.s. of $N$ D7 \\ \hline \hline
{\large $\frac{R_l^2  V_R}{R_I  l_p^{9} }$} &
{\large $   \frac{N^{1/3} V_s^{1/3} s_I^{2/3} }{g_{\rm YM}^2}$} &  7  & new
sector & D1 
\\    
{\large $\frac{R_l^2  R_I R_J V_R}{ l_p^{12} }$} &
{\large $   \frac{N^{1/3} V_s }{g^{8/3}_{\rm YM} (s_I s_J)^{2/3} }$} &  21  & new sector 
 & NS5 \\  
{\large $\frac{R_l^2   V_R^2}{ R_I R_J l_p^{15} }$} &
{\large $   \frac{N^{1/3} V_s (s_I s_J)^{2/3}}{g^{14/3}_{\rm YM}}$} &  21  & 
new  sector & 
$5_3^2$  \\ 
{\large $\frac{R_l^2 R_I  V_R^2}{  l_p^{18} }$} &
{\large $   \frac{N^{1/3} V_s^{5/3} }{ g^6_{\rm YM} s_I^{2/3}}$} &  7  & 
new  sector & $1_4^6$  \\ \hline  
\hline 
{\large $\frac{R_l   V_R}{  l_p^{9} }$} &
{\large $   \frac{ V_s}{g_{\rm YM}^4} $}  & 1  & new sector    & 7${}_2$  \\   
\hline \hline
{\large $\frac{R_l^3  V_R^2}{  l_p^{18} }$} &
{\large $   \frac{N^{1/2} V_s}{g_{\rm YM}^4} $}  & 1  & new  sector   & 7${}_3$  \\   
\hline 
\end{tabular}
\caption[Additional multiplets for $d=7$]{Additional multiplets for $d=7$: \irrep{56}, \irrep{1} and \irrep{1} of $E_7$.
\label{tra7} }
\end{center}
\end{table}

\clearpage

\end{document}